\documentclass[a4paper,12pt]{article}
\usepackage{graphicx} % Required for inserting images
\usepackage{amsmath}
\usepackage{amssymb}
\usepackage[top=25mm, bottom=25mm, left=15mm, right=15mm]{geometry}
\DeclareMathOperator\Log{Log}
\usepackage{caption}
\usepackage{subcaption}
\usepackage{authblk}
\usepackage{hyperref}
\usepackage[toc,page]{appendix}

\begin{document}

\title{On removing orders from amplitude equations}
\author[*]{David Juhasz}
\author[**]{Per Kristen Jakobsen}
\affil[*]{Department of Mathematics and Descriptive Geometry, Slovak University of Technology, 81005 Bratislava, Slovakia}
\affil[**]{Department of Mathematics and Statistics, the Arctic University of Norway, 9019 Troms\o, Norway}
\date{\today}

\maketitle
\begin{abstract}
In this paper, we introduce a modified version of the renormalization group (RG) method and test its numerical accuracy. It has been tested on numerous scalar ODEs and systems of ODEs. Our method is primarily motivated by the possibility of simplifying amplitude equations. The key feature of our method is the introduction of a new homogeneous function at each order of the perturbation hierarchy, which is then used to remove terms from the amplitude equations. We have shown that there is a limit to how many terms can be removed, as doing so beyond a certain point would reintroduce linear growth. There is thus a \textit{core} in the amplitude equation, which consists of the terms that cannot be removed while avoiding linear growth. Using our modified RG method, higher accuracy can also be achieved while maintaining the same level of complexity in the amplitude equation.
\end{abstract}

\section{Introduction}
Studying complex systems has been a major area of interest in mathematics. The most frequent questions addressed concern the long-term behavior and stability of the system, as well as how the behavior of one part affects other parts. Complex systems form and evolve through processes that require descriptions using multiple scales and mathematical tools such as scaling laws.

One broad family of mathematical tools based on different scales is perturbation methods. Nearly all perturbation methods share a common origin, as they were developed with the motivation to solve perturbed nonlinear equations whose solutions vary on different scales. However, one perturbation method, called renormalization, stands out. The origins of renormalization can be traced back to the 1930s and issues in quantum electrodynamics. At that time, computational methods struggled with handling  interactions of photons with extremely high momenta.

%Feynman in 1948 introduced a path integral formulation and Feynman diagrams that made the calculations easier. Particularly the diagrams can be used to help calculate the terms in perturbation expansions. However, it was discovered and later in the 1940's described (ref), that in the perturbative corrections many integrals were divergent.

%In interactions involving photons, vacuum polarizations are always present. These polarizations are due to virtual particles popping in and out of existence carrying energy. They are thus creating so called loops. The energies and momenta of incoming and outgoing particles do not uniquely determine the momentum of the particles involved in the loop. The energy of a loop-particle can be balanced by a change in the energy of another loop-particle while not affecting the incoming and outgoing particles. Depending on the number of particles, many variations are possible. In order to find the amplitude of the loop process, we have to integrate over all possible combinations of energy and momentum that could be involved in the loop. Feynman in 1948 introduced a path integral formulation and Feynman diagrams that made the calculations easier. Particularly the diagrams can be used to help calculate the terms in perturbation expansions. However, it was discovered and later in the 1940's described (ref), that in the perturbative corrections many integrals were divergent.

Around 1950, a breakthrough emerged, providing a solution for eliminating infinities, thanks to Julian Schwinger, Richard Feynman, Freeman Dyson, and Shinichiro Tomonaga. The key realization was that the quantities representing the electron's charge and mass, or the normalization constants of the quantum field appearing in the formulas, did not correspond to the physical constants measured in the laboratory. The main idea was to replace the calculated values, which could be infinite, with their finite, measured counterparts. The non-physical quantities, known as bare quantities, did not account for the effects of loop particles. These effects, like the integrals themselves, would be divergent, meaning that the measured finite quantities implied divergent bare quantities. The systematic procedure of replacing divergent quantities became known as renormalization and can be applied to arbitrary orders in the perturbation expansion \cite{RG1}.

%As a prime example for introducing renormalization of the bare quantities, a groundbreaking paper by Francis Low and Murray Gell-Mann from 1954 (ref) should be mentioned. It suffered a long period of neglect at first, from the year it was written until the early 1970s. It started by considering the old problem of Coulomb's force between two charges and asked how this force behaves at very short distances. If one goes to very high momentum transfer, the mass of the electron should become irrelevant which would imply that the potential should go as the reciprocal of length $1/r$. In other words, the authors realized that there is a scale variance in quantum field theory that is broken by particle masses which should be negligible at very high energy or very short distances. In the calculated radiative correction of the potential due to the electron loop in the photon exchange, the correction terms diverges as we let the electron mass go to zero. But in the corresponding Feynman diagram, the momentum transfer provides an infrared cutoff and there can be no singularity for zero electron mass. It was realized that the only thing breaking the scale invariance is the renormalization procedure. 

%We are dealing with a Lagrangian or a set of field equations with a finite number of constants where all the infinities can always be absorbed into a redefinition of the constants.

The first book to mention the issue of renormalization was written by Bogoliubov and Shirkov \cite{Bogoliubov}. In particular, it addressed the concept of invariance concerning the choice of normalization for the charge. The term "renormalization group" was introduced to express this invariance.

Since then, the renormalization group (RG) method has been used primarily in connection with quantum field theory \cite{Wilson1974}, \cite{Wilson1975}, \cite{Kadanoff} until the late 1980s, when Goldenfeld, Martin, and Oono \cite{Goldenfeld1989} found a way to extend RG ideas to the solutions of differential equations. Problems that were previously solved using various perturbation methods, such as WKB, matched asymptotic expansions, and multiple scales, have since been addressed using the RG method as well. It has been applied to problems in fluid dynamics \cite{RGinFluid}, general relativity \cite{RGinRelativity}, and kinetic processes \cite{RGinKinetic}, marking significant progress given its relatively recent development. Equally interesting is the relationship between the RG method and other techniques, such as the multiple scales method and center manifold theory \cite{RGonMMS}, \cite{RGonCMT}.

%In their paper (ref IA 1989), Goldenfeld et. al., intermediate asymptotics is used and the renormalization is performed through the introduction of phenomenological parameters. The inspiration came from Wilson as a set of transformations that alter only the microscopic parameters is considered, where the macroscopic universal features are unaltered. The argument is that one can obtain universal relations between phenomenological parameters if one can absorb the changes caused by the modification of microscopic parameters into the phenomenological parameters. If this is possible to do, then the system is renormalizable.  In some cases, the macroscopic description is insensitive to the microscopic details. Hence, there is no unique microscopic picture that gives rise to the same macroscopic description. As a consequence, one can choose the simplest microscopic model (or minimal model) to begin with. While Wilson was more concerned with reducing the degrees of freedom by a transformation in space, Goldenfeld introduced the same concept in time. The presence of a new phenomenological parameter in the transformation, however, presented a problem since the changing parameter should not change the macroscopic observables because it is introduced independent of the microscopic model. Thus, to express the invariance of the physics, this issue is remedied  by taking the derivative with respect to the phenomenological parameter and set it equal to zero. It is called the renormalization group equation. Solving this equation determines the macroscopic model.

The discussion in Goldenfeld’s paper \cite{Goldenfeld1989} demonstrates that the RG method is not restricted to field-theoretical problems. Its equivalence to the theory of intermediate asymptotics enables its application to partial differential equations. The full application of the original renormalization ideas to solving differential equations was introduced by Chen, Goldenfeld, and Oono \cite{CGO1994}, \cite{CGO1995}, \cite{CGO1996}.

Borrowing from conventional perturbation methods, the starting point of the RG method is primarily the removal of divergences from a perturbation series by replacing integration constants with slowly varying functions, resulting in a renormalized expansion. This aspect is inspired by its original purpose in quantum field theory, where it was used to eliminate divergences. The RG equation is interpreted as the amplitude equation.

One of the key features of the RG method is that it provides a unified approach, in contrast to the many different perturbation techniques. Another advantage is its starting point: a straightforward, naïve perturbation expansion that requires minimal prior knowledge. Its clear and systematic methodology allows for closed-form expressions of both the amplitude equation and the sought-after asymptotic solution.

%Given the mentioned similarities with the other perturbation methods, a natural question arises: what is the relation, if any, between conventional asymptotic methods such as the multiple scale method and the RG? In this paper, we try to answer (or at least partially) this question.

In this paper, we examine the RG method and the process of obtaining the naïve perturbation. When solving a perturbation hierarchy, only the particular solution is typically found at each order, while the homogeneous solution is ignored. This might be understandable, as extra homogeneous solutions would only add constants to the already existing ones from the 0th order. However, in our method, we realized that introducing homogeneous solutions at each order provides more freedom and flexibility, which can be used to achieve better results. These new constants are replaced by their renormalized counterparts and treated as functions, just like the main amplitudes. This approach results in new functions at each order, which can be used to remove terms from the amplitude equations. For each example, we will demonstrate that higher precision in the RG solution can be achieved while keeping the amplitude equation at the same level of complexity.

In the classical RG method, or any other perturbation method for that matter, achieving higher precision in the perturbation solution would mean adding more and more terms to the amplitude equation, making it more complex. With our new modification, the amplitude equation can remain the same while achieving higher precision. %Only the perturbation solution becomes more complex, but what we care about is the amplitude equation.

However, there is a limit to how much one can simplify the amplitude equation. In theory, it is possible to remove all orders, leaving only the fastest oscillating term $A'(t)=iA(t)$, with a solution $A=A_0e^{it}$. As shown in this paper, this leads to a problem. During the process of removing orders, the new homogeneous functions contain a term proportional to $\Log(A)$, which introduces a linearly growing term. %One can avoid this growth by not removing the next lowest-order term, thus avoiding an amplitude equation of the form $A'(t)=iA(t)$.
This suggests that there is a lowest nonlinear order in the amplitude equation that cannot be removed. We call this the \textit{core} of the amplitude equation. Sometimes, the same nonlinear terms found in the core can appear in higher-order terms, as in the case of the Van der Pol oscillator. In such cases, only the constant factor of the core changes, but no additional nonlinear terms are added.

Our defined "core" corresponds to the unique or hypernormal form of the system, obtained here constructively via a modified renormalization group (RG) approach. Classical normal form theory is used to study complex dynamical systems by applying near-identity coordinate transformations to remove non-essential terms, thereby simplifying the underlying equations. By stripping away terms that do not fundamentally alter the qualitative character of the dynamics, one is left with a normal form—a minimal, standard set of equations representing the system's behavioral class. A foundational result linking these methodologies is provided by DeVille et al. \cite{deVille}, who proved that the amplitude equations generated by the RG method are equivalent to classical Poincaré–Birkhoff normal forms. They demonstrated that the reductive step in the RG protocol removes non-resonant terms from the expansion in the exact same manner that normal-form coordinate changes eliminate non-resonant terms from a vector field. Within this context, our "core" consists precisely of the surviving resonant terms. In this paper, the "core" is defined as the lowest nonlinear term that cannot be removed without reintroducing $\Log(A)$ (linear growth). Since we identify this with the surviving resonant term and with the hypernormal form, one could state that "removable" means "non-resonant" in this construction. 

Another work that explicitly links normal form theory with renormalization group flows is that of Raju et al. \cite{Raju}. Utilizing polynomial coordinate transformations, the authors systematically eliminate as many nonlinear terms as possible directly from the RG flow equations, noting that non-removable terms yield universal logarithmic corrections. This represents the exact same mechanism underlying our core, albeit framed within the RG-flow setting. More recently, Kuniba and Motohashi \cite{Kuniba} focused on the exact algebraic and functional structures of the standard RG expansion to prove the absence of secular terms to all orders. While their approach follows a distinct mathematical trajectory, it shares a highly overlapping research territory with our goals.

The observation that algebraic simplification halts at an irreducible remainder is the foundational premise of hypernormal (or simplest) normal form theory. Because classical normal form transformations do not necessarily remove all mathematically non-essential terms, the resulting amplitude equation is often not the simplest achievable representation. If the kernel of the associated homological operator is non-trivial, the homological equation admits infinitely many solutions parameterized by arbitrary constants. These remaining free parameters can then be exploited to achieve additional simplifications in higher-order terms—a mechanism that directly mirrors our introduction of arbitrary homogeneous functions. Murdock \cite{Murdock} thoroughly investigates hypernormal forms, utilizing these free parameters to remove as many higher-order terms as mathematically possible. While conceptually aligned with our objective, Murdock's framework relies strictly on Lie algebras and homological equations. Thus, finding a "core" in our amplitude equation serves as the direct RG analogue to computing a hypernormal form. Similarly, Algaba et al. \cite{Algaba} seek a minimal "core" expression of a system; however, their techniques rely on quasi-homogeneous expansions and orbital equivalence within classical normal form theory, rather than a perturbative RG framework.

Frameworks that share our primary motivation—namely, leveraging the RG framework to simplify or reduce amplitude equations—include the works of Kirkinis \cite{Kirkinis1, Kirkinis2}. In his first paper, he demonstrates that the entire process of generating the amplitude equation can be reduced to a purely algebraic equation expressed in terms of Thiele semi-invariants (cumulants). In the follow-up work, this framework is applied to formalize the connections between the RG method and classical reduction techniques, such as the Method of Multiple Scales and the Bogoliubov-Krylov-Mitropolsky (BKM) averaging method. Critically, while Kirkinis utilizes an adapted RG framework to suppress secular and non-resonant terms during the generation of the amplitude equation, our work takes a different step: we introduce a new homogeneous function at each order to actively push the removal of higher-order terms to its absolute physical limit, thereby isolating the "core." In other words, whereas Kirkinis employs the RG method as a tool to yield a standard simplified amplitude equation, our modified protocol uses the RG architecture to systematically simplify the amplitude equation itself.

While our work shares significant conceptual similarities with hypernormal form theory, our methodological approach is fundamentally different. Hypernormal form theory relies on heavy mathematical machinery, utilizing Lie group theory on manifolds and vector spaces, which demands extensive knowledge of differential geometry. In contrast, our approach requires essentially no specialized mathematical background. Furthermore, while hypernormal form theory often draws conclusions in a generalized manner without providing concrete examples or numerical validation, the algebraic approach outlined in this paper prioritizes practicality. By offering specific examples and rigorous numerical assessments, we demonstrate the direct applicability of our method. Finally, the fact that this approach was developed independently—without prior knowledge of normal form theory—underscores that our method is a constructive, algebraic rediscovery of the theory.

We would like to demonstrate our new modified RG method on two types of equations: a scalar ODE and a system of ODEs. The paper is divided into two parts. In the first part, we examine a simple nonlinear second-order scalar equation: the Duffing equation. The second part deals with a systems of nonlinear second-order ODEs, thoroughly demonstrating our modified RG method.

\section{Scalar equations}
\subsection{Duffing equation (initial value problem)}
Consider a cubic oscillator also known as the Duffing equation with the initial conditions
\begin{align}
y''(t)+y(t)&=\varepsilon y^3(t),\quad t>0,\nonumber\\
y(0)&=1,\nonumber\\
y'(0)&=0,\label{eq1}
\end{align}
where $\varepsilon$ is the small perturbation parameter. In this section, we will demonstrate the modification of the RG method that is used for differential equations. Let us introduce some terminology that we will use throughout this paper. The original RG method will be referred to as the \textit{classical RG method}, and the one with our modifications will be referred to as the \textit{modified RG method}. We begin by presenting the solutions to (\ref{eq1}) from both methods in order to highlight the difference. The amplitude equation with the solution from the classical RG method reads
\begin{align}
A'(t)&=-\frac{3}{2} i \varepsilon  A^2A^* -\frac{15}{16} i \varepsilon ^2 A^3 \left(A^*\right)^2-\frac{123}{128} i \varepsilon ^3 A^4\left(A^*\right)^4,\label{eq53}\\
y_{c}(t)&=e^{i t} A-\frac{1}{8} e^{3 i t} \varepsilon  A^3+\varepsilon ^2 \left(\frac{1}{64} e^{5 i t} A^5-\frac{21}{64} e^{3 i t} A^4 A^*\right)\nonumber\\
&+\varepsilon ^3 \left(\frac{43}{512} e^{5 i t} A^6 A^*-\frac{417}{512} e^{3 i t} A^5\left(A^*\right)^2-\frac{1}{512} e^{7 i t} A^7\right)+(*).\label{eq54}
\end{align}
where $(*)$ means the complex conjugate of the terms standing before it. In comparison, the solution produced by the modified RG method is
\begin{align}
A'&=-\frac{3}{2} i \varepsilon  A^2 A^*,\label{eq1.1}\\
y_m(t)&=e^{i t} A+\varepsilon  \left(-\frac{5}{16} e^{i t} A^2A^* -\frac{1}{8} e^{3 i t} A^3\right)+\varepsilon ^2 \left(-\frac{27}{128} e^{3 i t} A^4A^*+\frac{11}{512} e^{i t} A^3\left(A^*\right)^2+\frac{1}{64} e^{5 i t} A^5\right)+\nonumber\\
&+\varepsilon ^3 \left(\frac{61 e^{5 i t} A^6A^*}{1024}-\frac{1419 e^{3 i t} A^5\left(A^*\right)^2}{4096}-\frac{1}{512} e^{7 i t} A^7\right)+(*).\label{eq1.2}
\end{align}
Both solution are valid up to the same order $\varepsilon^3$. The main difference is in the amplitude equation. In the modified RG method, all higher-order terms vanished, while the solution itself became more complicated. However, since what is solved is the amplitude equation, this complication is of less importance. The core of the amplitude equation (\ref{eq1.1}) is clearly identified as the term proportional to $A^2A^*$. This term cannot be removed without introducing linear growth.

Let us demonstrate our modified RG method on the Duffing equation (\ref{eq1}) and derive the solution \ref{eq1.1}), (\ref{eq1.2}). We begin with calculating the naive expansion
\begin{align}
y(t)=y_0(t)+\varepsilon y_1(t)+\varepsilon^2y_2(t)+\varepsilon^3y_3(t)+\ldots.\label{eq2}
\end{align}
Inserting (\ref{eq2}) into (\ref{eq1}), setting the expression at every order in $\varepsilon$ to zero, we get the hierarchy
\begin{align}
\text{order }\varepsilon^0:\;y_0''+y_0&=0,\nonumber\\
\text{order }\varepsilon^1:\;y_1''+y_1&=y_0^3,\nonumber\\
\text{order }\varepsilon^2:\;y_2''+y_2&=3y_0^2y_1,\nonumber\\
\text{order }\varepsilon^3:\;y_3''+y_3&=3 y_0 y_1^2+3 y_0^2 y_2.\label{eq3}
\end{align}
Order $\varepsilon^0$ has the following general solution
\begin{align}
y_0=A_0e^{it}+(*).\label{eq4}
\end{align}
Inserting (\ref{eq4}) into the the order $\varepsilon^1$ equation we obtain
\begin{align}
y_1''+y_1&=3A_0^2A_0^*e^{it}+A_0^3e^{i3t}+(*).\label{eq5}
\end{align}
At this point, conventionally, a particular solution is found to (\ref{eq5}) while ignoring the homogeneous solution. Our modified RG method, however, rests on introducing the homogeneous solution as well. The $e^{it}$ term is a resonant term for which we look for a solution in the form $(a+bt)e^{it}$, where $a$ and $b$ are constants. After some algebra we arrive at the solution
\begin{align}
y_1(t)=\left(\alpha_0-\frac{3i}{2}A_0^2A_0^*t\right) e^{it}-\frac{1}{8}A_0^3e^{i3t}+(*).\label{eq6}
\end{align}
Using both (\ref{eq4}) and (\ref{eq6}) in the order $\varepsilon^2$ equation, we get
\begin{align}
y_2''+y_2&=\left(-\frac{3}{8}A_0^3\left( A_0^*\right)^2-\frac{9}{2}iA_0^3\left( A_0^*\right)^2t+6A_0A_0^*\alpha_0+3A_0^2\alpha_0^*\right)e^{it}\nonumber\\
&+\left(-\frac{3}{4}A_0^4A_0^*-\frac{9}{2}iA_0^4A_0^*t+3A_0^2\alpha_0\right)e^{i3t}-\frac{3}{8}A_0^5e^{i5t}+(*).\label{eq7}
\end{align}
Taking both homogeneous and particular solution to (\ref{eq7}), we have
\begin{align}
y_2(t)&=\left(\beta_0-\left(3iA_0A_0^*\alpha_0+\frac{3}{2}iA^2\alpha_0^*+\frac{15i}{16}A_0^3\left(A_0^*\right)^2\right) t-\frac{9}{8}A_0^3\left(A_0^*\right)^2t^2\right)e^{it}\nonumber\\
&+\left(-\frac{21}{64}A_0^4A_0^*+\frac{9i}{16}A_0^4A_0^*t-\frac{3}{8}A_0^2\alpha_0\right)e^{i3t}+\frac{1}{64}A_0^5e^{i5t}+(*).\label{eq8}
\end{align}
We will do one more order. The $\varepsilon^3$ order equations becomes
\begin{align}
y_3''+y_3&=\left(-\frac{57 A_0^4 \left(A_0^*\right)^3}{64}-\frac{3}{4} A_0^3 \left(A_0^*\right) \alpha_0^*-\frac{9}{8} A_0^2 \left(A_0^*\right)^2 \alpha_0+3 A_0^2 \beta_0^*+6 A_0 A_0^* \beta_0+6 A_0 \alpha_0 \alpha_0^*+3 A_0^* \alpha_0^2\right.\nonumber\\
&\left.+t \left(-\frac{9}{4} i A_0^4 \left(A_0^*\right)^3-9 i A_0^3 A_0^* \alpha_0^*-\frac{27}{2} i A_0^2 \left(A_0^*\right)^2 \alpha_0\right)-\frac{27}{8} A_0^4 \left(A_0^*\right)^3 t^2\right)e^{it}\nonumber\\
&+\left(-\frac{81}{8} A_0^5 \left(A_0^*\right)^2 t^2-\frac{123 A_0^5 \left(A_0^*\right)^2}{64}-\frac{3 A_0^4 \alpha_0^*}{4}-3 A_0^3 A_0^* \alpha_0+3 A_0^2 \beta_0+3 A_0 \alpha_0^2\right.\nonumber\\
&\left.+t \left(\frac{9}{16}i A_0^5 \left(A_0^*\right)^2-\frac{9}{2} i A_0^4 \alpha_0^*-18 i A_0^3 A_0^* \alpha_0\right)\right)e^{3it}+\left(-\frac{27 A_0^6 A_0^*}{32}-\frac{15 A_0^4 \alpha_0}{8}+\frac{45}{16} i A_0^6 A_0^* t\right)e^{5it}\nonumber\\
&+\frac{3}{32} A_0^7 e^{7 i t}+(*),\label{eq9}
\end{align}
which results in the solution for this order
\begin{align}
y_3&=\left(\gamma_0+\right.\nonumber\\
&t \left(-\frac{123}{128}i A_0^4 \left(A_0^*\right)^3-\frac{15}{8} i A_0^3 A_0^* \alpha_0^*-\frac{45}{16} i A_0^2 \left(A_0^*\right)^2 \alpha_0-\frac{3}{2} i A_0^2 \beta_0^*-3 i A_0 A_0^* \beta_0-3 i A_0 \alpha_0 \alpha_0^*-\frac{3}{2} i A_0^* \alpha_0^2\right)\nonumber\\
&\left.t^2 \left(-\frac{45}{32}  A_0^4 \left(A_0^*\right)^3-\frac{9}{4} A_0^3 A_0^* \alpha_0^*-\frac{27}{8} A_0^2 \left(A_0^*\right)^2 \alpha_0\right)+\frac{9}{16} i A_0^4 \left(A_0^*\right)^3 t^3\right)e^{it}+\nonumber\\
&\left(-\frac{417}{512} A_0^5 \left(A_0^*\right)^2-\frac{21 A_0^4 \alpha_0^*}{64}-\frac{21}{16} A_0^3 A_0^* \alpha_0-\frac{3 A_0^2 \beta_0}{8}-\frac{3 A_0 \alpha_0^2}{8}\right.\nonumber\\
&\left. +t \left(\frac{117}{64} i A_0^5 \left(A_0^*\right)^2+\frac{9}{16} i A_0^4 \alpha_0^*+\frac{9}{4} i A_0^3 A_0^* \alpha_0\right)+\frac{81}{64} A_0^5 \left(A_0^*\right)^2 t^2\right)e^{3it}\nonumber\\
&+\left(\frac{43 A_0^6 A_0^*}{512}+\frac{5 A_0^4 \alpha_0}{64}-\frac{15}{128} i A_0^6 A_0^* t\right)e^{5it}-\frac{1}{512} A_0^7 e^{7 i t}+(*)\label{eq10}
\end{align}

The renormalization group method is based on introducing an arbitrary starting time $t_0$ at which the solution starts, i.e. $t\to t-t_0$. The overall naive solution to the Duffing equation up to order $\varepsilon^3$ is then
\begin{align}
y(t)&=y_0(t-t_0)+\varepsilon y_1(t-t_0)+\varepsilon^2 y_2(t-t_0)+\varepsilon^3 y_3(t-t_0).\label{eq11}
\end{align}
This naive perturbation breaks down for $\varepsilon (t-t_0)>1$ where the ordering of the terms is disrupted. To regularize the series (\ref{eq11}) we also introduce the time $\mu$ and split the interval $t-t_0$ as $t-\mu+\mu-t_0=\tau+\xi$, where we denoted $\tau=t-\mu$ and $\xi=\mu-t_0$. Usually, the constants $A_0,A_0^*$ are renormalized, but since homogeneous solutions were introduced, we need to include them in this process as well. Thus all the constants $A_0,\alpha_0,\beta_0,\gamma_0$ and their conjugate counterparts are renormalized using a multiplicative renormalization constants $Z,U,V,W$, respectively, such as
\begin{align}
A_0(t_0)=\sum_{n=0}^\infty Z_n(\xi,\mu)\varepsilon^nA(\mu)= Z(\xi,\mu)A(\mu),\label{eq12}\\
\alpha_0(t_0)=\sum_{n=0}^\infty U_n(\xi,\mu)\varepsilon^n\alpha(\mu)= U(\xi,\mu)\alpha(\mu),\label{eq13}\\
\beta_0(t_0)=\sum_{n=0}^\infty V_n(\xi,\mu)\varepsilon^n\beta(\mu)= V(\xi,\mu)\beta(\mu),\label{eq14}\\
\gamma_0(t_0)=\sum_{n=0}^\infty W_n(\xi,\mu)\varepsilon^n\gamma(\mu)= W(\xi,\mu)\gamma(\mu),\label{eq15}
\end{align}
where $Z(\xi,\mu)=\sum_{n=0}^\infty Z_n(\xi,\mu)\varepsilon^n$ etc. The terms containing $\xi$ will then be absorbed into the renormalized terms $A(\mu),\alpha(\mu),\beta(\mu),\gamma(\mu)$ through the quantities $Z_n,U_n,V_n$ and $W_n$. In our case, we use the sum in (\ref{eq12}) up to order $\varepsilon^3$, the sum (\ref{eq13}) up to $\varepsilon^2$, (\ref{eq14}) up to $\varepsilon^1$ and in (\ref{eq15}) only the first term. The reason is that for example $\alpha_0$ does not appear in the expansion until $\varepsilon^1$ order. Upon substituting (\ref{eq12})-(\ref{eq15}) into (\ref{eq11}) with their conjugate counterparts, collecting the terms with the same orders of $\varepsilon$, we obtain
\begin{align}
y(t)&=AZ_0e^{i(\tau+\xi)}+\varepsilon\left[AZ_1e^{i(\tau+\xi)}-\frac{3i}{2}A^2A^*Z_0^2Z_0^*(\tau-\xi)e^{i(\tau+\xi)}-\frac{1}{8}A^3Z_0^3e^{i3(\tau+\xi)}+U_0 \alpha e^{i (\xi +\tau )}\right]\nonumber\\
&+\ldots+(*),\label{eq16}
\end{align}
where the $\varepsilon^2$ and $\varepsilon^3$ order was omitted due to its length but we get back to it later. One of the goals is to choose the quantities $Z_{n},U_n,V_n$ and $W_n$ such that the solution $y(t)$ is $\xi$-free. Starting with the 0-th term, choosing $Z_0=e^{-i\xi}$ eliminates the $\xi$ from it. Likewise we get $Z^*_0=e^{i\xi}$ from its complex conjugate form. Using the newly calculated $Z_0$ in the $\varepsilon$ order, we get
\begin{align}
-\frac{3}{2} i e^{i \tau } \xi  A^2 A^*-\frac{3}{2} i e^{i \tau } \tau  A^2 A^*-\frac{1}{8} e^{3 i \tau } A^3+Z_1 A e^{i \tau +i \xi }+U_0 \alpha e^{i (\xi +\tau )}.\label{eq17}
\end{align}
In order to remove the $\xi$ variable from this expression, we must choose $Z_1=\frac{3}{2} i e^{-i \xi } \xi  A A^*$ and $U_0=e^{-i\xi}$. We then proceed the same way for the $\varepsilon^2$ order term, which, upon using $Z_0,Z_1$ turns to
\begin{align}
&-\frac{9}{8} e^{i \tau } \tau ^2 A^3 \left(A^*\right)^2+\frac{9}{8} e^{i \tau } \xi ^2 A^3 \left(A^*\right)^2-\frac{15}{16} i e^{i \tau } \xi  A^3 \left(A^*\right)^2-\frac{21}{64} e^{3 i \tau } A^4 A^*+\frac{9}{16} i e^{3 i \tau } \tau  A^4 A^*\nonumber\\
&-\frac{15}{16} i e^{i \tau } \tau  A^3 \left(A^*\right)^2+\frac{1}{64} e^{5 i \tau } A^5+Z_2 A e^{i \tau +i \xi }-3 i \xi  e^{i \tau } A A^* \alpha-3 i e^{i \tau } \tau  A A^* \alpha-\frac{1}{8} 3 e^{3 i \tau } A^2 \alpha-\frac{3}{2} i \xi  e^{i \tau } A^2 \alpha^*\nonumber\\
&-\frac{3}{2} i e^{i \tau } \tau  A^2 \alpha^*+U_1 \alpha e^{i \xi +i \tau }+V_0 \beta e^{i \xi +i \tau }.\label{eq18}
\end{align}
With $Z_2=\frac{15}{16} i e^{-i \xi } \xi  A^2 \left(A^*\right)^2-\frac{9}{8} e^{-i \xi } \xi ^2 A^2 \left(A^*\right)^2$, $U_1=3 i e^{-i \xi } \xi  A A^*+\frac{3 i e^{-i \xi } \xi  A^2 \alpha^*}{2 \alpha}$ and $V_0=e^{-i\xi}$ we get rid of the $\xi$ variable in the $\varepsilon^2$ order term. Here we note, that one can remove the $\xi$-terms in more than one way. For example the choice
\begin{align}
Z_2&=-\frac{9}{8} e^{-i \xi } \xi ^2 A^2 \left(A^*\right)^2+\frac{15}{16} i e^{-i \xi } \xi  A^2 \left(A^*\right)^2+\frac{3}{2} i e^{-i \xi } \xi  A \alpha+3 i e^{-i \xi } \xi  A^* \alpha,\nonumber\\
U_1&=0,\nonumber\\
V_0&=e^{-i\xi},\label{eq19}
\end{align}
also leads to the removal of the $\xi$-terms. This process is repeated also for the $\varepsilon^3$ terms where choose
\begin{align}
Z_3&=-\frac{9}{16} i e^{-i \xi } \xi ^3 A^3 \left(A^*\right)^3-\frac{45}{32} e^{-i \xi } \xi ^2 A^3 \left(A^*\right)^3+\frac{123}{128} i e^{-i \xi } \xi  A^3 \left(A^*\right)^3,\nonumber\\
U_2&=-\frac{9 e^{-i \xi } \xi ^2 A^3 A^* \alpha^*}{4 \alpha}+\frac{15 i e^{-i \xi } \xi  A^3 A^* \alpha^*}{8 \alpha}-\frac{27}{8} e^{-i \xi } \xi ^2 A^2 \left(A^*\right)^2+\frac{45}{16} i e^{-i \xi } \xi  A^2 \left(A^*\right)^2\nonumber\\
&+3 i e^{-i \xi } \xi  A \alpha^*+\frac{3}{2} i e^{-i \xi } \xi  A^* \alpha,\nonumber\\
V_1&=3 i e^{-i \xi } \xi  A A^*+\frac{3 i e^{-i \xi } \xi  A^2 \beta^*}{2 \beta},\nonumber\\
W_0&=e^{i\xi}.\label{eq20}
\end{align}
Using all the obtained renormalization constants and their complex conjugates in (\ref{eq16}), we get
\begin{align}
y(t)&=e^{i \tau } A+\varepsilon  \left(-\frac{3}{2} i e^{i \tau } \tau  A^2 A^*-\frac{1}{8} e^{3 i \tau } A^3+e^{i \tau } \alpha\right)\nonumber\\
&+\varepsilon ^2 \left(-\frac{9}{8} e^{i \tau } \tau ^2 A^3 \left(A^*\right)^2-\frac{21}{64} e^{3 i \tau } A^4 A^*+\frac{9}{16} i e^{3 i \tau } \tau  A^4 A^*-\frac{15}{16} i e^{i \tau } \tau  A^3 \left(A^*\right)^2+\frac{1}{64} e^{5 i \tau } A^5\right.\nonumber\\
&\left.-3 i e^{i \tau } \tau  A A^* \alpha-\frac{3}{8} e^{3 i \tau } A^2 \alpha-\frac{3}{2} i e^{i \tau } \tau  A^2 \alpha^*+e^{i \tau } \beta\right)+\nonumber\\
&\varepsilon^3\left[e^{i \tau } \gamma-\frac{21}{16} e^{3 i \tau } A^3 A^* \alpha+\frac{9}{16} i e^{i \tau } \tau ^3 A^4 \left(A^*\right)^3+\frac{43}{512} e^{5 i \tau } A^6 A^*-\frac{417}{512} e^{3 i \tau } A^5 \left(A^*\right)^2+\frac{5}{64} e^{5 i \tau } A^4 \alpha\right.\nonumber\\
&\left.-\frac{3}{8} e^{3 i \tau } A \alpha^2-\frac{21}{64} e^{3 i \tau } A^4 \alpha^*-\frac{1}{512} e^{7 i \tau } A^7+\tau  \left(\frac{9}{4} i e^{3 i \tau } A^3 A^* \alpha-\frac{45}{16} i e^{i \tau } A^2 \left(A^*\right)^2 \alpha-\frac{15}{8} i e^{i \tau } A^3 A^* \alpha^*\right.\right.\nonumber\\
&\left.\left.-3 i e^{i \tau } A A^* \beta-\frac{15}{128} i e^{5 i \tau } A^6 A^*+\frac{117}{64} i e^{3 i \tau } A^5 \left(A^*\right)^2-\frac{123}{128} i e^{i \tau } A^4 \left(A^*\right)^3-3 i e^{i \tau } A \alpha \alpha^*+\frac{9}{16} i e^{3 i \tau } A^4 \alpha^*\right.\right.\nonumber\\
&\left.\left.-\frac{3}{2} i e^{i \tau } A^2 \beta^*-\frac{3}{2} i e^{i \tau } A^* \alpha^2\right)+\tau ^2 \left(-\frac{27}{8} e^{i \tau } A^2 \left(A^*\right)^2 \alpha-\frac{9}{4} e^{i \tau } A^3 A^* \alpha^*+\frac{81}{64} e^{3 i \tau } A^5 \left(A^*\right)^2\right.\right.\nonumber\\
&\left.\left.-\frac{45}{32} e^{i \tau } A^4 \left(A^*\right)^3\right)\right]+(*).\label{eq21}
\end{align}
Observe that $\tau$ is defined as $\tau=t-\mu$. The original problem does not include the variable $\mu$, so the solution should not depend on $\mu$. The amplitude equation is derived by making sure that the solution (\ref{eq21}) does not depend on $\mu$. This is achieved by setting the partial derivative of (\ref{eq21}) with respect to $\mu$ equal to zero and letting $\mu\to t$ (or $\tau\to 0$). We observe that the $\mu$-derivative of the solution will contain the derivatives of the free amplitudes $\alpha,\beta$ and $\gamma$. We would like to obtain one amplitude equation for $A$, therefore we specify the new amplitudes as being functions directly of the amplitude $A$ and $A^*$: $\alpha(\mu)=\alpha(A(\mu),A^*(\mu))$, $\beta(\mu)=\beta(A(\mu),A^*(\mu))$, $\gamma(\mu)=\gamma(A(\mu),A^*(\mu))$. The amplitude equation then takes the form
\begin{align}
\frac{\partial y}{\partial\mu}&=A' e^{i (t-\mu )}-i A e^{i (t-\mu )}+\varepsilon  \left(-i e^{i (t-\mu )} \alpha+e^{i (t-\mu )} \left(\left(A^*\right)'\partial_{A^*}\alpha+A'\partial_{A}\alpha \right)+\frac{3}{8} i A^3 e^{3 i (t-\mu )}\right.\nonumber\\
&\left.-\frac{3}{8} A^2 A' e^{3 i (t-\mu )}-\frac{3}{2} i A^2 \left(A^*\right)' e^{i (t-\mu )} (t-\mu )+\frac{3}{2} i A^* A^2 e^{i (t-\mu )}-\frac{3}{2} A^* A^2 e^{i (t-\mu )} (t-\mu )\right.\nonumber\\
&\left.-3 i A^* A A' e^{i (t-\mu )} (t-\mu )\right)+\varepsilon ^2 \left(\frac{1}{64} (-5) e^{5 i (t-\mu )} i A^5+\frac{27}{64} e^{3 i (t-\mu )} i A^* A^4+\frac{27}{16} e^{3 i (t-\mu )} (t-\mu ) A^* A^4\right.\nonumber\\
&\left.+\frac{5}{64} e^{5 i (t-\mu )} A' A^4-\frac{21}{64} e^{3 i (t-\mu )} \left(A^*\right)' A^4+\frac{9}{16} e^{3 i (t-\mu )} i (t-\mu ) \left(A^*\right)' A^4+\frac{9}{8} e^{i (t-\mu )} i (t-\mu )^2 \left(A^*\right)^2 A^3\right.\nonumber\\
&\left.+\ldots\right)+\mathcal{O}(\varepsilon^3)+(*)=0.\label{eq22}
\end{align}
Before we let $\mu\rightarrow t$ we need to discuss how to extract a single amplitude equation for $A$ from the equation above. The derivative of $A$ and $A^*$ appear in multiple places, along with exponential terms and powers of $(t-\mu)^n$ for various $n$. These terms are linearly independent, which implies that all coefficients multiplying these distinct terms must vanish individually. This gives rise to a collection of equations involving various powers of $\varepsilon$. For instance, the coefficient of $e^{ i (t-\mu )}$  includes terms proportional to $(t-\mu)^0,(t-\mu),(t-\mu)^2$ and $(t-\mu)^3$. Each of these must vanish order-by-order in $\varepsilon\rightarrow 0$ up to the relevant level of approximation.

Let us focus on the coefficient that does not vanish in the limit $\varepsilon\rightarrow 0$. By setting this term to zero, we can isolate an expression for $A'(t)$. Any denominator appearing in this expression can be expanded in $\varepsilon$ using Taylor series. This results in a single amplitude equation for $A(t)$ that is valid up to a specified order in $\varepsilon$.

After substituting this expression for $A'$ back into the remaining terms of the equation, they will all turn out to be of higher order in $\varepsilon$, and hence vanish more rapidly as $\varepsilon\rightarrow 0$. This process confirms the independence of the various terms in equation (\ref{eq22}).

In the case of oscillatory behavior of the solution, the detailed process of isolating the amplitude equation by exploiting the independence of multiple terms becomes unnecessary. It turns out that the complex conjugate pairs in the expression can be treated as independent components. By setting the coefficients of each pair to zero, we recover the same amplitude equation as before. Therefore, we can directly let $\mu\to t$ in (\ref{eq22}), yielding
\begin{align}
\left.\frac{\partial y}{\partial\mu}\right|_{\mu=t}&=A'(t)-i A(t)+\varepsilon\left(A'\partial_{A}\alpha+A^{*'} \partial_{A^*}\alpha-i \alpha-\frac{3}{8} A^2 A'+\frac{3}{2} i A^2 A^*+\frac{3}{8} i A^3\right)\nonumber\\
&+\varepsilon^2\left(A' \partial_{A}\beta+A^{*'} \partial_{A^*}\beta-i \beta-\frac{3}{8} A^2 \left(A' \partial_{A}\alpha+A^{*'} \partial_{A^*}\alpha\right)-\frac{3}{4} A A' \alpha+\frac{9}{8} i A^2 \alpha+3 i A A^* \alpha\right.\nonumber\\
&\left.+\frac{3}{2} i A^2 \alpha^*-\frac{21}{16} A^3 A^* A'+\frac{5}{64} A^4 A'-\frac{21}{64} A^4 A^{*'}+\frac{27}{64} i A^4 A^*+\frac{15}{16} i A^3 A^{*2}-\frac{5}{64} i A^5\right)+\nonumber\\
&+\varepsilon^3\left(A'\partial_{A}\gamma+A^{*'} \partial_{A^*}\gamma-i \gamma+\ldots\right)+(*),\label{eq22a}
\end{align}
where the rest of the terms in $\varepsilon^3$ order were omitted. Observe that we only need to set the first part of (\ref{eq22a}) to zero, since its complex conjugate then automatically becomes zero. We see that the terms $A',A^{*'}$ appear only in a linear way. Setting both parts of (\ref{eq22a}) to zero, it is therefore possible to obtain a 2x2 system of linear equations for the unknowns $A',A^{*'}$ (the other equation is the complex conjugate of the first part). This system can be solved for the derivatives. The only issue with the solutions for the derivatives is that they include the determinant of the matrix in their denominator in the form $1/(c+x)$, where $x$ are the terms at least of order $\varepsilon$ and $c$ is some constant. To remove the denominator, we can use Taylor series and keep the terms up to $\varepsilon^3$. Doing so, we arrive at
\begin{align}
A'(t)&=iA+\varepsilon\left(-i A\partial_{A}\alpha +i A^* \partial_{A^*}\alpha+i \alpha-\frac{3}{2} i A^2 A^*\right)\nonumber\\
&+\varepsilon^2\left(-i A \partial_{A}\beta+i A^* \partial_{A^*}\beta+i \beta + i A \partial_{A^*}\alpha \partial_{A}\alpha^*-i A^* \partial_{A^*}\alpha \partial_{A^*}\alpha^*+i \partial_{A^*}\alpha \alpha^*+\frac{3}{2} i A^2 A^*\partial_{A}\alpha\right.\nonumber\\
&\left.+i A (\partial_{A}\alpha)^2-\frac{3}{2} i A A^{*2} \partial_{A^*}\alpha-i \alpha \partial_{A^*}\alpha-i A^* \partial_{A^*}\alpha \partial_{A}\alpha-3 i A A^* \alpha-\frac{3}{2} i A^2 \alpha^*-\frac{15}{16} i A^3 A^{*2}\right)+\nonumber\\
&+\varepsilon^3\left(-i A \partial_{A}\gamma+i A^* \partial_{A^*}\gamma+i \gamma+\ldots\right),\label{eq23}
\end{align}
where the rest of the terms in $\varepsilon^3$ order were omitted.

The overall solution is obtained from (\ref{eq21}) letting $\mu\to t$, or in other words $\tau\to 0$. The result then reads
\begin{align}
y(t)&=A+\varepsilon  \left(\alpha(A,A^*)-\frac{A^3}{8}\right)+\varepsilon ^2 \left(-\frac{3}{8} A^2 \alpha(A,A^*)+\beta(A,A^*)-\frac{21}{64} A^4 A^*+\frac{A^5}{64}\right)+\nonumber\\
&+\varepsilon ^3 \left(\frac{5}{64} A^4 \alpha(A,A^*)-\frac{21}{16} A^3 A^* \alpha(A,A^*)-\frac{3}{8} A \alpha^2(A,A^*)-\frac{21}{64} A^4 \alpha^*(A^*,A)-\frac{3}{8} A^2 \beta(A,A^*)\right.\nonumber\\
&\left.+\gamma(A,A^*)+\frac{43}{512} A^6 A^*-\frac{417}{512} A^5 \left(A^*\right)^2-\frac{1}{512} A^7\right)+(*).\label{eq24}
\end{align}

We have arrived at the RG method solution extended by arbitrary functions that came from the homogeneous solutions. If they were set to zero, we would get the original, unaltered solution. It is interesting to realize that, theoretically, these functions can be set to be anything as long as they don't break the order of the expansion. Also, the assumption that they depend explicitly on $A$ and $A^*$ was made by choice. In its most general form, these functions may only depend on $t$. Each suitable choice produces a different RG solution. This suggests that one equation can have a whole space of expansions, each equally valid. Of course, one may investigate the error each of them leads to, compared to the exact solution.

One might think of a number of ways of how to take advantage of these arbitrary functions. The most simple way is to use them to remove orders from the solution (\ref{eq24}). This can be done easily, since each order except the 0-th includes a new function. This will make sure that every order is removed. For example choosing $\alpha(A,A^*)=A^3/8$ removes the $\varepsilon^1$ order and $\beta(A,A^*)=\frac{21}{64} A^4 A^*+\frac{A^5}{32}$ removes the $\varepsilon^2$ order and so so on. This would then lead to a more complicated amplitude equation until the solution would be the amplitude itself $y(t)=A(t)$.

In this paper, we will take advantage of the functions $\alpha,\beta,\gamma$ in a different way. We are going to use them in the amplitude equation (\ref{eq23}) in order to remove orders. This is much more efficient as it reduces the computation time while keeping the order of the expansion. The solution is allowed to get more complicated while the amplitude equation is simplified significantly. Let us start with the $\varepsilon^1$ order in (\ref{eq23}). We set
\begin{align}
-i A\partial_{A}\alpha +i A^* \partial_{A^*}\alpha+i \alpha-\frac{3}{2} i A^2 A^*=0.\label{eq25}
\end{align}
This is a quasi-linear ODE which can be solved using the method of characteristics. Thus we are looking for curves $\Gamma(s)=(A(s),A^*(s))$ whose velocity at each point is equal to $\mathbf{f}(A,A^*)=(-iA,iA^*)$. This is true if
\begin{align}
\frac{\mathrm{d}}{\mathrm{d}s}A(s,\tau)&=-iA, A(0,\tau)=a(\tau),\label{eq26}\\
\frac{\mathrm{d}}{\mathrm{d}s}A^*(s,\tau)&=iA^*,A^*(0,\tau)=a^*(\tau),\label{eq27}
\end{align}
where the unknown functions are $A(s,\tau), A^*(s,\tau)$ for some initial conditions $a(\tau)$ and $a^*(\tau)$. Note, that the functions $A,A^*$ do not need to be strictly complex conjugates of each other as the main function of interest is $\alpha$. We are using them as independent variables. The solutions to (\ref{eq26}), (\ref{eq27}) are
\begin{align}
A(s,\tau)&=a(\tau)e^{-is},\label{eq28}\\
A^*(s,\tau)&=a^*(\tau)e^{is}.\label{eq29}
\end{align}
The third equation to solve is
\begin{align}
\frac{\mathrm{d}}{\mathrm{d}s}\alpha(s,\tau)=-i\alpha+\frac{3}{2}iA^2A^*=-i\alpha+\frac{3}{2}ia^2a^*e^{-is}.\label{eq30}
\end{align}
This equation has a homogeneous and a particular solution. The homogeneous solution reads
\begin{align}
\alpha(s,\tau)=b(\tau)e^{-is},\label{eq31}
\end{align}
for some function $b(\tau)$. The right hand side of (\ref{eq30}) includes the homogeneous solution (\ref{eq31}), hence we are looking for a particular solution in the form $\alpha_p=Bse^{-is}$ for some constant $B$. Substituting this function into (\ref{eq30}) we find $B=\frac{3}{2}ia^2a^*$, so that the full solution becomes
\begin{align}
\alpha(s,\tau)=\left[b(\tau)+\frac{3}{2}ia^2(\tau)a^*(\tau)s\right]e^{-is}.\label{eq32}
\end{align}
Following the theory, we need to invert the coordinate transform (\ref{eq28}), (\ref{eq29}) as $A(s,\tau),A^*(s,\tau)\rightarrow s(A,A^*),\tau(A,A^*)$ and use in (\ref{eq32}) to obtain the overall solution to (\ref{eq25}). To eliminate $s$ we multiply (\ref{eq28}) and (\ref{eq29}) to get $AA^*=aa^*$. Taking the inverse, we get
\begin{align}
\tau=\left(aa^*\right)^{-1}\left(AA^*\right)=a_1\left(AA^*\right),\label{eq33}
\end{align}
for some function $a_1$. Taking (\ref{eq28}) and with (\ref{eq33}) we get the other variable $s$.
\begin{align}
A=a\left(a_1\left(AA^*\right)\right)e^{-is}&=a_2\left(AA^*\right)e^{-is},\nonumber\\
\frac{A}{a_2\left(AA^*\right)}&=e^{-is},\nonumber\\
Aa_3\left(AA^*\right)&=e^{-is},\nonumber\\
\Log(A)+\Log\left(a_3\left(AA^*\right)\right)&=-is,\nonumber\\
\Log(A)+a_4\left(AA^*\right)&=-is,\label{eq35}
\end{align}
for some functions $a_{2,3,4}$. With $e^{-is}=A/a(\tau)=A/a_2\left(AA^*\right)$, $AA^*=aa^*$ and (\ref{eq33}), (\ref{eq35}) we finally get
\begin{align}
\alpha(A,A^*)&=\left[b(\tau)+\frac{3}{2}ia^2(\tau)a^*(\tau)s\right]\frac{A}{a(\tau)}=b\left(a_2\left(AA^*\right)\right)\frac{A}{a_2\left(AA^*\right)}+\frac{3}{2}iAa(\tau)a^*(\tau)s\nonumber\\
&=Aa_5\left(AA^*\right)+\frac{3}{2}iA^2A^*s=Aa_5\left(AA^*\right)-\frac{3}{2}A^2A^*\Log(A)-\frac{3}{2}A^2A^*a_4\left(AA^*\right)\nonumber\\
&=Aa_5\left(AA^*\right)-\frac{3}{2}A^2A^*\Log(A)-\frac{3}{2}Aa_6\left(AA^*\right)=c\left(AA^*\right)A-\frac{3}{2}A^2A^*\Log(A),\label{eq36}
\end{align}
for some functions $a_{5,6},c$. We could continue by plugging the above solution into the $\varepsilon^2$ order in (\ref{eq23}) and solve for $\beta(AA^*)$ in a similar fashion, and so on. his could be done in theory, but we must remember that the functions $\alpha,\beta,\ldots$ cannot break the ordering of the terms in the expansion. This is exactly what would happen if we continued, because after removing all orders from the amplitude equations, we end up with $A'=iA$ which gives $A(t)=A_0e^{it}$. The function $\alpha$ would then include the term $\Log(A)=\Log(A_0)+it$, which is a growing term, thus breaking the order and rendering the renormalization useless. It would have been too good to be true, though, if we could remove all the terms from the amplitude equation, essentially reducing the amplitude to a simple oscillating term. The amplitude seems to be protecting itself against being reduced to a simple oscillatory behavior. So, we have learned that we cannot allow $\Log(A)$ in (\ref{eq35}). The solution (\ref{eq35}) is divided into the homogeneous $c\left(AA^*\right)A$ and particular part $\frac{3}{2}A^2A^*\Log(A)$. It is easy to see that the particular part comes from the term $\frac{3}{2}iA^2A^*$ in the equation (\ref{eq25}) which can be written in terms of the homogeneous part choosing $c\left(AA^*\right)=\frac{3}{2}iAA^*$. Thus the term
\begin{align}
\frac{3}{2}iA^2A^*,\label{eq37}
\end{align}
in (\ref{eq25}) is a resonant term which, as we now know, cannot be removed. With this term included, the amplitude is no longer a simple oscillation, thus any possible future $\Log$-s in the free functions $\beta,\gamma$ would not pose a problem anymore. This feature of the amplitude equation will be present in other examples as well. Let us therefore call terms like (\ref{eq37}) in the amplitude equation, \textit{the core}. By core, we mean the first term (in the increasing orders) in the amplitude equation that is nonlinear, essentially causing the amplitude not to be of the form $A_0e^{\lambda t}$, with $\lambda$ being some eigenvalue of the problem.

Keeping the core (\ref{eq37}) in (\ref{eq25}) we will only have
\begin{align}
\alpha\left(AA^*\right)=c\left(AA^*\right)A,\label{eq38}
\end{align}
for some arbitrary function $c$. We will now aspire to remove all the terms in all orders in (\ref{eq23}). The amplitude $A$, which we treat as a variable, appears only in powers or polynomial forms. Inspired by this, we look for the function $c$ of the form
\begin{align}
c\left(AA^*\right)&=\sum_{i=0}^1 c_i\left(AA^*\right)^i=c_0+c_1AA^*,\label{eq39}\\
&\Downarrow\nonumber\\
\alpha\left(AA^*\right)&=c_0A+c_1A^2A^*,\label{eq40}
\end{align}
for some complex constants $c_0,c_1$. We could, in theory, go to arbitrary high power. This power is chosen to be sufficiently high to remove the terms in the $\varepsilon^2$ order of (\ref{eq23}).

The $\varepsilon^2$ order in (\ref{eq23}) then becomes
\begin{align}
-i A \partial_A\beta+i A^* \partial_{A^*}\beta+i\beta +A^3 \left(A^*\right)^2 \left(-3 i\text{Re}[c_1]-\frac{15}{16} i\right)-3 i A^2 A^*\text{Re}[c_0].\label{eq41}
\end{align}
We can use the constants $c_0,c_1$ to remove the terms not containing $\beta$. We get
\begin{align}
\text{Re}[c_1]&=-\frac{5}{16},\nonumber\\
\text{Re}[c_0]&=0.\label{eq42}
\end{align}
The imaginary part of the constants are arbitrary, so we set them to be zero. The function $\alpha$ then takes the form
\begin{align}
\alpha\left(AA^*\right)=-\frac{5}{16}A^2A^*.\label{eq43}
\end{align}
The $\varepsilon^2$ order now reads $-i A \partial_A\beta+i A^* \partial_{A^*}\beta+i\beta$. After setting this expression to zero, it has the same form as (\ref{eq25}) containing $\alpha$. It has then the solution
\begin{align}
\beta(A,A^*)=d\left(AA^*\right)A,\label{eq44}
\end{align}
for some arbitrary function $d$. As before, we are looking for the function $d$ of the form
\begin{align}
d\left(AA^*\right)&=\sum_{i=0}^2 d_i\left(AA^*\right)^i=d_0+d_1AA^*+d_2\left(AA^*\right)^2,\label{eq45}\\
&\Downarrow\nonumber\\
\beta\left(AA^*\right)&=d_0A+d_1A^2A^*+d_2A^3\left( A^*\right)^2.\label{eq46}
\end{align}
The highest power of 2 was chosen such that it is high enough to remove all the terms in the $\varepsilon^3$ order.

The $\varepsilon^3$ order then simplifies to
\begin{align}
-i A\partial_A\gamma+i A^* \partial_{A^*}+i\gamma+A^4\left(A^*\right)^3 \left(\frac{33 i}{512}-3 i\text{Re}[d_2]\right)-3 i A^3\left(A^*\right)^2\text{Re}[d_1]-3 i A^2 A^*\text{Re}[d_0].\label{eq47}
\end{align}
With the following choice of the constants $d_i$ the polynomic terms in $A,A^*$ vanish.
\begin{align}
\text{Re}[d_2]&=\frac{11}{512},\nonumber\\
\text{Re}[d_1]=\text{Re}[d_0]&=0.\label{eq48}
\end{align}
Again, the imaginary parts are set to be zero, since they are arbitrary. The function $\beta$ becomes
\begin{align}
\beta(A,A^*)=\frac{11}{512}A^3\left(A^*\right)^2.\label{eq49}
\end{align}
The $\varepsilon^3$ order turns to the already well-known form of $-i A\partial_A\gamma+i A^* \partial_{A^*}+i\gamma$. The solution is $\gamma(A,A^*)=e\left(AA^*\right)A$, for some function $e$. If we also had the $\varepsilon^4$ order we would proceed in a similar manner, but we terminate our expansion at this point and choose the primitive solution $\gamma=0$.

After the dust has settled down, the amplitude equation reads
\begin{align}
A'(t)=i A-\frac{3}{2} i \varepsilon  A^2 A^*,\label{eq50}
\end{align}
as expected. Note the term $iA$ being of order $\varepsilon^0$ means that $A$ varies fast and solving (\ref{eq50}) numerically could bring about unwanted errors and long running time due to high discretizations. Therefore we introduce the following transform $A=\tilde{A}e^{it}$. The amplitude equation (\ref{eq50}) then turns to
\begin{align}
A'=-\frac{3}{2} i \varepsilon  A^2 A^*,\label{eq51}
\end{align}
where we dropped the tilde signs.

The overall solution (\ref{eq24}) takes a particular form using the newly calculated functions $\alpha,\beta,\gamma$. We also need to do the transform we introduced for the amplitude. The final result then reads
\begin{align}
y(t)&=e^{i t} A+\varepsilon  \left(-\frac{5}{16} e^{i t} A^2A^* -\frac{1}{8} e^{3 i t} A^3\right)+\varepsilon ^2 \left(-\frac{27}{128} e^{3 i t} A^4A^*+\frac{11}{512} e^{i t} A^3\left(A^*\right)^2+\frac{1}{64} e^{5 i t} A^5\right)+\nonumber\\
&+\varepsilon ^3 \left(\frac{61 e^{5 i t} A^6A^*}{1024}-\frac{1419 e^{3 i t} A^5\left(A^*\right)^2}{4096}-\frac{1}{512} e^{7 i t} A^7\right)+(*).\label{eq52}
\end{align}

What we have done is obtain a solution up to order $\varepsilon^3$ using a perturbation method called renormalization with an amplitude equation only up to order $\varepsilon^1$. The precision of the solution is preserved but the amplitude equation remained as simple as possible, which is a significant modification.

\subsubsection{Numerical results}
We present some numerical test results where we compare a high-precision numerical solution to the Duffing equation (\ref{eq1}), the RG solution obtained using the homogeneous functions (modified RG), and the same solution but without the homogeneous functions (classical RG). In this test, the perturbation parameter will be $\varepsilon=0.1$.

\begin{figure}[h]
  \centering
\captionsetup{width=0.85\textwidth}
  \includegraphics[scale=0.65]{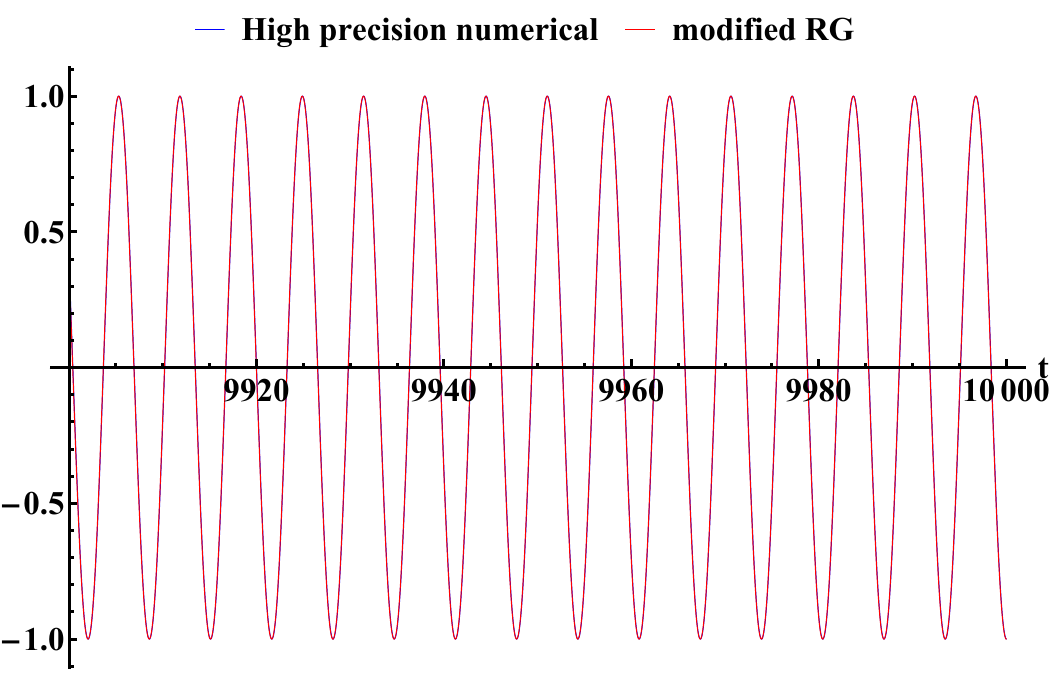}
  \caption{Plots of a high precision numerical solution (\ref{eq1}) and the modified RG solution (\ref{eq51}), (\ref{eq52}) for $t\in[9900,10000]$.}
\label{fig1}
\end{figure}

\begin{figure}[h]
  \centering
\captionsetup{width=0.85\textwidth}
  \includegraphics[scale=0.65]{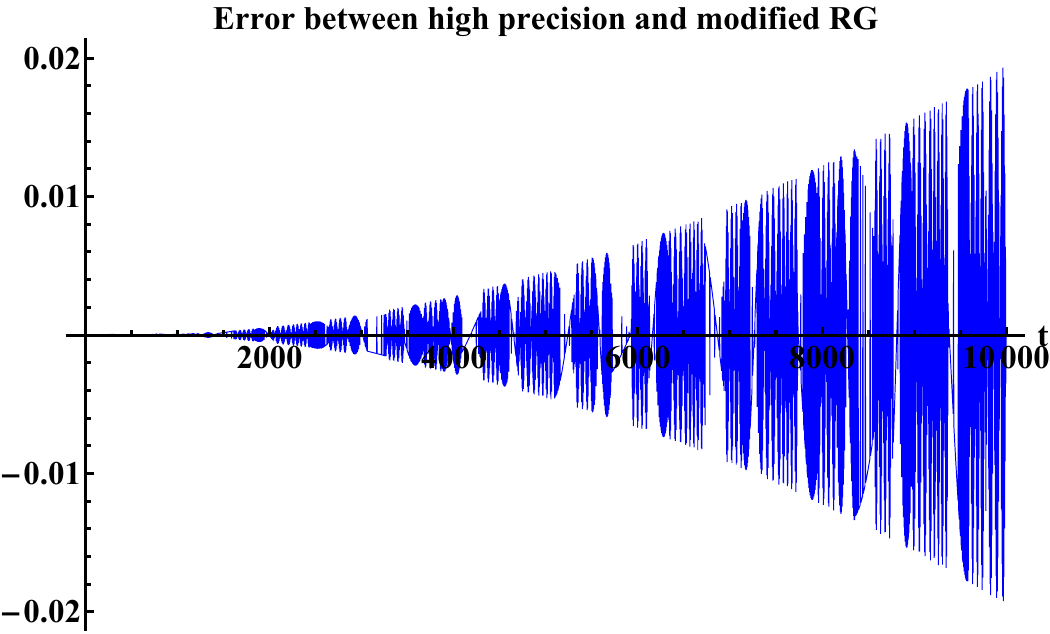}
  \caption{Plot of the difference between the high precision numerical solution to (\ref{eq1}) and the RG solution (\ref{eq51}), (\ref{eq52}) using the homogeneous functions for $t\in[0,10^4]$.}
\label{fig2}
\end{figure}

In figure (\ref{fig1}) we can see an almost perfect overlap between the two solutions up to $t=10^4$ which is $t=1/\varepsilon^4$. The secular terms were removed up to $\varepsilon^3$ order so we expect an error of magnitude at most $\varepsilon$ up to order $1/\varepsilon^4$. From figures (\ref{fig2}) we can see that the asymptotic solution does perform as we would expect. The error grows in a nonlinear fashion and is approximately 0.02 at $t=1/\varepsilon^4$.

Let us now compare the errors of classical and modified RG solutions. The amplitude equation with the overall solution for the classical RG solution are (\ref{eq1.1}) and (\ref{eq1.2}).

\begin{figure}[ht]
  \centering
\captionsetup{width=0.85\textwidth}
\begin{subfigure}{.5\textwidth}
  \centering
  \includegraphics[scale=0.45]{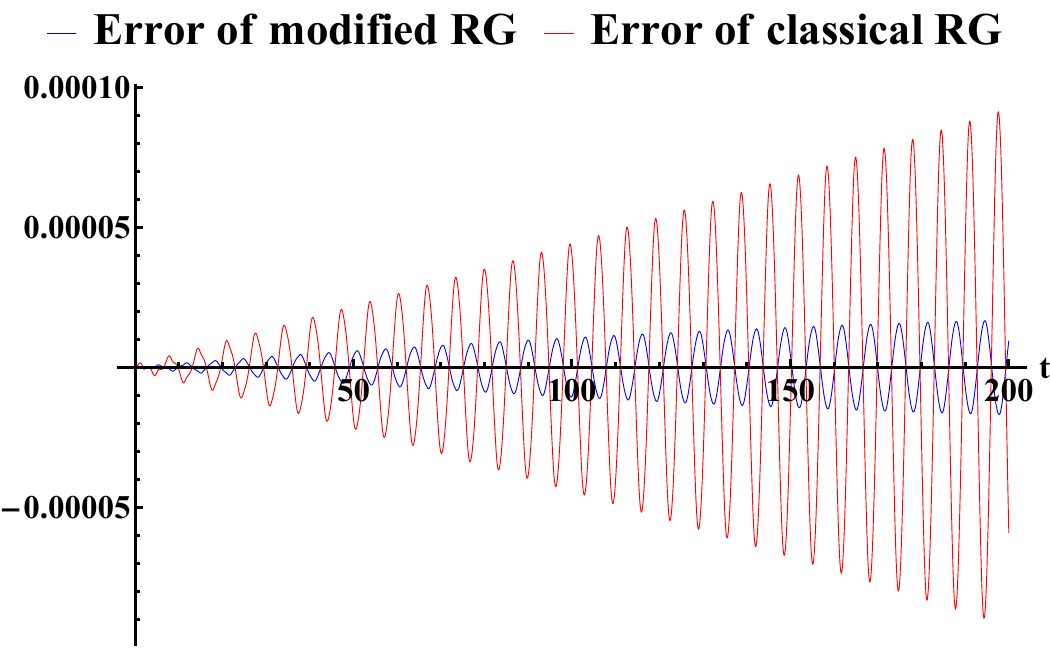}
  \caption{}
  \label{fig3a}
\end{subfigure}%
\begin{subfigure}{.5\textwidth}
  \centering
  \includegraphics[scale=0.45]{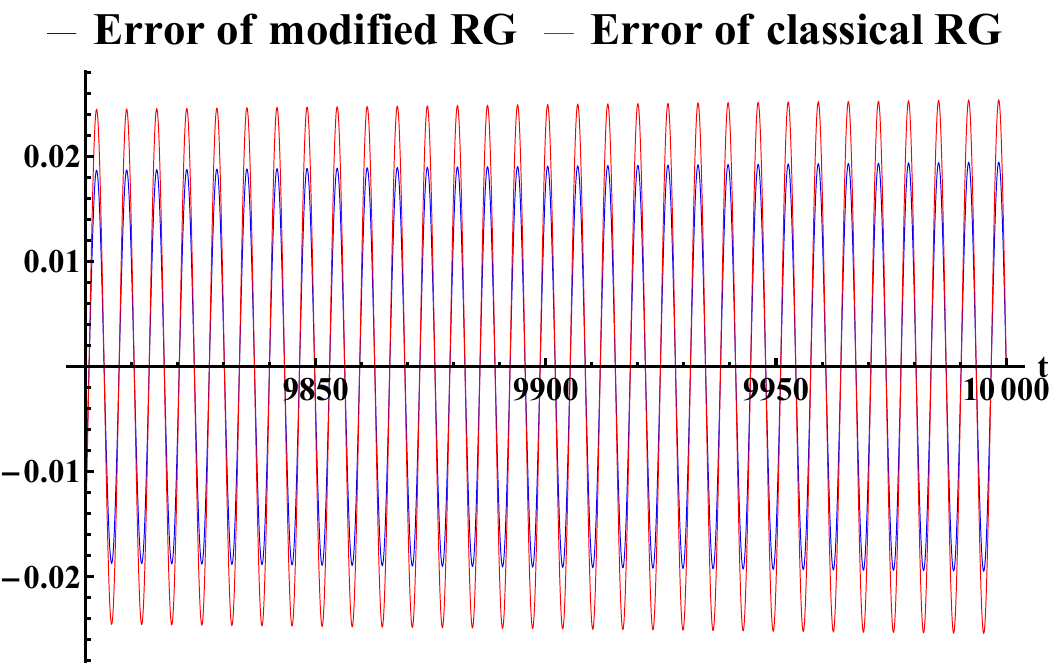}
  \caption{}
  \label{fig3b}
\end{subfigure}
\caption{Comparing the errors of the modified RG solution (\ref{eq50}), (\ref{eq51}) fig. \ref{fig3a} and the classical RG solution (\ref{eq53}), (\ref{eq54}) fig. \ref{fig3b} against high precision numerical solution. On the left, the timeline spans through $t\in [0,200]$ and on the right, it goes through $t\in [9800,10^4]$.}
\label{fig3}
\end{figure}

In figure (\ref{fig3}) we see the comparison of the error for both RG solutions: with (\ref{eq50}), (\ref{eq51}) and without the use of homogeneous functions (\ref{eq53}), (\ref{eq54}). On the left image, the variable $t$ goes from 0 to 200, and on the right, it goes from 9800 to $10^4$. In both cases, the error of the solution with homogeneous functions is smaller compared to the classical solution with no homogeneous functions. Of course, our choice of these functions was driven by the simplification of the amplitude equation, but it is also good to see that it leads to a lowering of the error.

\subsection{Van der Pol oscillator}
In this section we will look at the Van der Pol oscillator. The equation is of the form
\begin{align}
y''(t)+y(t)&=\varepsilon\left(1-y^2(t)\right)y'(t),\quad t>0,\label{eq55}\\
y(0)&=1,\nonumber\\
y'(0)&=0.\nonumber
\end{align}
We present again, both solutions from the classical RG and the modified RG method. The former reads
\begin{align}
A'(t)&=\varepsilon\left(\frac{A}{2}-\frac{A^2 A^*}{2}\right)  +\varepsilon ^2\left(-\frac{7}{16} i \left(A^*\right)^2 A^3+\frac{1}{2} i A^* A^2-\frac{i A}{8}\right) +\varepsilon ^3\left(-\frac{37}{128} \left(A^*\right)^3 A^4+\frac{35}{64} \left(A^*\right)^2 A^3\right.\nonumber\\
&\left.-\frac{A^* A^2}{4}\right) +\varepsilon ^4\left(\frac{497 i \left(A^*\right)^4 A^5}{3072}-\frac{211}{512} i \left(A^*\right)^3 A^4+\frac{85}{256} i \left(A^*\right)^2 A^3-\frac{1}{16} i A^* A^2-\frac{i A}{128}\right),\label{eq91}\\
y_c(t)&=A e^{i t}+\frac{1}{8} i A^3 e^{3 i t} \varepsilon+\varepsilon ^2 \left(-\frac{1}{192} 5 A^5 e^{5 i t}-\frac{1}{64} A^4 A^* e^{3 i t}-\frac{1}{32} A^3 e^{3 i t}\right)\nonumber\\
&+\varepsilon ^3 \left(-\frac{7 i A^7 e^{7 i t}}{1152}-\frac{5 i A^6 A^* e^{5 i t}}{1536}+\frac{29}{512} i A^5 \left( A^*\right)^2 e^{3 i t}-\frac{35 i A^5 e^{5 i t}}{2304}-\frac{21}{256} i A^4 A^* e^{3 i t}+\frac{1}{128} i A^3 e^{3 i t}\right)\nonumber\\
&+\varepsilon ^4 \left(\frac{61 A^9 e^{9 i t}}{40960}+\frac{133 A^8 A^* e^{7 i t}}{221184}-\frac{2521 A^7 \left(A^*\right)^2 e^{5 i t}}{110592}+\frac{623 A^7 e^{7 i t}}{110592}+\frac{989 A^6 \left(A^*\right)^3 e^{3 i t}}{12288}+\frac{197 A^6 A^* e^{5 i t}}{6144}\right.\nonumber\\
&\left.-\frac{1103 A^5 \left(A^*\right)^2 e^{3 i t}}{6144}+\frac{5 A^5 e^{5 i t}}{27648}+\frac{113 A^4 A^* e^{3 i t}}{1024}-\frac{3}{512} A^3 e^{3 i t}\right)+(*).\label{eq55.1}
\end{align}
Here, $(*)$ means again the complex conjugate of the terms standing before it. Also, the real quantities, for example $A^*A$ are conjugated and added, so the full function will include $2A^*A$. The modified RG method yields the following solution
\begin{align}
A'(t)&=A \left(-\frac{i \varepsilon ^4}{128}-\frac{i \varepsilon ^2}{8}+\frac{\varepsilon }{2}\right)+A^* A^2 \left(-\frac{7}{16} \varepsilon ^4 \text{Im}\left(d_0\right)-\varepsilon ^4 \text{Re}\left(e_0\right)-\varepsilon ^4 \text{Re}\left(e_1\right)+\frac{5 i \varepsilon ^4}{192}-\frac{\varepsilon ^3}{4}+\frac{i \varepsilon ^2}{16}-\frac{\varepsilon }{2}\right),\label{eq55.2}\\
y_m(t)&=A e^{i t}+\varepsilon\left(\frac{1}{8} e^{3 i t} i A^3+\alpha \right) +\varepsilon ^2\left(-\frac{1}{192} 5 e^{5 i t} A^5-\frac{1}{64} e^{3 i t} A^* A^4-\frac{1}{32} e^{3 i t} A^3+\frac{3}{8} e^{2 i t} i \alpha  A^2+\beta \right)\nonumber\\
&+\varepsilon ^3\left(-\frac{7 i e^{7 i t} A^7}{1152}-\frac{5 i e^{5 i t} A^* A^6}{1536}+\frac{29}{512} e^{3 i t} i \left(A^*\right)^2 A^5-\frac{35 i e^{5 i t} A^5}{2304}-\frac{25}{192} e^{4 i t} \alpha  A^4-\frac{21}{256} i e^{3 i t} A^* A^4\right.\nonumber\\
&\left.-\frac{1}{64} e^{4 i t} \alpha ^* A^4+\frac{1}{128} e^{3 i t} i A^3-\frac{1}{16} e^{2 i t} \alpha  A^* A^3-\frac{3}{32} e^{2 i t} \alpha  A^2+\frac{3}{8} e^{2 i t} i \beta  A^2+\frac{3}{8} e^{i t} i \alpha ^2 A+\gamma \right)\nonumber\\
&+\varepsilon^4f(A,\alpha,\beta,\gamma,\delta)+(*),\label{eq55.3}
\end{align}
together with the functions
\begin{align}
\alpha(A,A^*)&=\frac{7}{16} i A^2 A^*,\nonumber\\
\beta(A,A^*)&=A \left(\frac{91}{512} A^2 \left(A^*\right)^2+i \text{Im}\left(d_0\right)\right),\nonumber\\
\gamma(A,A^*)&=\frac{1}{64} i \left(A^*\right)^3 A^4 \Log (A)+A \left(A A^* \left(\text{Re}\left(e_1\right)-\frac{1}{192} (17 i)\right)+\frac{15551 i A^3 \left(A^*\right)^3}{73728}+\frac{355 i A^2 \left(A^*\right)^2}{1536}\right.\nonumber\\
&\left.+i \text{Im}\left(e_0\right)+\text{Re}\left(e_0\right)\right),\nonumber\\
\delta(A,A^*)&=\frac{5}{256} \left(A^*\right)^4 A^5 \Log ^2(A)-\frac{1}{256} \left(A^*\right)^4 A^5 \Log ^2\left(A^*\right)-\frac{3}{128} \left(A^*\right)^3 A^4 \Log ^2(A).\label{eq55.3.1}
\end{align}
The four free parameters $\text{Im}(d_0),\text{Im}(e_0),\text{Re}(e_0)$ and $\text{Re}(e_1)$ will be determined in the next section. All these functions are derived in Appendix A.

Clearly, the amplitude equation is much simpler in the modified RG method solution just as it was the case of the Duffing equation. The solution itself $y_m(t)$ is, however, much more complicated as it contains more terms. In this example, the core of the amplitude equation (\ref{eq55.2}) are two terms proportional to $A$ and $A^2A^*$ from which only one is nonlinear. The point is to have at most one nonlinear term in the core as the linear once don't render the equation more difficult to solve. In the next chapter, we are going to use them to reduce the error of the asymptotic solution.

\subsubsection{Numerical results}
Since we have four free real parameters in our solution (\ref{eq55.3.1}), at the start, we don't know what value they should have. Let us therefore start with the trivial choice where we set them all to zero and compare it to the classical RG method solution. The latter has the amplitude equation (\ref{eq91}).

\begin{figure}[ht]
  \centering
\captionsetup{width=0.85\textwidth}
\begin{subfigure}{.5\textwidth}
  \centering
  \includegraphics[scale=0.5]{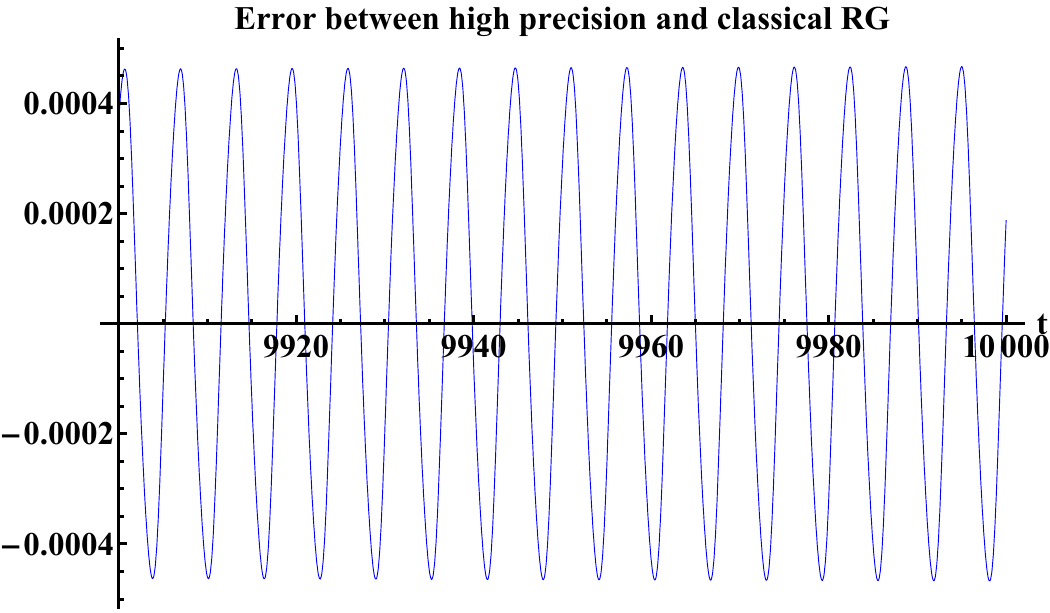}
  \caption{}
  \label{fig4a}
\end{subfigure}%
\begin{subfigure}{.5\textwidth}
  \centering
  \includegraphics[scale=0.5]{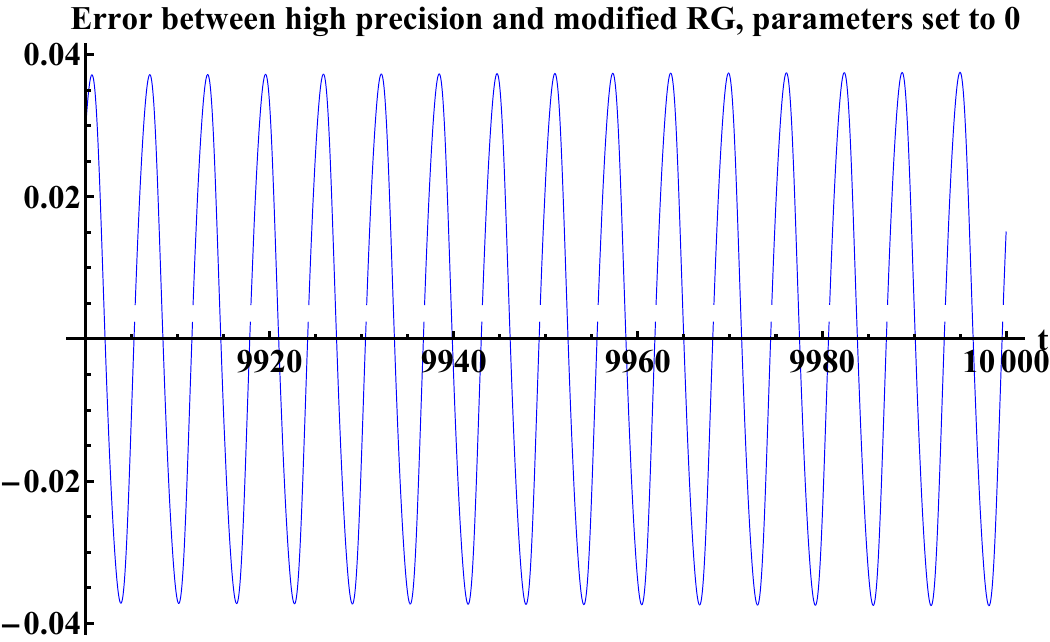}
  \caption{}
  \label{fig4b}
\end{subfigure}
\caption{Comparing the errors of two RG solutions of the Van der Pol oscillator (\ref{eq55}), the classical (\ref{eq91}) (left) and modified (\ref{eq55.2}) (right) with all parameters set to zero. The error is against the high precision numerical solution. In both cases, the timeline spans through $t\in [9900,10^4]$.}
\label{fig4}
\end{figure}

The numerical results of this comparison in figure (\ref{fig4}) tells us that the modified RG solution with all free parameters set to zero has a 100 times larger error than the classical RG solution with the amplitude equation (\ref{eq91}). In order to correct this we can find such values of the free parameters so that the error is minimized. A special function $F$ can be constructed whose variables are the free parameters $\text{Im}(d_0),\text{Im}(e_0),\text{Re}(e_0),\text{Re}(e_1)$. In pseudo-programming language, we can write
\begin{align}
&F\left(\text{Im}(d_0),\text{Im}(e_0),\text{Re}(e_0),\text{Re}(e_1)\right):=\nonumber\\
&- \text{define the functions }\alpha,\beta,\gamma,\delta \text{ accorgind to (\ref{eq55.3.1}) and the function }y_m(t) \text{ from (\ref{eq55.3}) using the input},\nonumber\\
&-\text{calculate the initial conditions and solve the amplitude equation (\ref{eq55.2})},\nonumber\\
&-\text{use the calculated amplitude back in (\ref{eq55.3}), then the output is }\nonumber\\
&\max_{t\in(9800,10^4)}y_{\text{exact}}(t)-y_{m\text{ with this input}}(t).\label{eq92}
\end{align}
The task is then to minimize $F$ using some optimalization method. In this case we used gradient descent method starting from $(0,0,0,0)$. This optimalization method converges and terminates at the point
\begin{align}
\text{Im}(d_0)&\approx-0.301134\nonumber\\
\text{Im}(e_0)&\approx-0.000414268\nonumber\\
\text{Re}(e_0)&\approx-0.683162\nonumber\\
\text{Re}(e_1)&\approx-0.68445.\label{eq93}
\end{align}

\begin{figure}[h]
  \centering
\captionsetup{width=0.85\textwidth}
  \includegraphics[scale=0.65]{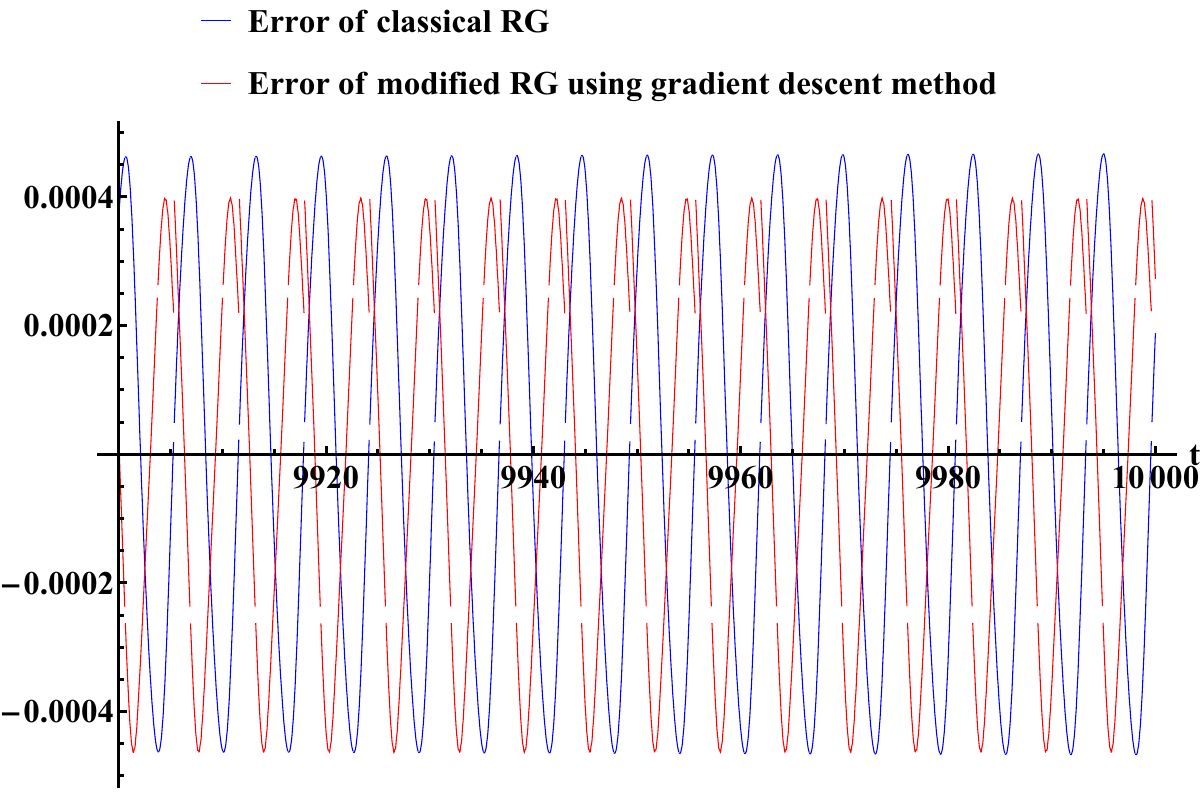}
  \caption{Comparing the error of the classical RG solution using (\ref{eq91}) with the modified RG solution using (\ref{eq55.3}), where the free parameter are (\ref{eq93}).}
\label{fig5}
\end{figure}
In figure (\ref{fig5}) we see the results. Using the calculated values (\ref{eq93}), the error got a 100 times smaller and the solution became just as good as using (\ref{eq91}) (classical RG). Thus we demonstrated that even if we are not able to remove all the non-core terms without introducing $\Log$ functions, we can still achieve a satisfying precision.

\section{System of equations}
We proceed to demonstrate our modified method on three systems of ODEs, each of them being somewhat specific. We start with a special case of a system known as the Lotka-Volterra system, which is a fixed-point dynamical planar system. Here, our method is fully presented up to $\varepsilon^3$ order. The other two examples will be presented in less detail with respect to the method deployment while giving the differences from the other examples adequate attention.

\subsection{Lotka-Volterra system}
The Lotka-Volterra system also known as the predator-prey model is a system of nonlinear ODE's often used in biology to describe the dynamics between two different species, one as predator and the other as prey. We take the system with parameters set to simple constants. The form of the equations are
\begin{align}
x'(t)+y(t)&=-\varepsilon x(t)y(t),\nonumber\\
y'(t)-x(t)&=\varepsilon x(t)y(t),\nonumber\\
x(0)=0,y(0)&=1,\label{eq94}
\end{align}
where $\varepsilon$ is a small parameter. The classical RG solution for this system is
\begin{align}
A'&=-\varepsilon^2\frac{1}{3} i A^* A^2-\varepsilon^3\frac{2}{3} i A^* A^2,\nonumber\\
x_c&=i A e^{i t}+\varepsilon  \left(i A e^{i t}-\left(\frac{2}{3}-\frac{i}{3}\right) A^2 e^{2 i t}\right)+\varepsilon ^2 \left(-\left(\frac{1}{2}+\frac{i}{4}\right) A^3 e^{3 i t}+\left(-\frac{4}{3}+\frac{2 i}{3}\right) A^2 e^{2 i t}\right.\nonumber\\
&\left.+\frac{1}{3} A^* A^2 e^{i t}\right)+\varepsilon ^3 \left(-\left(\frac{14}{135}+\frac{56 i}{135}\right) A^4 e^{4 i t}-\left(\frac{3}{2}+\frac{3 i}{4}\right) A^3 e^{3 i t}+\left(-\frac{2}{3}+\frac{i}{3}\right) A^2 e^{2 i t}\right.\nonumber\\
&\left.+\left(\frac{2}{27}+\frac{19 i}{54}\right) A^* A^3 e^{2 i t}+A^* A^2 e^{i t}\right),\nonumber\\
y_c&=A e^{i t}+\varepsilon  \left(\left(\frac{2}{3}+\frac{i}{3}\right) A^2 e^{2 i t}+A e^{i t}\right)+\varepsilon ^2 \left(\left(\frac{1}{4}+\frac{i}{2}\right) A^3 e^{3 i t}+\left(\frac{4}{3}+\frac{2 i}{3}\right) A^2 e^{2 i t}\right)\nonumber\\
&+\varepsilon ^3 \left(-\left(\frac{14}{135}-\frac{56 i}{135}\right) A^4 e^{4 i t}+\left(\frac{3}{4}+\frac{3 i}{2}\right) A^3 e^{3 i t}+\left(\frac{2}{3}+\frac{i}{3}\right) A^2 e^{2 i t}+\left(\frac{4}{27}-\frac{5 i}{54}\right) A^* A^3 e^{2 i t}\right).\label{eq136}
\end{align}
The amplitude equation above has the $\varepsilon^3$ order term proportional to the previous order, so it seems in this case that our method does not need to be used since only the factor of the core term has changed with higher order. No other nonlinear terms were introduced. But as we will see, our method is not only great for simplifying the amplitude equation but also for lowering the error of the solution. The modified RG solution we find to be
\begin{align}
A'(t)&=-\varepsilon^2\frac{1}{3} i A^* A^2,\nonumber\\
x_m(t)&=i A e^{i t}-\left(\frac{2}{3}-\frac{i}{3}\right) A^2 e^{2 i t} \varepsilon+\varepsilon ^2 \left(\left(-\frac{1}{2}-\frac{i}{4}\right) A^3 e^{3 i t}+\frac{1}{3} A^* A^2 e^{i t}\right)\nonumber\\
& +\varepsilon ^3 \left(\left(-\frac{14}{135}-\frac{56 i}{135}\right) A^4 e^{4 i t}+\left(\frac{2}{27}+\frac{19 i}{54}\right) A^* A^3 e^{2 i t}\right)+(*),\nonumber\\
y_m(t)&=A e^{i t}+\left(\frac{2}{3}+\frac{i}{3}\right) A^2 e^{2 i t} \varepsilon+\left(\frac{1}{4}+\frac{i}{2}\right) A^3 e^{3 i t} \varepsilon ^2\nonumber\\
& +\varepsilon ^3 \left(\left(-\frac{14}{135}+\frac{56 i}{135}\right) A^4 e^{4 i t}+\left(\frac{4}{27}-\frac{5 i}{54}\right) A^* A^3 e^{2 i t}\right)+(*).\label{eq94.1}
\end{align}
As expected by now, the $\varepsilon^3$ order term disappeared from the amplitude equation, but something else is different than before. The solution itself got simpler compared with the classical RG solution. Before, the modified RG solution had a more complicated solution due to the extra homogeneous functions.

In the next section, we look at some numerical results and compare the errors of the different solutions.

\subsubsection{Numerical results}
\begin{figure}[ht]
  \centering
\captionsetup{width=0.85\textwidth}
\begin{subfigure}{.5\textwidth}
  \centering
  \includegraphics[scale=0.5]{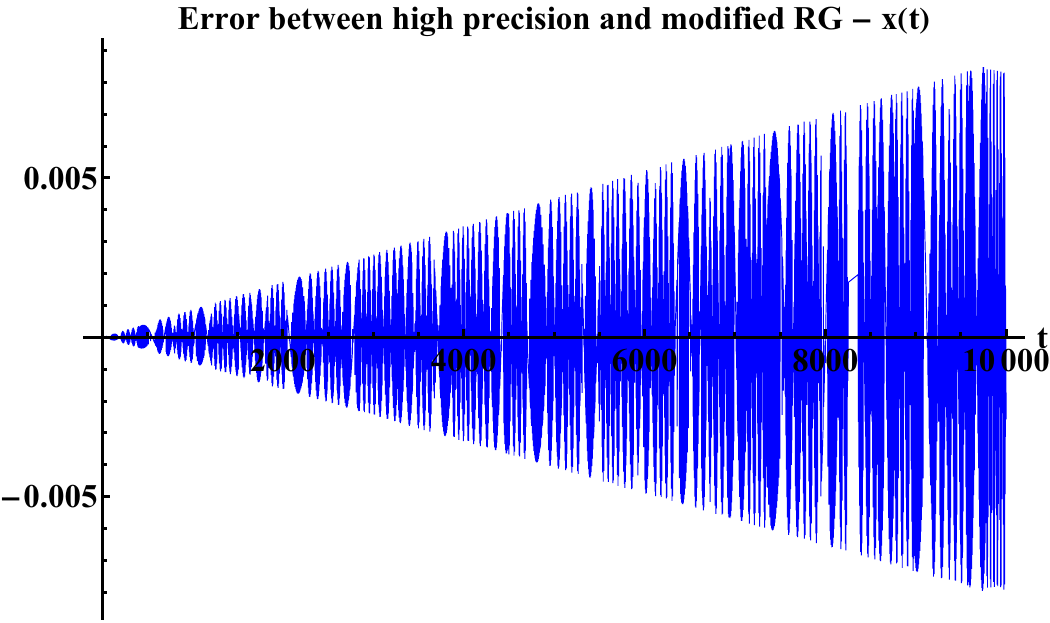}
  \caption{}
  \label{fig6a}
\end{subfigure}%
\begin{subfigure}{.5\textwidth}
  \centering
  \includegraphics[scale=0.5]{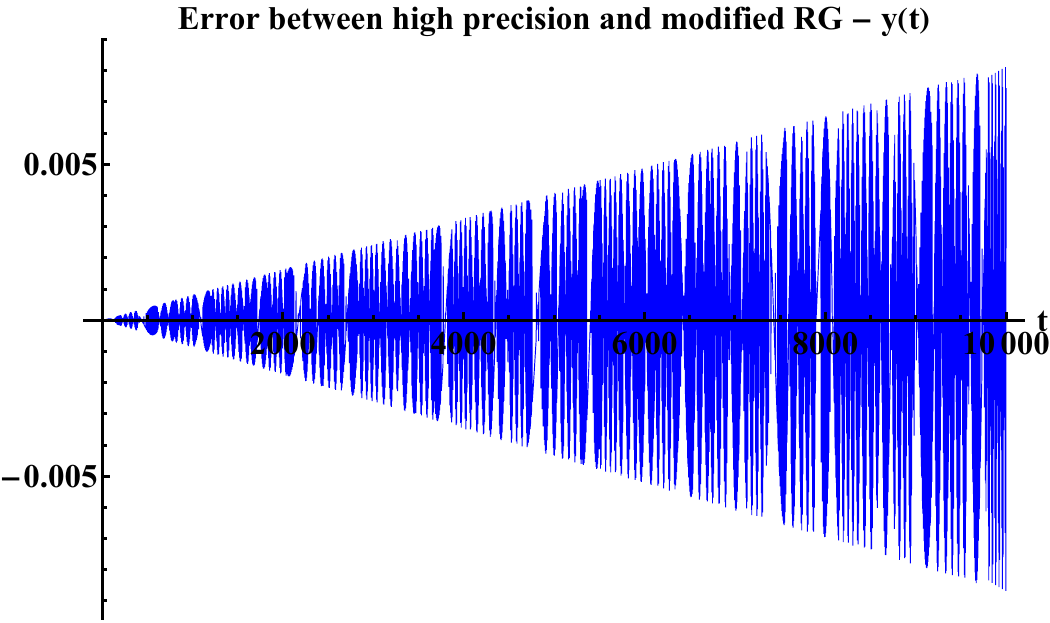}
  \caption{}
  \label{fig6b}
\end{subfigure}
\caption{Plot of the difference between the high precision numerical solution to (\ref{eq94}) and the modified RG solution (\ref{eq94.1}) for $t\in[0,10^4]$.}
\label{fig6}
\end{figure}

In these numerical results we will set the perturbation parameter to be $\varepsilon=0.1$. In figure (\ref{fig6}) we display the error between a high precision numerical solution and the modified RG solution (\ref{eq94.1}).

The order of the solution is up to $\varepsilon^3$ order. Having $\varepsilon=0.1$ we should expect error of the order 1 for $\varepsilon^4t\approx 1\Rightarrow t\approx 10^4$. In this numerical test we went up to this order in time and as we can see, we found the error to be much better than expected.

\begin{figure}[ht]
  \centering
\captionsetup{width=0.85\textwidth}
\begin{subfigure}{.5\textwidth}
  \centering
  \includegraphics[scale=0.5]{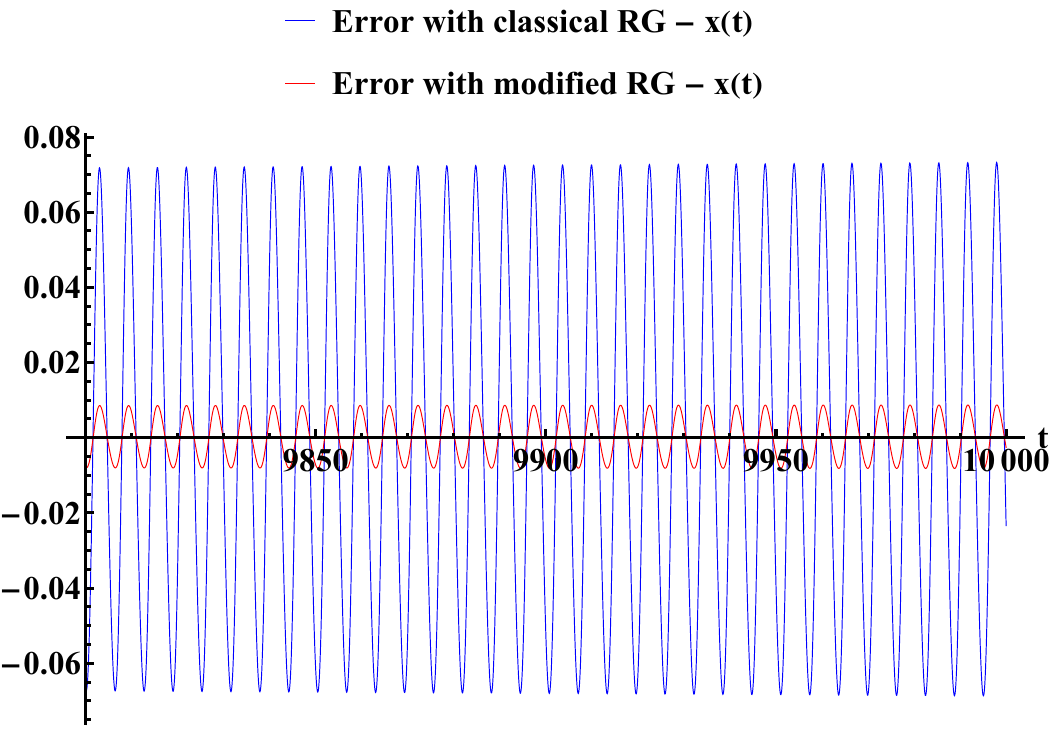}
  \caption{}
  \label{fig7a}
\end{subfigure}%
\begin{subfigure}{.5\textwidth}
  \centering
  \includegraphics[scale=0.5]{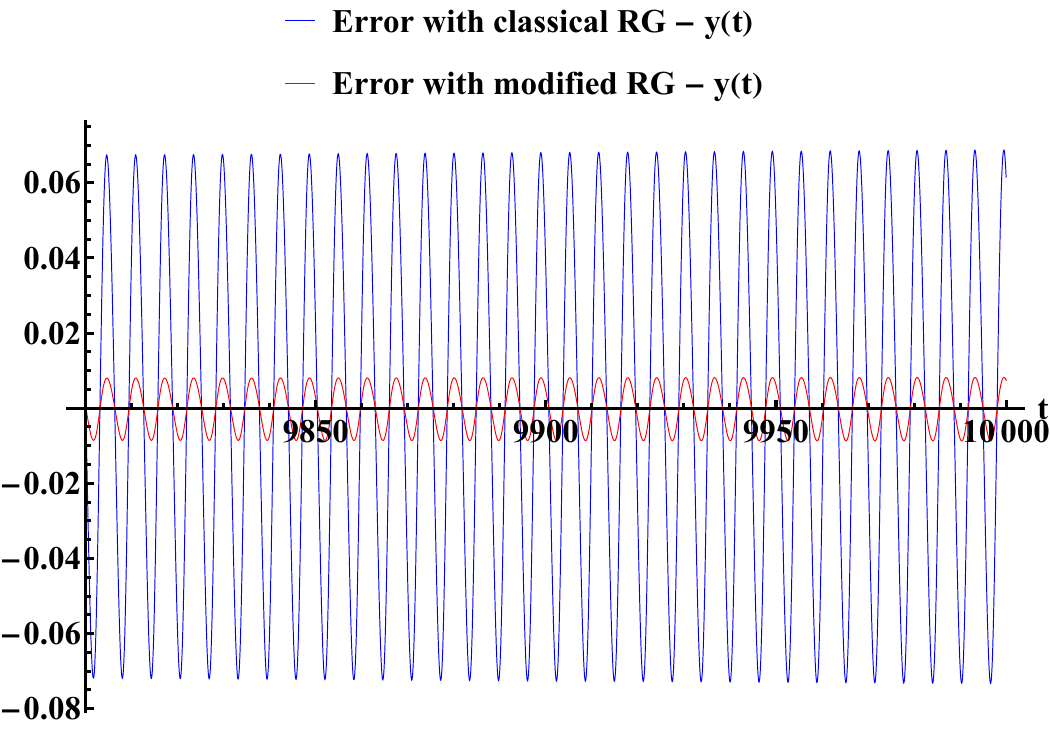}
  \caption{}
  \label{fig7b}
\end{subfigure}
\caption{Comparing the error of the classical RG solution with (\ref{eq136}) and the error of the modified RG solution (\ref{eq94.1}) against high precision numerical solution. The timeline spans through $t\in [9800,10^4]$.}
\label{fig7}
\end{figure}

Let us look at the classical RG solution to our system (\ref{eq94}) where we don't introduce any homogeneous solutions (\ref{eq136}). We compare the errors of the modified RG solution with the errors for the classical RG solution (\ref{eq136}). In figure (\ref{fig7}) are the results. Not only is the amplitude equation of the modified RG solution simpler, but the error of this solution is nearly 10 times smaller than the error of the classical RG solution. As mentioned previously, the amplitude equation for both solutions is of minor difference, but the main difference is in the errors. This example showed that the modified RG method prevails in one way or another.

\subsection{System of second order ODE's}
In this section we would like to demonstrate the modified RG method on a system with two amplitudes. In order for a system of ODE's have two oscillating amplitudes, the system must be of a second order with 4 pure complex eigenvalues. Let us consider the following second order system
\begin{align}
x''(t)+2x(t)-y(t)&=\varepsilon x(t)y(t),\nonumber\\
y''(t)-2x(t)+3y(t)&=\varepsilon x(t)y(t),\nonumber\\
x(0)=1,y(0)&=-1,\nonumber\\
x'(0)=0,y'(0)&=0,\label{eq137}
\end{align}
where $\varepsilon$ is a small parameter. We first present the classical RG solution.
\begin{align}
A'&=-\frac{1}{2} i A^* B \varepsilon+\varepsilon ^2 \left(\frac{67}{16} i A B B^*-\frac{7}{4} i A^2 A^*\right),\nonumber\\
B'&=-\frac{1}{6} i A A^* B \varepsilon ^2,\nonumber\\
x_c(t)&=A e^{i t}-B e^{2 i t}+\varepsilon  \left(-\frac{1}{2} A^2 e^{2 i t}+A^* A-\frac{1}{8} A B e^{3 i t}+\frac{2}{15} B^2 e^{4 i t}-2 B B^*\right)+\varepsilon ^2 \left(\frac{1}{16} A^3 e^{3 i t}\right.\nonumber\\
&\left.-\frac{1}{2} A^2 B^*+\frac{1}{12} A^2 B e^{4 i t}-\frac{77}{960} A^* B^2 e^{3 i t}-\frac{37}{24} A^* A B e^{2 i t}-\frac{1}{2}\left(A^*\right)^2 B-\frac{17 A B^2 e^{5 i t}}{2880}-\frac{2}{525} B^3 e^{6 i t}\right.\nonumber\\
&\left.+\frac{29}{15} B^2 B^* e^{2 i t}\right)+(*),\nonumber\\
y_c(t)&=A e^{i t}-B e^{2 i t}+\varepsilon  \left(-\frac{1}{2} A^2 e^{2 i t}+A^* A-\frac{1}{8} A B e^{3 i t}+\frac{2}{15} B^2 e^{4 i t}-2 B B^*\right)+\varepsilon ^2 \left(\frac{1}{16} A^3 e^{3 i t}\right.\nonumber\\
&\left.-\frac{1}{2} A^2 B^*+\frac{1}{12} A^2 B e^{4 i t}-\frac{77}{960} A^* B^2 e^{3 i t}-\frac{37}{24} A^* A B e^{2 i t}-\frac{1}{2}\left(A^*\right)^2 B-\frac{17 A B^2 e^{5 i t}}{2880}-\frac{2}{525} B^3 e^{6 i t}\right.\nonumber\\
&\left.+\frac{29}{15} B^2 B^* e^{2 i t}\right)+(*).\label{eq186}
\end{align}
Notice there are many nonlinear terms in the amplitude equations which we expect to vanish with the modified method except for the core. The modified RG solution reads
\begin{align}
A'&=-\frac{1}{2} i \varepsilon  A^* B,\nonumber\\
B'&=-\frac{10}{3} i \varepsilon ^2 AA^*B,\nonumber\\
x_m(t)&=A e^{i t}-B e^{2 i t}+\varepsilon  \left(3 A^2 e^{2 i t}+\frac{67}{16} A^* B e^{i t}+A^* A-\frac{1}{8} A B e^{3 i t}+\frac{2}{15} B^2 e^{4 i t}-2 B B^*\right)\nonumber\\
&+\varepsilon ^2 \left(\frac{1}{2} A^3 e^{3 i t}+\frac{171 A^2 B^*}{16}-\frac{17}{20} A^2 B e^{4 i t}-\frac{1159 A^* B^2 e^{3 i t}}{1920}-\frac{275}{48} A^* A B e^{2 i t}+\frac{171}{16} \left(A^*\right)^2 B\right.\nonumber\\
&\left.-\frac{17 A B^2 e^{5 i t}}{2880}-\frac{2}{525} B^3 e^{6 i t}+\frac{29}{15} B^2 B^* e^{2 i t}\right)+(*),\nonumber\\
y_m(t)&=A e^{i t}+2 B e^{2 i t}+\varepsilon  \left(-7 A^2 e^{2 i t}+\frac{67}{16} A^* B e^{i t}+A^* A-\frac{1}{8} A B e^{3 i t}+\frac{2}{15} B^2 e^{4 i t}-2 B B^*\right)\nonumber\\
&+\varepsilon ^2 \left(\frac{1}{2} A^3 e^{3 i t}+\frac{171 A^2 B^*}{16}-\frac{17}{20} A^2 B e^{4 i t}-\frac{1159 A^* B^2 e^{3 i t}}{1920}+\frac{171}{16} \left(A^*\right)^2 B-\frac{17 A B^2 e^{5 i t}}{2880}\right.\nonumber\\
&\left.-\frac{2}{525} B^3 e^{6 i t}\right)+(*).\label{eq137.1}
\end{align}
Indeed, all the non-core terms disappeared and the amplitude equations became much simpler. The $B$ equation had only the core term before the modification, but the $A$ equation got significantly simpler.

\subsubsection{Numerical results}
As before, we will set our perturbation parameter to be $\varepsilon=0.1$. The classical RG solution (\ref{eq186}) will be compared to a high precision numerical solution of the problem.

\begin{figure}[ht]
  \centering
\captionsetup{width=0.85\textwidth}
\begin{subfigure}{.5\textwidth}
  \centering
  \includegraphics[scale=0.5]{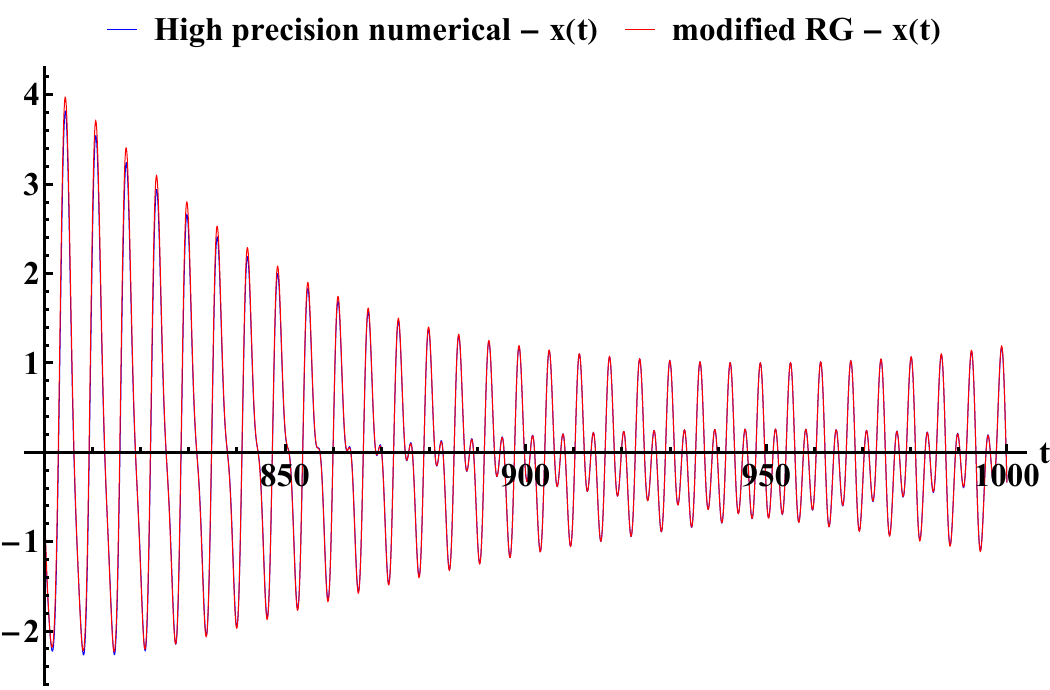}
  \caption{}
  \label{fig8a}
\end{subfigure}%
\begin{subfigure}{.5\textwidth}
  \centering
  \includegraphics[scale=0.5]{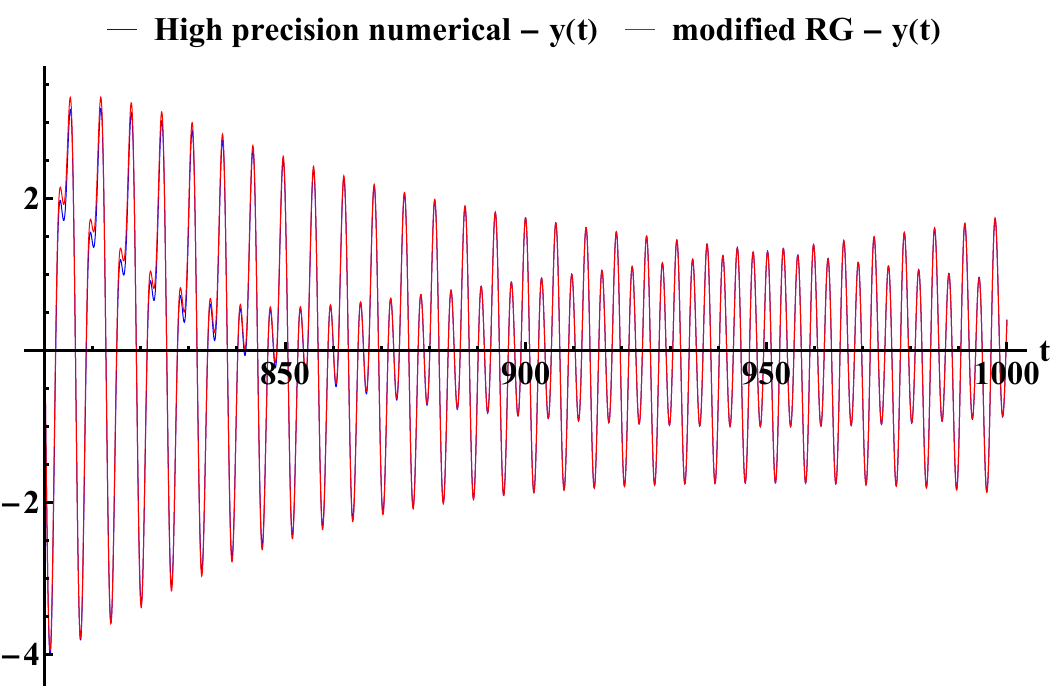}
  \caption{}
  \label{fig8b}
\end{subfigure}
\caption{Comparison of the high precision numerical solution to (\ref{eq137}) and the modified RG solution (\ref{eq137.1}) for $t\in[800,1000]$.}
\label{fig8}
\end{figure}
\begin{figure}[ht]
  \centering
\captionsetup{width=0.85\textwidth}
\begin{subfigure}{.5\textwidth}
  \centering
  \includegraphics[scale=0.5]{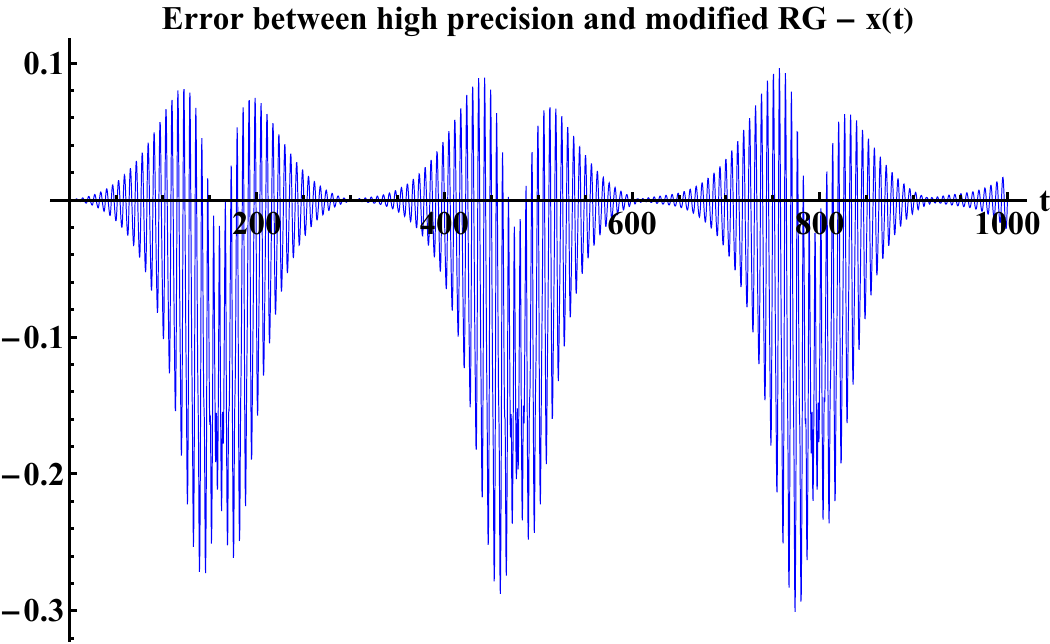}
  \caption{}
  \label{fig9a}
\end{subfigure}%
\begin{subfigure}{.5\textwidth}
  \centering
  \includegraphics[scale=0.5]{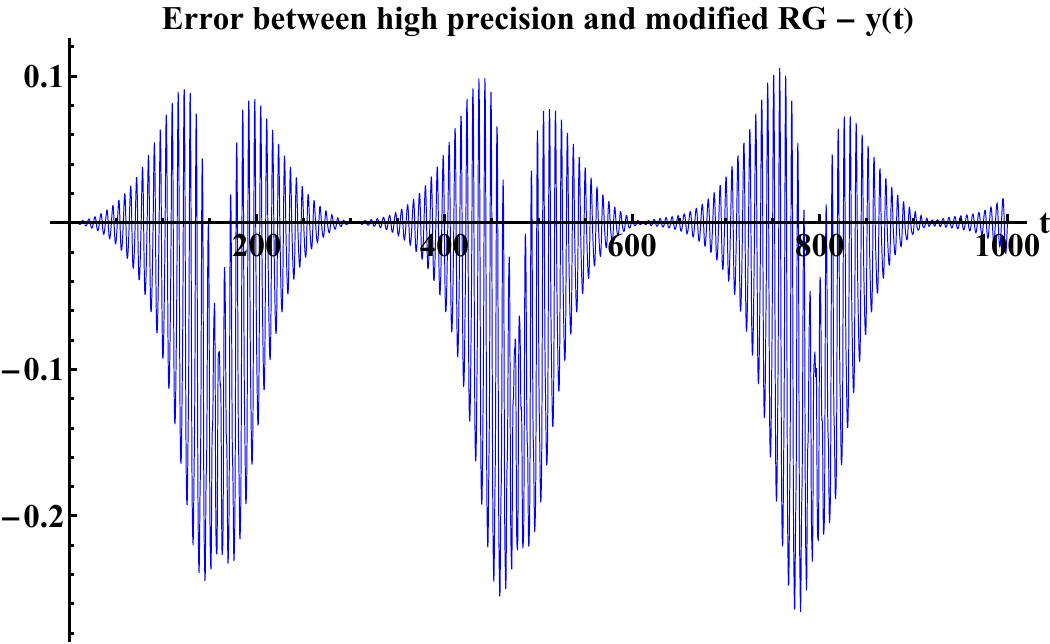}
  \caption{}
  \label{fig9b}
\end{subfigure}
\caption{Error between the high precision numerical solution to (\ref{eq137}) and the modified RG solution (\ref{eq184}) for $t\in[0,1000]$.}
\label{fig9}
\end{figure}

Figure (\ref{fig8}) shows the comparison of the modified RG solution to the high precision numerical one. They overlap rather nicely even for $t\sim 1000$. To see the error better, we plot the difference between these two solutions. On figure (\ref{fig9}) we can see this error.
The highest order of the solution is $\varepsilon^2$. With $\varepsilon=0.1$ we should expect error of the order 1 for $\varepsilon^3t\approx 1\Rightarrow t\approx 10^3$. The maximum value of the error is approximately 0.25 which is much better than expected.

\begin{figure}[ht]
  \centering
\captionsetup{width=0.85\textwidth}
\begin{subfigure}{.5\textwidth}
  \centering
  \includegraphics[scale=0.5]{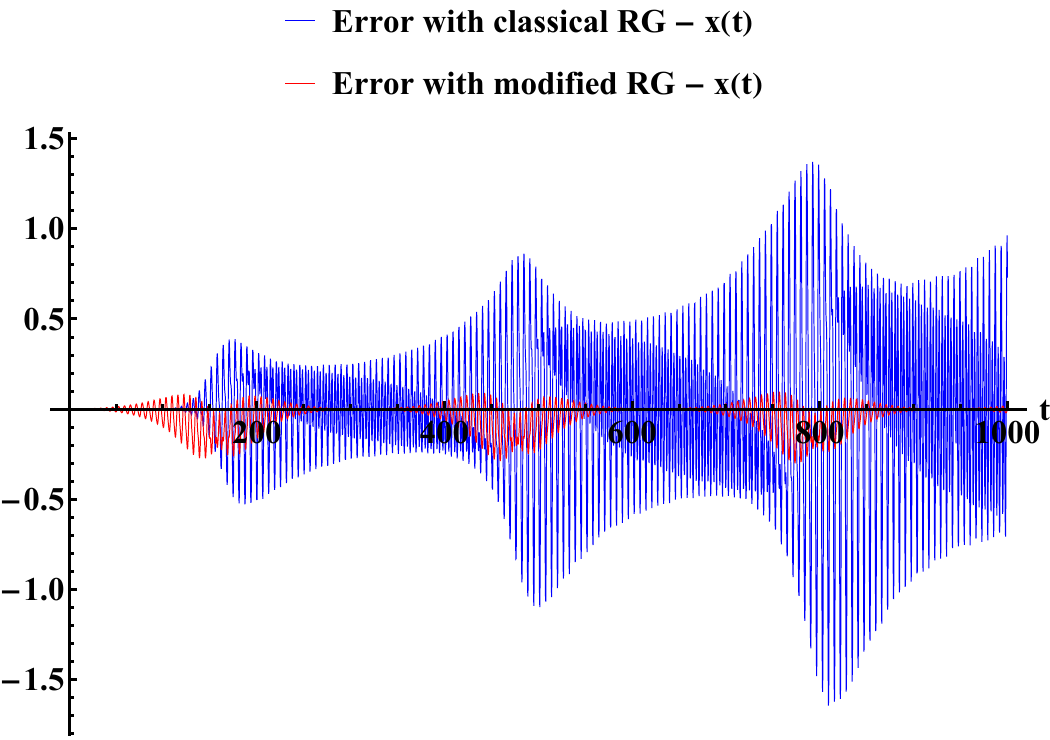}
  \caption{}
  \label{fig10a}
\end{subfigure}%
\begin{subfigure}{.5\textwidth}
  \centering
  \includegraphics[scale=0.5]{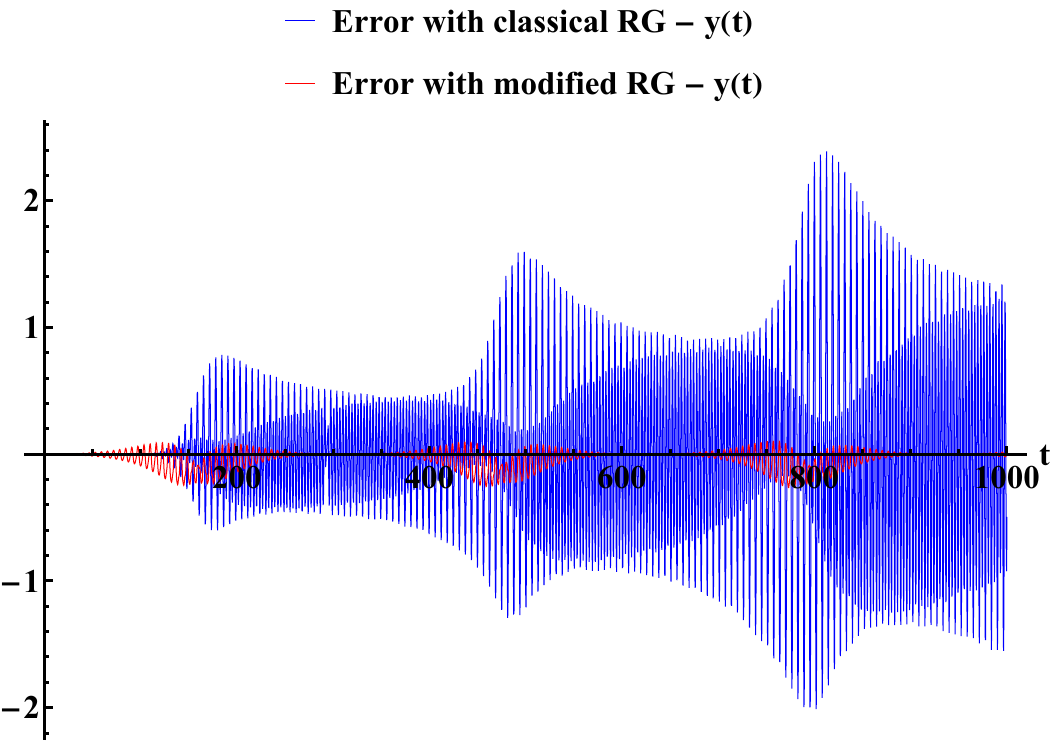}
  \caption{}
  \label{fig10b}
\end{subfigure}
\caption{Comparing error of the classical RG solution to (\ref{eq137}) with the modified RG solution (\ref{eq137.1}) error for $t\in[0,1000]$.}
\label{fig10}
\end{figure}

Next, let us investigate if our modified RG solution does better than the classical one (\ref{eq186}). The results are shown in figure (\ref{fig10}). As we can see, the error of the classical RG solution behaves more or less as we would expect from the reasoning of the error amplitude we did earlier. The error of the modified RG solution is, however, much better. In addition, the amplitude equations are also simpler in the modified RG case, having no $\varepsilon^2$ order term in the $A$-equation, while the classical $A$-equation (\ref{eq186}) has a complicated $\varepsilon^2$ order term. Once again, we have shown that the modified RG solution not only simplifies the amplitude equations but also has a smaller error.

\subsection{Selkov model (Hopf bifurcation)}
In this section we will apply our modified RG method to a problem with a Hopf bifurcation. The problem we consider is the Selkov model of the form
\begin{align}
x'&=-x+ay+x^2y,\nonumber\\
y'&=b-ay-x^2y,\label{eq187}
\end{align}
for some real, positive constants $a,b>0$, where the bifurcation parameter is $b$. As one changes slowly this parameter, there is a point, where stability of the solution starts to change from a periodic behaviour to a exponential decay or growth. The stability of the solution to a system depends on the eigenvalues of the underlying matrix of the linearized problem around a fix-point. It is straightforward to find this critical point for the parameter $b$. First, one needs to find the fix-point of the system (\ref{eq187}) and it turns out to be
\begin{align}
\bar{x}=b,\quad \bar{y}=\frac{b}{a+b^2}.\label{eq188}
\end{align}
Next, one finds the matrix $M$ of the linearized system at the fix-point (\ref{eq188}) using $x(t)=\bar{x}+x_0(t),y(t)=\bar{y}+y_0(t)$ and obtains the characteristic equation. Then critical bifurcation point $b_c$ is the smallest value of $b$ such that the eigenvalues are purely imaginary. This leads to a quadratic equation for $b$ where one chooses the smallest root. One finds
\begin{align}
b_c=\left(\frac{1-\sqrt{1-8a}-2a}{2}\right)^\frac{1}{2}.\label{eq189}
\end{align}
The Hopf bifurcation point $b_c$ must be a real number which puts some constraints on the other parameter, $0<a<\frac{1}{8}$. This ensures that both the square roots in (\ref{eq189}) are real. With this bifurcation point, the eigenvalues of $M$ become $\lambda_{\pm}=\pm i\omega$ where
\begin{align}
\omega=\left(\frac{1-\sqrt{1-8a}}{2}\right)^\frac{1}{2},\label{eq190}
\end{align}
with $0\leq\omega\leq \frac{1}{\sqrt{2}}$. We can invert (\ref{eq190}) and express the parameter $a$ in terms of $\omega$ and get $a=\frac{1}{2}(\omega^2-\omega^4)$. Then we can also write the matrix of the linearized system $M$ in terms of $\omega$. Next, we wish to investigate the dynamics close to the fix-point for $b\gtrsim b_c$. To start, we shift the origin to the position of the fix-point (\ref{eq188}) by the transformation $x=\bar{x}+z,y=\bar{y}+w$. In vector form our system for $(z,w)$ becomes
\begin{align}
\begin{pmatrix}
z\\
w
\end{pmatrix}'&=\mathbf{B}\begin{pmatrix}
z\\
w
\end{pmatrix}+\left(2\bar{x}zw+\bar{y}z^2+z^2w\right)\boldsymbol{\xi},\nonumber\\
\text{where }\mathbf{B}&=\begin{pmatrix}
2\bar{x}\bar{y}-1 & a+\bar{x}^2\\
-2\bar{x}\bar{y} & -\left( a+\bar{x}^2\right)
\end{pmatrix},\quad \boldsymbol{\xi}=\begin{pmatrix}1\\-1\end{pmatrix}.\label{eq191}
\end{align}
The matrix $\mathbf{B}$ is identical to the matrix of the linearized system of (\ref{eq187}).
For this system (\ref{eq191}), we are only concerned what happens at the origin, since the origin is the fix-point for the original system (\ref{eq187}). Therefore new variables are introduced $(u,v)$ such that their size is assumed to be of order one, by $z=\delta u,w=\delta v$ for $0<\delta\ll 1$. We would also like to investigate the system close to the critical value $b_c$. This is facilitated by introducing a small parameter $\varepsilon$, by $b=b_c+\varepsilon$ for $0<\varepsilon\ll 1$. Note that $\mathbf{B}(\varepsilon),\bar{x}(\varepsilon),\bar{y}(\varepsilon)$ depend on $\varepsilon$ through the parameter $b$. This dependence is anything but linear, therefore we use Taylor expansion in $\varepsilon$ around zero for example $\mathbf{B}(\varepsilon)\approx \mathbf{B}(0)+\varepsilon\mathbf{B}'(0)+\varepsilon^2\frac{1}{2}\mathbf{B}''(0)$ and similar expressions for $\bar{x}(\varepsilon)$ and $\bar{y}(\varepsilon)$. The system for $(u,v)$ then becomes
\begin{align}
&\begin{pmatrix}
u\\v
\end{pmatrix}'-\mathbf{B}_0\begin{pmatrix}
u\\v
\end{pmatrix}=\varepsilon\mathbf{B}_1\begin{pmatrix}
u\\v
\end{pmatrix}+\varepsilon^2\mathbf{B}_2\begin{pmatrix}
u\\v
\end{pmatrix}+\delta\left(2\bar{x}_0uv+\bar{y}_0u^2\right)\boldsymbol{\xi}+\varepsilon\delta\left(2\bar{x}_0'uv+\bar{y}_0'u^2\right)\boldsymbol{\xi}+\delta^2u^2v\boldsymbol{\xi},\nonumber\\
&\text{where}\quad\mathbf{B}_0=\mathbf{B}(0),\mathbf{B}_1=\mathbf{B}'(0),\mathbf{B}_2=\frac{1}{2}\mathbf{B}''(0),\bar{x}_0=\bar{x}(0),\bar{x}_0'=\bar{x}'(0),\bar{y}_0=\bar{y}(0),\bar{y}_0'=\bar{y}'(0)\label{eq192}
\end{align}

The last thing one needs to do is find a relation between the two small parameters $\delta$ and $\varepsilon$. We know that for $\varepsilon=0$ we have the Hopf bifurcation point $b=b_c$ for which the system $\mathbf{B}_0$ has pure imaginary eigenvalues. It is also straightforward to see that for $\varepsilon>0$ and thus for $b\gtrsim b_c$, the term $\varepsilon\mathbf{B}_1$ will add a positive real part to the eigenvalues so that the solution will be exponentially growing. In order to preserve periodic solutions, the nonlinear terms in the system must balance out the growth of the linear ones in the limit $\varepsilon\to 0$. The only way this can happen is when $\delta=O(\varepsilon)$. In fact, we can assume without loss of generality $\delta(\varepsilon)=\varepsilon$. Given this functional dependence, our system (\ref{eq192}) takes the form
\begin{align}
&\begin{pmatrix}
u\\v
\end{pmatrix}'-\mathbf{B}_0\begin{pmatrix}
u\\v
\end{pmatrix}=\varepsilon\mathbf{B}_1\begin{pmatrix}
u\\v
\end{pmatrix}+\varepsilon^2\mathbf{B}_2\begin{pmatrix}
u\\v
\end{pmatrix}+\varepsilon\left(2\bar{x}_0uv+\bar{y}_0u^2\right)\boldsymbol{\xi}+\varepsilon^2\left(2\bar{x}_0'uv+\bar{y}_0'u^2+u^2v\right)\boldsymbol{\xi}.\label{eq193}
\end{align}
All the parameters and matrices can be expressed in terms of $\omega$ (\ref{eq190}) as
\begin{align}
\mathbf{B}_0&=\left(
\begin{array}{cc}
 \omega ^2 & \omega ^2 \\
 -\omega ^2-1 & -\omega ^2 \\
\end{array}
\right),\mathbf{B}_1=\frac{\sqrt{2} \sqrt{\omega ^2+1}}{\omega }\left(
\begin{array}{cc}
 1-\omega ^2 & \omega ^2 \\
 \omega ^2-1 & -\omega ^2 \\
\end{array}
\right),\nonumber\\
\mathbf{B}_2&=\frac{1}{\omega^2}\left(
\begin{array}{cc}
 2 \omega ^4-\omega ^2-1 & \omega ^2 \\
 -2 \omega ^4+\omega ^2+1 & -\omega ^2 \\
\end{array}
\right),\bar{x}_0=\frac{\omega  \sqrt{\omega ^2+1}}{\sqrt{2}},\bar{y}_0=\frac{\sqrt{\omega ^2+1}}{\sqrt{2} \omega },\bar{x}_0'=1,\bar{y}_0'=-1.\label{eq194}
\end{align}
Note that only by determining the value of the parameter $0<a<\frac{1}{8}$ we have completely determined all the matrices and parameters in the system above.

We are now ready to deploy the RG method on the system (\ref{eq194}). The solution we get is going to be in terms of the parameter $\omega$. In this case, we will go to the 4th order of $\varepsilon$ in the perturbation hierarchy. The classical RG solution is
\begin{align}
A'&=\varepsilon\frac{A \left(-2 \omega ^2+i \omega +1\right) \sqrt{\omega ^2+1}}{\sqrt{2} \omega }+\varepsilon^2\left(Ap_1(\omega)+A^2A^*p_2(\omega)\right)+\varepsilon^3\left(Ap_3(\omega)+A^2A^*p_4(\omega)\right)\nonumber\\
&+\varepsilon^4\left(Ap_5(\omega)+A^2A^*p_6(\omega)+A^3\left(A^*\right)^2p_7(\omega)\right),\label{eq208}
\end{align}
where $p_i(\omega)$ are some rational functions of $\omega$, with the functions $u_c$ and $v_c$
\begin{align}
u_c&=\omega  Ae^{i t \omega }+\varepsilon\left[\frac{1}{3} \sqrt{2} \left(2 i \omega ^2+2 \omega -i\right) \sqrt{\omega ^2+1} A^2 e^{2 i t \omega }+\frac{\left(2 \omega ^2-i \omega -1\right) \sqrt{\omega ^2+1} A e^{i t \omega }}{\sqrt{2} \omega  (\omega -i)}\right]\nonumber\\
&+\varepsilon^2\left[\frac{A e^{i t \omega }}{12 \omega ^3 (\omega -i) \sqrt{\omega ^2+1}}\left(2 A A^* \left(8 i \omega ^6-6 \omega ^5+2 i \omega ^4+6 \omega ^3-7 i \omega ^2+3 \omega +2 i\right) \sqrt{\omega ^2+1} \omega ^2\right.\right.\nonumber\\
&\left.\left.+3 \left(-4 i \omega ^6-4 \omega ^5+3 i \omega ^4-2 \omega ^3-2 i \omega ^2+4 \omega +i\right) \sqrt{\omega ^2+1}\right)\right.\nonumber\\
&\left.+\frac{A^2 \left(4 \omega ^6+2 i \omega ^5+7 i \omega ^3+15 \omega ^2-i \omega -5\right) e^{2 i t \omega }}{9 \omega ^2}\right.\nonumber\\
&\left.+\frac{A^3 \left(-8 \omega ^6+14 i \omega ^5+6 \omega ^4+10 i \omega ^3+15 \omega ^2-7 i \omega -2\right) e^{3 i t \omega }}{8 \omega }\right]+\varepsilon^3(\ldots)+\varepsilon^4(\ldots)+(*).\nonumber\\
v_c&=A (-\omega +i) e^{i t \omega }+\varepsilon\left[\frac{A^2 \sqrt{\omega ^2+1} \left(-4 i \omega ^3-6 \omega ^2+4 i \omega +1\right) e^{2 i t \omega }}{3 \sqrt{2} \omega }\right]\nonumber\\
&+\varepsilon^2\left[\frac{A^2 \left(-4 \omega ^7+6 i \omega ^6+8 \omega ^5-10 i \omega ^4-14 \omega ^3+i \omega ^2+\omega -i\right) e^{2 i t \omega }}{9 \omega ^3}\right.\nonumber\\
&\left.+\frac{A^3 \left(24 \omega ^7-50 i \omega ^6-32 \omega ^5-24 i \omega ^4-55 \omega ^3+36 i \omega ^2+13 \omega -2 i\right) e^{3 i t \omega }}{24 \omega ^2}\right]+\varepsilon^3(\ldots)+\varepsilon^4(\ldots)+(*),\label{eq194.1}
\end{align}
where the $\varepsilon^3$ and the $\varepsilon^4$ order terms were omitted.

The modified RG solution is found to be
\begin{align}
A'(t)&=\varepsilon\frac{A \left(-2 \omega ^2+i \omega +1\right) \sqrt{\omega ^2+1}}{\sqrt{2} \omega }+\varepsilon^2\left(Aq_1(\omega)+A^2A^*q_2(\omega)\right)+\varepsilon^3\left(Aq_3(\omega)+A^2A^*q_4(\omega,c_0,d_1)\right)\nonumber\\
&+\varepsilon^4A^2A^*q_5(\omega,c_0,d_0,d_1,e_1),\label{eq194.2}
\end{align}
where the functions $q_i(\omega)$ are some rational functions of $\omega$ and/or the complex constants $c_0,d_0,d_1,e_0,e_1$. The corresponding functions $u_m$ and $v_m$ are
\begin{align}
u_m&=\omega  Ae^{i t \omega }+\varepsilon\left[\frac{1}{3} \sqrt{2} \left(2 i \omega ^2+2 \omega -i\right) \sqrt{\omega ^2+1} A^2 e^{2 i t \omega }+\frac{\left(2 \omega ^2-i \omega -1\right) \sqrt{\omega ^2+1} A e^{i t \omega }}{\sqrt{2} \omega  (\omega -i)}+\omega  \alpha\right]\nonumber\\
&+\varepsilon^2\left[\frac{A e^{i t \omega }}{12 \omega ^3 (\omega -i) \sqrt{\omega ^2+1}}\left(8 \sqrt{2} \alpha  (\omega -i)^2 \left(2 i \omega ^3+i \omega +1\right) \omega ^3\right.\right.\nonumber\\
&\left.\left.+2 A A^* \left(8 i \omega ^6-6 \omega ^5+2 i \omega ^4+6 \omega ^3-7 i \omega ^2+3 \omega +2 i\right) \sqrt{\omega ^2+1} \omega ^2\right.\right.\nonumber\\
&\left.\left.+3 \left(-4 i \omega ^6-4 \omega ^5+3 i \omega ^4-2 \omega ^3-2 i \omega ^2+4 \omega +i\right) \sqrt{\omega ^2+1}\right)\right.\nonumber\\
&\left.+\frac{A^2 \left(4 \omega ^6+2 i \omega ^5+7 i \omega ^3+15 \omega ^2-i \omega -5\right) e^{2 i t \omega }}{9 \omega ^2}\right.\nonumber\\
&\left.+\frac{A^3 \left(-8 \omega ^6+14 i \omega ^5+6 \omega ^4+10 i \omega ^3+15 \omega ^2-7 i \omega -2\right) e^{3 i t \omega }}{8 \omega }\right]+\varepsilon^3(\ldots)+\varepsilon^4(\ldots)+(*).\nonumber\\
v_m&=A (-\omega +i) e^{i t \omega }+\varepsilon\left[\frac{A^2 \sqrt{\omega ^2+1} \left(-4 i \omega ^3-6 \omega ^2+4 i \omega +1\right) e^{2 i t \omega }}{3 \sqrt{2} \omega }-\alpha  (\omega -i)\right]\nonumber\\
&+\varepsilon^2\left[\frac{\sqrt{2} A \sqrt{\omega ^2+1} \left(\alpha ^* \left(6 \omega ^2-3\right)+\alpha  \left(-4 i \omega ^3-6 \omega ^2+4 i \omega +1\right)\right) e^{i t \omega }}{3 \omega }\right.\nonumber\\
&\left.+\frac{A^2 \left(-4 \omega ^7+6 i \omega ^6+8 \omega ^5-10 i \omega ^4-14 \omega ^3+i \omega ^2+\omega -i\right) e^{2 i t \omega }}{9 \omega ^3}\right.\nonumber\\
&\left.+\frac{A^3 \left(24 \omega ^7-50 i \omega ^6-32 \omega ^5-24 i \omega ^4-55 \omega ^3+36 i \omega ^2+13 \omega -2 i\right) e^{3 i t \omega }}{24 \omega ^2}\right]+\varepsilon^3(\ldots)+\varepsilon^4(\ldots)+(*),\label{eq194.3}
\end{align}
with
\begin{align}
\alpha(A,A^*)&=Ac_0,\nonumber\\
\beta(A,A^*)&=A\left(d_0+AA^*d_1\right),\nonumber\\
\gamma(A,A^*)&=A\left(e_0+AA^*e_1\right)+\frac{A^3\left(A^*\right)^2}{864 \sqrt{2} \omega ^2 \left(\omega ^2+1\right)^{3/2} \left(2 \omega ^2-1\right)}\left[d_1\left(576 \omega ^{10}-432 i \omega ^9+720 \omega ^8-360 \omega ^6\right.\right.\nonumber\\
&\left.\left.+648 i \omega ^5-360 \omega ^4+216 i \omega ^3+144 \omega ^2\right)+d_1^*\left(576 \omega ^{10}+432 i \omega ^9+720 \omega ^8-360 \omega ^6-648 i \omega ^5\right.\right.\nonumber\\
&\left.\left.-360 \omega ^4-216 i \omega ^3+144 \omega ^2\right)+320 \omega ^{14}+1920 i \omega ^{13}+2964 \omega ^{12}+480 i \omega ^{11}-848 \omega ^{10}-4272 i \omega ^9\right.\nonumber\\
&\left.-2560 \omega ^8-1344 i \omega ^7-1147 \omega ^6+1992 i \omega ^5+394 \omega ^4+264 i \omega ^3+669 \omega ^2-240 i \omega -76\right],\nonumber\\
\delta(A,A^*)&=0,\label{eq194.4}
\end{align}
where the $\varepsilon^3$ and the $\varepsilon^4$ order terms were omitted. Each order of $\varepsilon$ in (\ref{eq194.3}) contains one homogeneous function. The parameters $c_0,d_0,d_1,e_0$ and $e_1$ will be determined in the next section

\subsubsection{Numerical results}
In this section we look at how well does our modified RG solution perform. Throughout the section we will use $a=\frac{1}{10}$ which gives us $\omega=0.5257$.

\begin{figure}[ht]
  \centering
\captionsetup{width=0.85\textwidth}
\begin{subfigure}{.5\textwidth}
  \centering
  \includegraphics[scale=0.5]{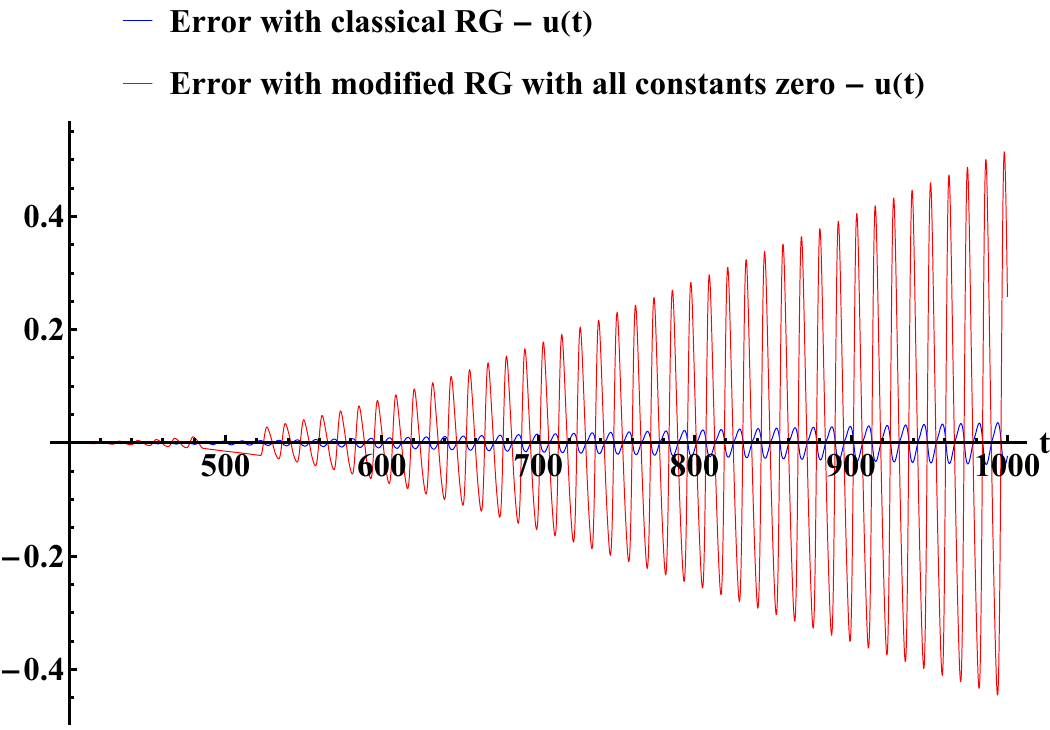}
  \caption{}
  \label{fig11a}
\end{subfigure}%
\begin{subfigure}{.5\textwidth}
  \centering
  \includegraphics[scale=0.5]{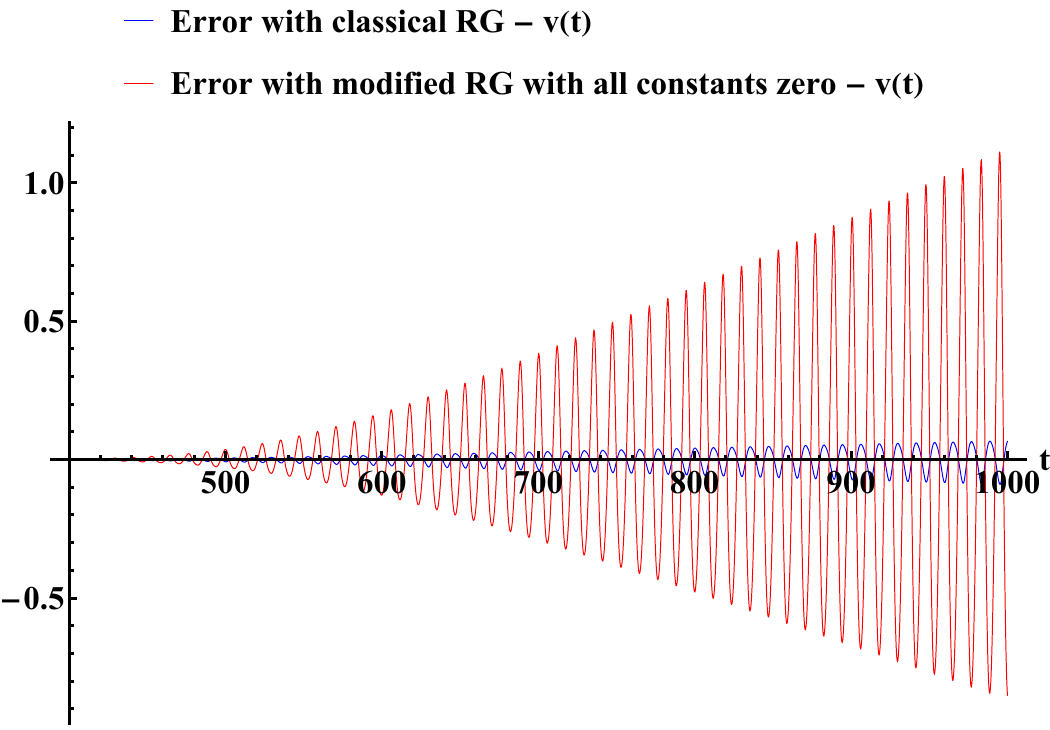}
  \caption{}
  \label{fig11b}
\end{subfigure}
\caption{Comparing the error of the classical RG solution with (\ref{eq208}) to the error of the modified RG solution (\ref{eq194.2}), (\ref{eq194.3}) with $c_0=c_1=d_0=d_1=e_1=0$ against high precision numerical solution. The timeline spans through $t\in [400,1000]$.}
\label{fig11}
\end{figure}

Recall that we have 5 free complex constants in (\ref{eq194.2}), (\ref{eq194.3}) $c_0,c_1,d_0,d_1,e_1$ that we can still choose. We are first going to compare the error of the classical RG solution (\ref{eq194.1}) with the error produced by our modified RG solution (\ref{eq194.3}) where we set all the parameters to zero. \iffalse
The amplitude equation simplifies to
\begin{align}
A'(t)&=A\left[-\varepsilon\frac{\sqrt{\omega ^2+1} \left(2 \omega ^2-i \omega -1\right)}{\sqrt{2} \omega }-\varepsilon^2\frac{i \left(4 \omega ^6+4 i \omega ^5+\omega ^4-4 i \omega ^3-4 \omega ^2-2 i \omega +1\right)}{4 \omega ^3}\right.\nonumber\\
&\left.+\varepsilon^3\frac{i\left(12 \omega ^8+\omega ^6-15 \omega ^4-\omega ^2+3\right)}{4 \sqrt{2} \omega ^4 \sqrt{\omega ^2+1}}-\varepsilon^4\frac{i \left(16 \omega ^{12}+72 \omega ^{10}-59 \omega ^8-60 \omega ^6+30 \omega ^4+8 \omega ^2+1\right)}{32 \omega ^7}\right]\nonumber\\
&+A^2 A^*\left[\varepsilon^2\frac{ \left(-8 i \omega ^6+6 \omega ^5-2 i \omega ^4-6 \omega ^3+7 i \omega ^2-3 \omega -2 i\right)}{6 \omega }+\varepsilon^3\left(-\frac{1}{18 \sqrt{2} \omega ^3 \sqrt{\omega ^2+1}}\left(208 \omega ^{10}\right.\right.\right.\nonumber\\
&\left.\left.\left.-336 i \omega ^9+264 \omega ^8+18 i \omega ^7-332 \omega ^6+279 i \omega ^5-167 \omega ^4+15 i \omega ^3+183 \omega ^2-54 i \omega -38\right)\right)\right.\nonumber\\
&\left.+\varepsilon^4\left(-\frac{i}{216 \omega ^5 \left(\omega ^2+1\right)}\left(8672 \omega ^{14}+9024 i \omega ^{13}+5872 \omega ^{12}-2664 i \omega ^{11}-6110 \omega ^{10}\right.\right.\right.\nonumber\\
&\left.\left.\left.-16788 i \omega ^9-9003 \omega ^8-1044 i \omega ^7+171 \omega ^6+7830 i \omega ^5+1567 \omega ^4+258 i \omega ^3+901 \omega ^2-924 i \omega\right.\right.\right.\nonumber\\
&\left.\left.\left. +14\right)\right)\right].\label{eq209}
\end{align}
\fi
We can see the results in figure (\ref{fig11}). The left plot displays the error of the function $u(t)$, while the right plot shows the error of $v(t)$. It is evident that the classical RG solution performs significantly better. In fact, its error is approximately ten times lower than that of the modified RG solution. However, this issue can be addressed rather easily. Recall that the free constants were set to zero; in reality, they can be adjusted to minimize the solution's error.

The free constants play a role in the $A^2A^*$ term of the amplitude equation (\ref{eq194.2}) through the functions $g$ and $h$. These functions depend on the constants in a polynomial manner, meaning that only one of them can be utilized to minimize the error. The other can be assigned an arbitrary value, such as zero.

To determine which constant should be treated as a variable, we should select the one that appears the fewest times in the equations. This choice impacts both computational efficiency in the code and the simplification of expressions, as setting the rest to zero helps eliminate unnecessary terms. The constants introduced later in the hierarchy tend to appear less frequently; in this case, it is $e_1$.

Thus, we set $c_0=d_0=d_1=e_0=0$ and keep $e_1=\text{Re}(e_1)+i\text{Im}(e_1)$ as a variable. This variable will be used to minimize the error of the solution through gradient descent, just as we did in the case of the Van der Pol oscillator (\ref{eq92}). The homogeneous functions then simplify to
\begin{align}
\alpha(A,A^*)&=0,\nonumber\\
\beta(A,A^*)&=0,\nonumber\\
\gamma(A,A^*)&=A^2A^*e_1+\frac{A^3\left(A^*\right)^2}{864 \sqrt{2} \omega ^2 \left(\omega ^2+1\right)^{3/2} \left(2 \omega ^2-1\right)}\left[320 \omega ^{14}+1920 i \omega ^{13}+2964 \omega ^{12}+480 i \omega ^{11}\right.\nonumber\\
&\left.-848 \omega ^{10}-4272 i \omega ^9-2560 \omega ^8-1344 i \omega ^7-1147 \omega ^6+1992 i \omega ^5+394 \omega ^4+264 i \omega ^3+669 \omega ^2\right.\nonumber\\
&\left.-240 i \omega -76\right],\nonumber\\
\delta(A,A^*)&=0.\label{eq210}
\end{align}

\noindent
The amplitude equation in this case simplifies to
\begin{align}
A'(t)&=A\left[-\varepsilon\frac{\sqrt{\omega ^2+1} \left(2 \omega ^2-i \omega -1\right)}{\sqrt{2} \omega }-\varepsilon^2\frac{i \left(4 \omega ^6+4 i \omega ^5+\omega ^4-4 i \omega ^3-4 \omega ^2-2 i \omega +1\right)}{4 \omega ^3}\right.\nonumber\\
&\left.+\varepsilon^3\frac{i\left(12 \omega ^8+\omega ^6-15 \omega ^4-\omega ^2+3\right)}{4 \sqrt{2} \omega ^4 \sqrt{\omega ^2+1}}-\varepsilon^4\frac{i \left(16 \omega ^{12}+72 \omega ^{10}-59 \omega ^8-60 \omega ^6+30 \omega ^4+8 \omega ^2+1\right)}{32 \omega ^7}\right]\nonumber\\
&+A^2 A^*\left[\varepsilon^2\frac{ \left(-8 i \omega ^6+6 \omega ^5-2 i \omega ^4-6 \omega ^3+7 i \omega ^2-3 \omega -2 i\right)}{6 \omega }+\varepsilon^3\left(-\frac{1}{18 \sqrt{2} \omega ^3 \sqrt{\omega ^2+1}}\left(208 \omega ^{10}\right.\right.\right.\nonumber\\
&\left.\left.-336 i \omega ^9+264 \omega ^8+18 i \omega ^7-332 \omega ^6+279 i \omega ^5-167 \omega ^4+15 i \omega ^3+183 \omega ^2-54 i \omega -38\right)\right)\nonumber\\
&\left.+\varepsilon^4\left(-\frac{i}{216 \omega ^5 \left(\omega ^2+1\right)}\left(8672 \omega ^{14}+9024 i \omega ^{13}+5872 \omega ^{12}-2664 i \omega ^{11}-6110 \omega ^{10}\right.\right.\right.\nonumber\\
&\left.\left.\left.-16788 i \omega ^9-9003 \omega ^8-1044 i \omega ^7+171 \omega ^6+7830 i \omega ^5+1567 \omega ^4+258 i \omega ^3+901 \omega ^2-924 i \omega\right.\right.\right.\nonumber\\
&\left.\left.\left. +14\right)+e_1\frac{\sqrt{2} \sqrt{\omega ^2+1} \left(2 \omega ^2-1\right)}{\omega }\right)\right].\label{eq211}
\end{align}
Our new variable $e_1$ appears in the $\varepsilon^4$ order $A^2A^*$ term.

Let us set up the function to be minimized.
\begin{align}
&F\left(\text{Re}(e_1),\text{Im}(e_1)\right):=\nonumber\\
&- \text{define the functions } \text{ accorgind to (\ref{eq210}) and the functions }u_m(t)\text{ and }v_m(t) \text{ from (\ref{eq194.3}) using the input},\nonumber\\
&-\text{calculate the initial conditions and solve the amplitude equation (\ref{eq211})},\nonumber\\
&-\text{use the calculated amplitude back in (\ref{eq194.3}), then the output is }\nonumber\\
&\max_{t\in(950,1000)}\left( u_{\text{exact}}(t)-u_{m\text{ with this input}}(t)\right)^2+\left( v_{\text{exact}}(t)-v_{m\text{ with this input}}(t)\right)^2.\label{eq212}
\end{align}
The function $F$ is then minimized using an optimalization method, for example gradient descent method starting from the point $(0,0)$. We find that this method converges to the following value
\begin{align}
e_1=-9.4786+i69.6637.\label{eq213}
\end{align}

\begin{figure}[ht]
  \centering
\captionsetup{width=0.85\textwidth}
\begin{subfigure}{.5\textwidth}
  \centering
  \includegraphics[scale=0.5]{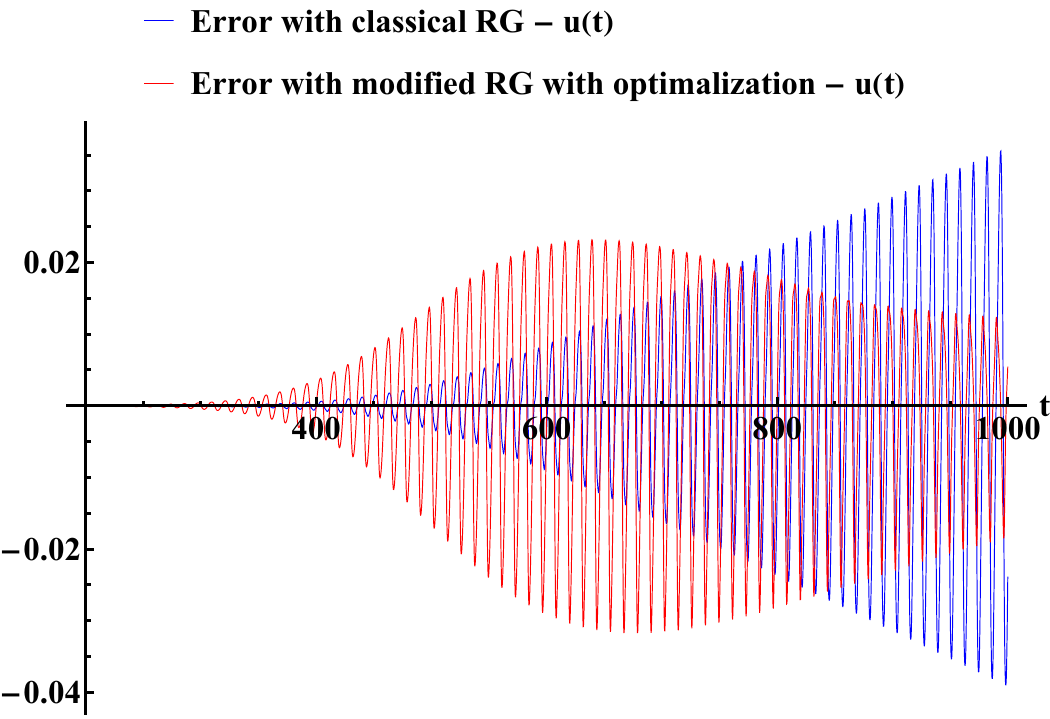}
  \caption{}
  \label{fig12a}
\end{subfigure}%
\begin{subfigure}{.5\textwidth}
  \centering
  \includegraphics[scale=0.5]{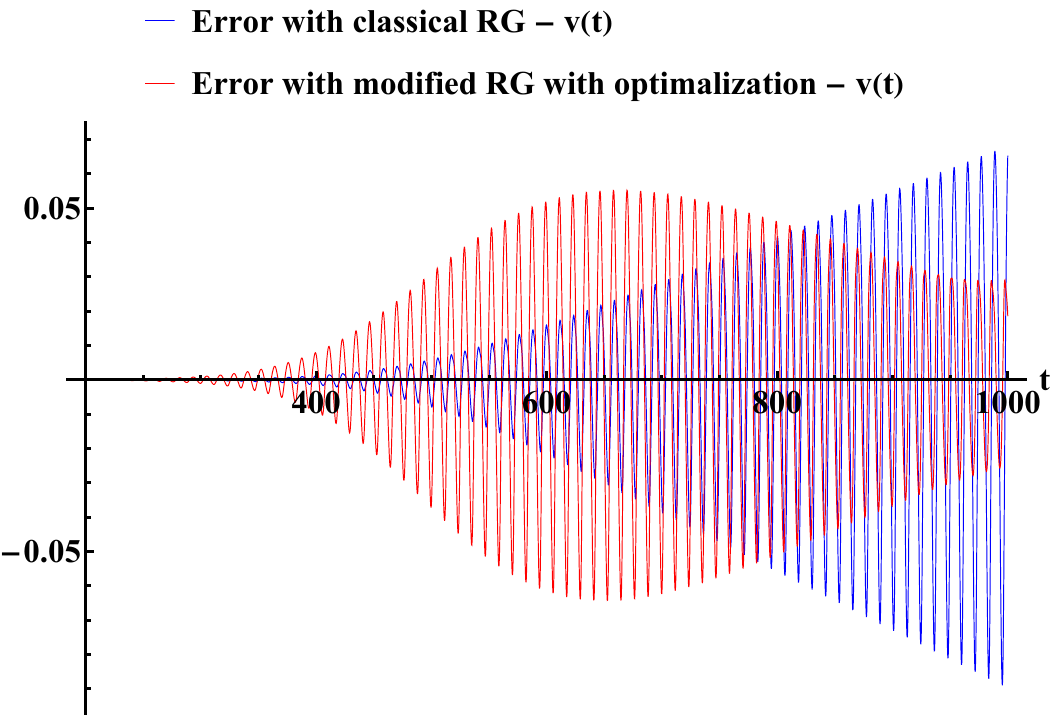}
  \caption{}
  \label{fig12b}
\end{subfigure}
\caption{Comparing the error of the classical RG solution with (\ref{eq208}) and the error of the modified RG solution (\ref{eq194.2}), (\ref{eq194.3}) with $c_0=c_1=d_0=d_1=0$ and optimized constant value for $e_1$ being (\ref{eq213}) against high precision numerical solution. The timeline spans through $t\in [200,1000]$.}
\label{fig12}
\end{figure}

In figure (\ref{fig12}), we compare the errors of the solutions. Our modified RG solution performs much better with the optimized value of $e_1$ (\ref{eq213}) and with $c_0=c_1=d_0=d_1=0$ than before. However, it is not immediately clear which solution is superior. For $t\in[200,750]$, the error of the classical RG solution is lower, but for the rest of the range, the modified RG solution performs better. In any case, both solutions are within the same order of magnitude, which is, at the very least, acceptable.

\section{Conclusion}
In this paper, we have successfully demonstrated a modified RG method, showcasing its improved accuracy and simpler amplitude equation compared to the classical RG method. To remove orders from the amplitude equation, we solved a quasilinear PDE for the additional homogeneous functions, which are central to our new method. We found that the amplitude equation cannot be simplified indefinitely; there is a limit, and the terms that cannot vanish are referred to as \textit{the core} of the amplitude equation. 

The first section presents a scalar ODE: the Duffing equation, which offers a straightforward demonstration of the advantages of our method. However, when applied to other examples—such as the Van der Pol oscillator—some minor differences arise. In that case, the core of the amplitude equation was modified by introducing additional terms as higher-order corrections were considered. These new terms included complex constants originating from the homogeneous functions, which were subsequently tuned using a gradient descent method to minimize the error of the solution.

Our method was further demonstrated on a system of second-order ODEs. 
This system introduced two amplitudes, complicating the quasilinear equation for the homogeneous functions. Nevertheless, this challenge could be successfully navigated, and our method resulted in a solution with significant improvements in both accuracy and simplicity of the amplitude equations. As in the case of the scalar ODE, our method was tested on several other systems, each presenting unique characteristics with different aspects of our approach. The first, the Lotka-Volterra system, involved a single amplitude, allowing for a straightforward application of our method.  Finally, the Selkov model, widely used in the study of dynamical systems, produced a system with one amplitude, but the core of the amplitude equation gained additional factor terms containing free constants. As demonstrated in the Van der Pol case, these constants could be utilized to minimize the error. The details of all these calculations are presented and can be viewed in \cite{juhasz2025}.

It should be noted that the presence of these free functions ($\alpha, \beta, \dots$) implies that the final perturbation solution is not unique. Here, these functions are utilized to eliminate all non-essential terms from the amplitude equation. In contrast, for the Van der Pol oscillator, they were chosen to minimize the approximation error, demonstrating the versatility of this approach.

We propose that this modified RG method can be generalized to nonlinear PDEs as well. While the RG method has already been applied to PDEs, its success has been limited \cite{CGO1995}, \cite{CGO1996}, \cite{RGonPDE}, \cite{RGonPDE2}. One potential application of our modified RG method is in solving the light propagation equation with cubic nonlinearity, such as the Kerr effect. This equation leads to the nonlinear Schr\"{o}dinger equation, which cannot be solved analytically when higher-order terms are included. Our method would address this challenge by effectively removing the higher-order terms, simplifying the solution process.

\begin{appendices}
\setcounter{equation}{0}
\numberwithin{equation}{section}
\section*{Appendix A, Van der Pol oscillator}
\renewcommand{\theequation}{A.\arabic{equation}}
In this appendix, we derive the solution to the Van der Pol oscillator (\ref{eq55}) using the modified RG method. Starting with the naive expansion (\ref{eq2}), we get the perturbation hierarchy
\begin{align}
\text{order }\varepsilon^0:\;y_0''+y_0&=0,\nonumber\\
\text{order }\varepsilon^1:\;y_1''+y_1&=y_0'-y_0^2y'_0,\nonumber\\
\text{order }\varepsilon^2:\;y_2''+y_2&=y_1'-2y_0y_1y_0' -y_0^2y_1',\nonumber\\
\text{order }\varepsilon^4:\;y_4''+y_4&=y_2'-y_1^2y_0'-2y_0y_2y_0'-2y_0y_1y_1'-y_0^2y_2',\nonumber\\
\text{order }\varepsilon^4:\;y_3''+y_3&=y_3'-2y_1y_2y_0'-2y_0y_3y_0'-2 y_0y_2y_1'-2y_0y_1y_2'-y_0^2y_3'-y_1^2y_1'.\label{eq56}
\end{align}
With this example we shall go up to $\varepsilon^4$ order, as our method will be applied in a slightly different manner at the end. The solution to the hierarchy (\ref{eq56}) is straightforward, and in an illustrative way, one obtains
\begin{align}
y_0(t)&=A_0e^{it}+A_0^*e^{-it},\nonumber\\
y_1(t)&=\frac{1}{8} i A_0^3 e^{3 i t}+e^{i t} \left(\alpha_0 -\frac{1}{2} A_0^2 A_0^* t+\frac{A_0 t}{2}\right)+(*),\nonumber\\
y_2(t)&=-\frac{5}{192} A_0^5 e^{5 i t}+e^{3 i t} \left(-\frac{1}{16} 3 i A_0^4A_0^*t-\frac{A_0^4 A_0^*}{64}+\frac{3}{16} i A_0^3 t-\frac{A_0^3}{32}+\frac{3}{8} i \alpha_0  A_0^2\right)\nonumber\\
&+e^{i t} \left(\beta_0+\frac{3}{8} A_0^3\left(A_0^*\right)^2 t^2-\frac{7}{16} i A_0^3\left(A_0^*\right)^2 t-\frac{1}{2} A_0^2 A_0^*t^2+\frac{1}{2} i A_0^2A_0^*t-\frac{1}{2} A_0^2\alpha_0^* t-\alpha_0  A_0A_0^*t\right.\nonumber\\
&\left.+\frac{A_0 t^2}{8}-\frac{i A_0 t}{8} +\frac{\alpha_0  t}{2}\right)+(*),\nonumber\\
y_3(t)&=e^{it}\left(\gamma_0+P_3(t)\right)+e^{3it}P_2(t)+e^{5it}P_1(t)-e^{7 i t}\frac{7 i A_0^7 }{1152}+(*),\nonumber\\
y_4(t)&=e^{it}\left(\delta_0+P_4(t)\right)+e^{3it}P_3(t)+e^{5it}P_2(t)-e^{7 i t}P_1(t)+e^{9 i t}\frac{61 A_0^9 }{40960}+(*),\label{eq57}
\end{align}
where we included also the homogeneous solutions $\alpha_0,\beta_0,\gamma_0,\delta_0$ and $P_n(t)$ are some polynomials of order $n$. The naive solution to the perturbation hierarchy then reads
\begin{align}
y(t)=\sum_{n=0}^4\varepsilon^ny_n(t-t_0),\label{eq58}
\end{align}
where we shifted the starting point from 0 to $t_0$. The next step is to split the interval $t-t_0$ into $\tau+\xi$ with $\tau=t-\mu$ and $\xi=\mu-t_0$ for some point in time $\mu$. The amplitude $A_0$ and all the homogeneous solutions are now renormalized as
\begin{align}
A_0(t_0)\approx\sum_{n=0}^4 Z_n(\xi,\mu)\varepsilon^nA(\mu)= Z(\xi,\mu)A(\mu),\label{eq59}\\
\alpha_0(t_0)\approx\sum_{n=0}^3 U_n(\xi,\mu)\varepsilon^n\alpha(\mu)= U(\xi,\mu)\alpha(\mu),\label{eq60}\\
\beta_0(t_0)\approx\sum_{n=0}^2 V_n(\xi,\mu)\varepsilon^n\beta(\mu)= V(\xi,\mu)\beta(\mu),\label{eq61}\\
\gamma_0(t_0)\approx\sum_{n=0}^1 W_n(\xi,\mu)\varepsilon^n\gamma(\mu)= W(\xi,\mu)\gamma(\mu),\label{eq62}\\
\delta_0(t_0)\approx\sum_{n=0}^0 X_n(\xi,\mu)\varepsilon^n\delta(\mu)= X(\xi,\mu)\delta(\mu),\label{eq63}
\end{align}
where the maximum exponent for $\varepsilon$ for each quantity was determined with respect to the order for which they appear in the expansion (\ref{eq58}). Substituting these renormalization groups into (\ref{eq58}) we get
\begin{align}
y(t)&=Z_0 A(\mu ) e^{i \xi +i \tau }+\varepsilon\left(-\frac{1}{2} Z_0^2 Z_0^* A(\mu )^2 A^*(\mu) e^{i \xi +i \tau } (\xi +\tau )+\frac{1}{8} i Z_0^3 A(\mu )^3 e^{3 i \xi +3 i \tau }\right.\nonumber\\
&\left.+\frac{1}{2} Z_0 A(\mu ) e^{i \xi +i \tau } (\xi +\tau )+Z_1 A(\mu ) e^{i \xi +i \tau }+U_0 \alpha (\mu ) e^{i \xi +i \tau }\right)+\ldots+(*),\label{eq64}
\end{align}
where the higher orders terms were omitted due to their extensive lengths. As before, we use $Z_n,U_n,\ldots$ to remove the terms including the $\xi$-variable. Thus we get $Z_0=e^{-i\xi}$. From the $\varepsilon^1$ order term we get
\begin{align}
Z_1&=\frac{1}{2} e^{-i \xi } \xi  A(\mu ) A^*(\mu )-\frac{1}{2} e^{-i \xi } \xi,\nonumber\\
U_0&=e^{-i\xi},\label{eq65}
\end{align}
with their respective conjugate counterparts. The $\varepsilon^2$ becomes $\xi$-free if we choose
\begin{align}
Z_2&=-\frac{e^{-i \xi } \xi  \alpha}{2 A}+\frac{1}{2} e^{-i \xi } \xi  A \alpha^*+\frac{3}{8} e^{-i \xi } \xi ^2 A^2 \left( A^*\right)^2-\frac{1}{2} e^{-i \xi } \xi ^2 A A^*+\frac{7}{16} i e^{-i \xi } \xi  A^2 \left(A^*\right)^2-\frac{1}{2} i e^{-i \xi } \xi  A A^*\nonumber\\
&+e^{-i \xi } \xi  \alpha A^*+\frac{1}{8} e^{-i \xi } \xi ^2+\frac{1}{8} i e^{-i \xi } \xi,\nonumber\\
U_1&=0,\nonumber\\
V_0&=e^{-i\xi}.\label{eq66}
\end{align}
We continue in the same way for the higher orders and determine all the normalization quantities. After renormalizing the naive solution (\ref{eq58}) we get
\begin{align}
y(\mu)&=A e^{i \tau }+\varepsilon\left(\frac{1}{8} e^{3 i \tau } i A^3-\frac{1}{2} e^{i \tau } \tau  A^* A^2+\frac{1}{2} e^{i \tau } \tau  A+e^{i \tau } \alpha \right)+\varepsilon ^2\left(-\frac{1}{192} 5 e^{5 i \tau } A^5-\frac{1}{64} e^{3 i \tau } A^* A^4\right.\nonumber\\
&-\frac{3}{16} i e^{3 i \tau } \tau  A^* A^4-\frac{1}{32} e^{3 i \tau } A^3+\frac{3}{8} e^{i \tau } \tau ^2 \left(A^*\right)^2 A^3-\frac{7}{16} i e^{i \tau } \tau  \left(A^*\right)^2 A^3+\frac{3}{16} e^{3 i \tau } i \tau  A^3+\frac{3}{8} e^{3 i \tau } i \alpha  A^2\nonumber\\
&\left.-\frac{1}{2} e^{i \tau } \tau ^2 A^* A^2+\frac{1}{2} e^{i \tau } i \tau  A^* A^2-\frac{1}{2} e^{i \tau } \tau  \alpha ^* A^2+\frac{1}{8} e^{i \tau } \tau ^2 A-\frac{1}{8} i e^{i \tau } \tau  A-e^{i \tau } \alpha  \tau  A^* A+e^{i \tau } \beta +\frac{1}{2} e^{i \tau } \alpha  \tau \right)\nonumber\\
&+\varepsilon ^3(\ldots)+\varepsilon^4(\ldots)+(*),\label{eq67}
\end{align}
where $\tau=t-\mu$ and the amplitude $A$ depends on $\mu$ and we assume the dependence\\ $\alpha(A,A^*),\beta(A,A^*)\ldots$. Next, the amplitude equation is determined by the equation $\partial_\mu y\Big|_{\mu\rightarrow t}=0$. Splitting the function $y(\mu)$ into to parts that are complex conjugate of each other and equating them to zero, we solve them for $A'$ and $(A^*)'$. The solutions are then simplified as in the case of the Duffing equation, and only the terms with order $\varepsilon^4$ or lower are kept. The amplitude equation then reads
\begin{align}
A'(t)&=i A+\varepsilon  \left(-i A \partial_A\alpha +i A^* \partial_{A^*}\alpha+i \alpha -\frac{1}{2} A^* A^2+\frac{A}{2}\right)+\varepsilon ^2 \left(-i A \partial_A\beta +i A^* \partial_{A^*}\beta+i \beta\right.\nonumber\\
&+i A \partial_{A^*}\alpha \partial_{A^*}\alpha^* -i A^* \partial_{A^*}\alpha \partial_A\alpha^* +i A \left(\partial_A\alpha\right) ^2+\frac{1}{2} \left(A^*\right)^2 A \partial_{A^*}\alpha -\frac{1}{2} A \partial_A\alpha -\frac{1}{2} A^* \partial_{A^*}\alpha \nonumber\\
&-i \alpha  \partial_A\alpha-i A^* \partial_{A^*}\alpha  \partial_A\alpha +i \alpha ^* \partial_{A^*}\alpha  +\frac{1}{2} A^* A^2 \partial_A\alpha+\frac{\alpha }{2}-\frac{\alpha ^* A^2}{2}-\alpha  A^* A-\frac{1}{16} 7 i \left(A^*\right)^2 A^3\nonumber\\
&\left.+\frac{1}{2} i A^* A^2-\frac{i A}{8} \right)+\varepsilon^3\left(-i A \partial_A\gamma +i A^* \partial_{A^*}\gamma+i \gamma+\ldots\right)+\varepsilon^4\left(-i A \partial_A\delta +i A^* \partial_{A^*}\delta+i \delta+\ldots\right).\label{eq68}
\end{align}
The overall solution is determined from (\ref{eq67}) by taking $\tau\rightarrow 0$ and $\mu\rightarrow t$. We arrive at
\begin{align}
y(t)&=A+\varepsilon  \left(\alpha +\frac{i A^3}{8}\right)+\varepsilon ^2 \left(\beta-\frac{5 A^5}{192}-\frac{A^3}{32}+\frac{3}{8} i \alpha  A^2-\frac{A^* A^4}{64} \right)+\varepsilon ^3 \left(\gamma-\frac{7 i A^7}{1152}-\frac{35 i A^5}{2304}\right.\nonumber\\
&\left.-\frac{\alpha ^* A^4}{64}-\frac{25 \alpha  A^4}{192}+\frac{i A^3}{128}-\frac{3 \alpha  A^2}{32}+\frac{3}{8} i A^2 \beta -\frac{5 i A^* A^6}{1536}+\frac{29}{512} i \left(A^*\right)^2 A^5-\frac{21}{256} i A^* A^4-\frac{1}{16} \alpha  A^* A^3\right.\nonumber\\
&\left.+\frac{3}{8} i \alpha ^2 A \right)+\varepsilon ^4 \left(\delta+\frac{i \alpha ^3}{8}+\frac{61 A^9}{40960}+\frac{623 A^7}{110592}-\frac{5 i \alpha ^* A^6}{1536}-\frac{49 i \alpha  A^6}{1152}+\frac{5 A^5}{27648}-\frac{21}{256} i \alpha ^* A^4\right.\nonumber\\
&\left.-\frac{175 i \alpha  A^4}{2304}-\frac{A^4 \beta ^*}{64}-\frac{25 A^4 \beta }{192}-\frac{25 \alpha ^2 A^3}{96}-\frac{1}{16} \alpha  \alpha ^* A^3-\frac{3 A^3}{512}+\frac{3}{128} i \alpha  A^2-\frac{3 A^2 \beta }{32}+\frac{3}{8} i A^2 \gamma\right.\nonumber\\
&\left. +\frac{133 A^* A^8}{221184}-\frac{2521 \left(A^*\right)^2 A^7}{110592}+\frac{989 \left(A^*\right)^3 A^6}{12288}+\frac{197 A^* A^6}{6144}+\frac{29}{256} i \alpha ^* A^* A^5-\frac{5}{256} i \alpha  A^* A^5\right.\nonumber\\
&\left.-\frac{1103 \left(A^*\right)^2 A^5}{6144}+\frac{145}{512} i \alpha  \left(A^*\right)^2 A^4+\frac{113 A^* A^4}{1024}-\frac{21}{64} i \alpha  A^* A^3-\frac{1}{16} A^* A^3 \beta -\frac{3}{32} \alpha ^2 A^* A^2\right.\nonumber\\
&\left.-\frac{3 \alpha ^2 A}{32}+\frac{3}{4} i \alpha  A \beta  \right)+(*),\label{eq69}
\end{align}

Let us deploy our method to simplify the amplitude equation (\ref{eq68}). We observe that the homogeneous bit of the equation for $\alpha$ in the $\varepsilon^1$ order is exactly the same as in (\ref{eq25}). Therefore the homogeneous solution is
\begin{align}
\alpha(A,A^*)=c(AA^*)A,\label{eq70}
\end{align}
for some function $c$. The terms in the particular part of the equation for $\alpha$ are identified as the core, namely $A/2-A^2A^*/2$. Next, we assume a polynomial function $c$ of the form
\begin{align}
c(AA^*)=c_0+c_1AA^*,\label{eq71}
\end{align}
where $c_0$ and $c_1$ are complex constants. Then the $\varepsilon^2$ order in (\ref{eq68}) becomes
\begin{align}
&-i A \partial_A\beta +i A^* \partial_{A^*}\beta +i \beta +\left(A^*\right)^2 A^3 \left(i\text{Im}[c_1]-i\frac{7}{16}\right)+A^* A^2 \left(-i\text{Im}[c_1] -\text{Re}[c_0]-\text{Re}[c_1]+\frac{i}{2}\right)\nonumber\\
&-\frac{i A}{8}.\label{eq72}
\end{align}
According to our method, we don't need to be concerned about the core terms $A,A^2A^*$. The term we would like to remove is $A^3(A^*)^2$. This can be easily done choosing
\begin{align}
\text{Im}[c_1]=\frac{7}{16}.\label{eq73}
\end{align}
We observe that with this choice, the factor at $A^2A^*$ cannot be removed because the imaginary part is a fixed constant, and the only free parameters we have there are in the real part. The other core term, $A$, does not have any free parameters, therefore it cannot be removed either. In summary, we have core terms in both $\varepsilon^1$ and $\varepsilon^2$ order. The rest of the free parameters is left untouched for now. The $\varepsilon^2$ equation is then solved for $\beta$ taking only the homogeneous part and we get
\begin{align}
\beta(A,A^*)=d(AA^*)A=A\left(d_0+d_1AA^*+d_2(AA^*)^2\right),\label{eq74}
\end{align}
where we assumed a polynomial form of the function $d$ for some complex constants $d_{0,1,2}$.

Using both $\alpha$ and $\beta$, the $\varepsilon^3$ term turns to
\begin{align}
&-i A \gamma ^{(1,0)}\left(A,A^*\right)+i A^* \gamma ^{(0,1)}\left(A,A^*\right)+i \gamma \left(A,A^*\right)+\left(A^*\right)^3 A^4 \left(-\frac{1}{2} 3 \text{Re}\left(c_1\right){}^2-\frac{21}{8} i \text{Re}\left(c_1\right)\right.\nonumber\\
&\left.+2 i \text{Im}\left(d_2\right)+\text{Re}\left(d_2\right)-\frac{99}{512}\right)+\left(A^*\right)^2 A^3 \left(-i \text{Im}\left(c_0\right) \text{Re}\left(c_1\right)+3 \text{Re}\left(c_1\right){}^2+\frac{11}{4} i \text{Re}\left(c_1\right)-\frac{21}{16} i \text{Re}\left(c_0\right)\right.\nonumber\\
&\left.+i \text{Im}\left(d_1\right)-2 i \text{Im}\left(d_2\right)-2 \text{Re}\left(d_2\right)+\frac{91}{256}\right)+A^* A^2 \left(i \text{Im}\left(c_0\right) \text{Re}\left(c_1\right)-\frac{1}{2} \text{Im}\left(c_0\right){}^2-\frac{7 \text{Im}\left(c_0\right)}{16}\right.\nonumber\\
&\left.-\frac{1}{2} \text{Re}\left(c_0\right){}^2+\frac{23}{16} i \text{Re}\left(c_0\right)+\text{Re}\left(c_0\right) \text{Re}\left(c_1\right)-i \text{Im}\left(d_1\right)-\text{Re}\left(d_0\right)-\text{Re}\left(d_1\right)-\frac{1}{4}\right)\label{eq75}
\end{align}
As before, we look away from the core terms and focus on the other two which are $A^3(A^*)^2$ and $A^4(A^*)^3$. The factors at these terms are complex so we need to remove both the real and imaginary parts. If we collect the real part from the factors at $A^3(A^*)^2$ and $A^4(A^*)^3$ we get a system of algebraic equations
\begin{align}
-\frac{1}{2} 3 \text{Re}\left(c_1\right){}^2+\text{Re}\left(d_2\right)-\frac{99}{512}&=0,\nonumber\\
3 \text{Re}\left(c_1\right){}^2-2 \text{Re}\left(d_2\right)+\frac{91}{256}&=0.\label{eq76}
\end{align}
It is easy to see that this system doesn't have a solution. This means that we cannot remove both $A^3(A^*)^2$ and $A^4(A^*)^3$ simultaneously. Thus we can decide which one we remove, so we choose $A^3(A^*)^2$. This is achieved with
\begin{align}
\text{Re}\left(d_2\right)&=\frac{91}{512},\nonumber\\
\text{Re}\left(c_1\right)&=0.\label{eq77}
\end{align}
To remove also the imaginary part of the factor of $A^3(A^*)^2$, we choose
\begin{align}
\text{Re}\left(c_0\right)=\text{Re}\left(c_1\right)=\text{Im}\left(d_1\right)=\text{Im}\left(d_2\right)=0.\label{eq78}
\end{align}
After these choices, the $\varepsilon^3$ order becomes
\begin{align}
-i A \partial_A\gamma +i A^* \partial_{A^*}\gamma +i \gamma -\frac{1}{64} \left(A^*\right)^3 A^4+A^* A^2 \left(-\frac{1}{2} \text{Im}\left(c_0\right){}^2-\frac{7 \text{Im}\left(c_0\right)}{16}-\text{Re}\left(d_0\right)-\text{Re}\left(d_1\right)-\frac{1}{4}\right).\label{eq79}
\end{align}
Equating this expression to zero, we solve it for $\gamma$ taking only the non-core term $A^4(A^*)^3$.
\begin{align}
\gamma(A,A^*)&=\frac{1}{64}iA^4(A^*)^3\Log(A)+e(AA^*)A\nonumber\\
&=\frac{1}{64}iA^4(A^*)^3\Log(A)+A\left(e_0+e_1AA^*+e_2(AA^*)^2+e_3(AA^*)^3\right).\label{eq80}
\end{align}
Here, we had to remove also the term $-\frac{1}{64} \left(A^*\right)^3 A^4$, that is why we get the $\Log$ function. Also, we again assumed a polynomial form of the function $e$ which we will use in the next order terms. The $\Log$ function could be dangerous if we didn't have the core of the amplitude equation, because otherwise it would cause growing. The rest of the parameters in (\ref{eq79}) we leave untouched.

We are ready for the last order $\varepsilon^4$ in the amplitude equation (\ref{eq68}) which now reads
\begin{align}
&-i A \partial_A\delta +i A^* \partial_{A^*}\delta +i \delta +\left(A^*\right)^4 A^5 \left(\frac{1}{128} i \Log \left(A^*\right)+\frac{5}{128} i \Log (A)+3 i \text{Im}\left(e_3\right)+2 \text{Re}\left(e_3\right)\right.\nonumber\\
&\left.-\frac{15551 i}{24576}\right)+\left(A^*\right)^3 A^4 \left(-\frac{3}{64} i \Log (A)-\frac{389}{256} i \text{Im}\left(c_0\right)-\frac{21}{8} i \text{Re}\left(d_1\right)+2 i \text{Im}\left(e_2\right)-3 i \text{Im}\left(e_3\right)+\text{Re}\left(e_2\right)\right.\nonumber\\
&\left.-3 \text{Re}\left(e_3\right)+\frac{1397 i}{8192}\right)+\left(A^*\right)^2 A^3 \left(-i \text{Im}\left(c_0\right) \text{Re}\left(d_1\right)-\frac{1}{8} 7 i \text{Im}\left(c_0\right){}^2+\frac{399}{256} i \text{Im}\left(c_0\right)-\frac{21}{16} i \text{Re}\left(d_0\right)\right.\nonumber\\
&\left.+\frac{11}{4} i \text{Re}\left(d_1\right)+i \text{Im}\left(e_1\right)-2 i \text{Im}\left(e_2\right)-2 \text{Re}\left(e_2\right)+\frac{141 i}{256}\right)+A^* A^2 \left(i \text{Im}\left(c_0\right) \text{Re}\left(d_1\right)-\text{Im}\left(c_0\right) \text{Im}\left(d_0\right)\right.\nonumber\\
&\left.+\frac{15}{16} i \text{Im}\left(c_0\right){}^2-\frac{7 \text{Im}\left(d_0\right)}{16}+\frac{23}{16} i \text{Re}\left(d_0\right)-i \text{Im}\left(e_1\right)-\text{Re}\left(e_0\right)-\text{Re}\left(e_1\right)-\frac{i}{16}\right)-\frac{i A}{128}.\label{eq81}
\end{align}
Even though we have $\Log$ functions in the factors of $A^5(A^*)^4$ and so on, which we cannot remove, we will try to simplify that factor so that only the $\Log$ functions are left. The term $A^3(A^*)^2$ can be removed completely. The equations to solve are
\begin{align}
3 i \text{Im}\left(e_3\right)-\frac{15551 i}{24576}&=0,\nonumber\\
\frac{1397 i}{8192}+2 i \text{Im}\left(e_2\right)-3 i \text{Im}\left(e_3\right)&=0,\nonumber\\
\frac{141 i}{256}+i \text{Im}\left(e_1\right)-2 i \text{Im}\left(e_2\right)&=0,\label{eq82}
\end{align}
which has the solution
\begin{align}
\text{Im}\left(e_3\right)&=\frac{15551}{73728},\nonumber\\
\text{Im}\left(e_2\right)&=\frac{355}{1536},\nonumber\\
\text{Im}\left(e_1\right)&=-\frac{17}{192}.\label{eq83}
\end{align}
To remove the $A^3(A^*)^2$ term completely, we choose
\begin{align}
\text{Im}\left(c_0\right)=\text{Re}\left(d_0\right)=\text{Re}\left(d_1\right)=\text{Re}\left(e_2\right)=\text{Re}\left(e_3\right)=0.\label{eq84}
\end{align}
With all these choices, the $\varepsilon^4$ term becomes
\begin{align}
&-i A \partial_A\delta +i A^* \partial_{A^*}\delta +i \delta +\left(A^*\right)^4 A^5 \left(\frac{1}{128} i \Log \left(A^*\right)+\frac{5}{128} i \Log (A)\right)-\frac{3}{64} i \left(A^*\right)^3 A^4 \Log (A)\nonumber\\
&+A^* A^2 \left(-\frac{7 \text{Im}\left(d_0\right)}{16}-\text{Re}\left(e_0\right)-\text{Re}\left(e_1\right)+\frac{5 i}{192}\right)-\frac{i A}{128}.\label{eq85}
\end{align}
We solve (\ref{eq85}) for $\delta$ ignoring the core terms $A$ and $A^2A^*$. Such solution is
\begin{align}
\delta(A,A^*)&=\frac{5}{256} \left(A^*\right)^4 A^5 \Log ^2(A)-\frac{1}{256} \left(A^*\right)^4 A^5 \Log ^2\left(A^*\right)-\frac{3}{128} \left(A^*\right)^3 A^4 \Log ^2(A).\label{eq86}
\end{align}
The homogeneous solution was not taken because we don't have more orders to solve for. The final form of the amplitude equation (\ref{eq68}) is
\begin{align}
A'(t)&=A \left(-\frac{i \varepsilon ^4}{128}-\frac{i \varepsilon ^2}{8}+\frac{\varepsilon }{2}+i\right)+A^* A^2 \left(-\frac{7}{16} \varepsilon ^4 \text{Im}\left(d_0\right)-\varepsilon ^4 \text{Re}\left(e_0\right)-\varepsilon ^4 \text{Re}\left(e_1\right)+\frac{5 i \varepsilon ^4}{192}-\frac{\varepsilon ^3}{4}\right.\nonumber\\
&\left.+\frac{i \varepsilon ^2}{16}-\frac{\varepsilon }{2}\right).\label{eq87}
\end{align}
Performing the transformation $A=\tilde{A}e^{it}$ we get rid of the fastest varying terms and we get
\begin{align}
A'(t)&=A \left(\frac{-i \varepsilon ^4}{128}-\frac{i \varepsilon ^2}{8}+\frac{\varepsilon }{2}\right)+A^* A^2 \left(-\frac{7}{16} \varepsilon ^4 \text{Im}\left(d_0\right)-\varepsilon ^4 \text{Re}\left(e_0\right)-\varepsilon ^4 \text{Re}\left(e_1\right)+\frac{5 i \varepsilon ^4}{192}-\frac{\varepsilon ^3}{4}+\frac{i \varepsilon ^2}{16}-\frac{\varepsilon }{2}\right),\label{eq88}
\end{align}
where we dropped the tilde sign. The overall solution turns under the same transformation to
\begin{align}
y(t)&=A e^{i t}+\varepsilon\left(\frac{1}{8} e^{3 i t} i A^3+\alpha \right) +\varepsilon ^2\left(-\frac{1}{192} 5 e^{5 i t} A^5-\frac{1}{64} e^{3 i t} A^* A^4-\frac{1}{32} e^{3 i t} A^3+\frac{3}{8} e^{2 i t} i \alpha  A^2+\beta \right)\nonumber\\
&+\varepsilon ^3\left(-\frac{7 i e^{7 i t} A^7}{1152}-\frac{5 i e^{5 i t} A^* A^6}{1536}+\frac{29}{512} e^{3 i t} i \left(A^*\right)^2 A^5-\frac{35 i e^{5 i t} A^5}{2304}-\frac{25}{192} e^{4 i t} \alpha  A^4-\frac{21}{256} i e^{3 i t} A^* A^4\right.\nonumber\\
&\left.-\frac{1}{64} e^{4 i t} \alpha ^* A^4+\frac{1}{128} e^{3 i t} i A^3-\frac{1}{16} e^{2 i t} \alpha  A^* A^3-\frac{3}{32} e^{2 i t} \alpha  A^2+\frac{3}{8} e^{2 i t} i \beta  A^2+\frac{3}{8} e^{i t} i \alpha ^2 A+\gamma \right)\nonumber\\
&  +\varepsilon ^4\left(\frac{61 e^{9 i t} A^9}{40960}+\frac{133 e^{7 i t} A^* A^8}{221184}+\frac{623 e^{7 i t} A^7}{110592}-\frac{2521 e^{5 i t} \left(A^*\right)^2 A^7}{110592}+\frac{989 e^{3 i t} \left(A^*\right)^3 A^6}{12288}\right.\nonumber\\
&\left.+\frac{197 e^{5 i t} A^* A^6}{6144}-\frac{49 i e^{6 i t} \alpha  A^6}{1152}-\frac{5 i e^{6 i t} \alpha ^* A^6}{1536}+\frac{5 e^{5 i t} A^5}{27648}-\frac{5}{256} i e^{4 i t} \alpha  A^* A^5+\frac{29}{256} e^{4 i t} i A^* \alpha ^* A^5\right.\nonumber\\
&\left.-\frac{1103 e^{3 i t} \left(A^*\right)^2 A^5}{6144}+\frac{145}{512} e^{2 i t} i \alpha  \left(A^*\right)^2 A^4-\frac{25}{192} e^{4 i t} \beta  A^4+\frac{113 e^{3 i t} A^* A^4}{1024}-\frac{21}{256} i e^{4 i t} \alpha ^* A^4\right.\nonumber\\
&\left.-\frac{1}{64} e^{4 i t} \beta ^* A^4-\frac{175 i e^{4 i t} \alpha  A^4}{2304}-\frac{3}{512} e^{3 i t} A^3-\frac{25}{96} e^{3 i t} \alpha ^2 A^3-\frac{21}{64} i e^{2 i t} \alpha  A^* A^3-\frac{1}{16} e^{2 i t} \beta  A^* A^3\right.\nonumber\\
&\left.-\frac{1}{16} e^{3 i t} \alpha  \alpha ^* A^3+\frac{3}{128} e^{2 i t} i \alpha  A^2-\frac{3}{32} e^{2 i t} \beta  A^2+\frac{3}{8} e^{2 i t} i \gamma  A^2-\frac{3}{32} e^{i t} \alpha ^2 A^* A^2-\frac{3}{32} e^{i t} \alpha ^2 A\right.\nonumber\\
&\left.+\frac{3}{4} e^{i t} i \alpha  \beta  A+\frac{i \alpha ^3}{8}+\delta \right)+(*),\label{eq89}
\end{align}
together with the functions
\begin{align}
\alpha(A,A^*)&=\frac{7}{16} i A^2 A^*,\nonumber\\
\beta(A,A^*)&=A \left(\frac{91}{512} A^2 \left(A^*\right)^2+i \text{Im}\left(d_0\right)\right),\nonumber\\
\gamma(A,A^*)&=\frac{1}{64} i \left(A^*\right)^3 A^4 \Log (A)+A \left(A A^* \left(\text{Re}\left(e_1\right)-\frac{1}{192} (17 i)\right)+\frac{15551 i A^3 \left(A^*\right)^3}{73728}+\frac{355 i A^2 \left(A^*\right)^2}{1536}\right.\nonumber\\
&\left.+i \text{Im}\left(e_0\right)+\text{Re}\left(e_0\right)\right),\nonumber\\
\delta(A,A^*)&=\frac{5}{256} \left(A^*\right)^4 A^5 \Log ^2(A)-\frac{1}{256} \left(A^*\right)^4 A^5 \Log ^2\left(A^*\right)-\frac{3}{128} \left(A^*\right)^3 A^4 \Log ^2(A).\label{eq90}
\end{align}
Notice the four free parameters $\text{Im}(d_0),\text{Im}(e_0),\text{Re}(e_0)$ and $\text{Re}(e_1)$.

\setcounter{equation}{0}
\numberwithin{equation}{section}
\section*{Appendix B, Lotka-Volterra system}
\renewcommand{\theequation}{B.\arabic{equation}}
We will go through the modified renormalization method on the Lotka-Volterra system (\ref{eq94}). It begins with finding the naive expansion for both functions
\begin{align}
x(t)&=x_0(t)+\varepsilon x_1(t)+\varepsilon^2 x_2(t)+\varepsilon^3 x_3(t)\ldots,\nonumber\\
y(t)&=y_0(t)+\varepsilon y_1(t)+\varepsilon^2 y_2(t)+\varepsilon^3 y_3(t)\ldots.\label{eq95}
\end{align}
Inserting (\ref{eq95}) into (\ref{eq94}) and by collecting the terms with different orders of $\varepsilon$, we get the naive expansion hierarchy
\begin{align}
\text{order }\varepsilon^0:\;x_0'+y_0&=0\nonumber\\
y'_0-x_0&=0,\nonumber\\
\text{order }\varepsilon^1:\;x_1'+y_1&=-x_0 y_0\nonumber\\
y'_1-x_1&=x_0y_0,\nonumber\\
\text{order }\varepsilon^2:\;x_2'+y_2&=-x_0y_1-x_1y_0\nonumber\\
y'_2-x_2&=x_0y_1+x_1y_0,\nonumber\\
\text{order }\varepsilon^3:\;x_3'+y_3&=-x_0y_2-x_1y_1-x_2y_0\nonumber\\
y'_3-x_3&=x_0y_2+x_1y_1+x_2y_0.\label{eq96}
\end{align}
Let us start with the general solution to the $\varepsilon^0$ order equation. It can be written in a matrix notation
\begin{align}
\frac{\mathrm{d}}{\mathrm{d}x}
\begin{pmatrix}
x_0\\y_0
\end{pmatrix}=
\begin{pmatrix}
0 & -1\\
1 & 0
\end{pmatrix}
\begin{pmatrix}
x_0\\y_0
\end{pmatrix}.\label{eq97}
\end{align}
We look for a solution in the form
\begin{align}
\begin{pmatrix}
x_0\\y_0
\end{pmatrix}=\boldsymbol{\alpha} e^{\omega t},\label{eq98}
\end{align}
where $\boldsymbol{\alpha}$ is an unknown vector and $\omega$ is a complex number. We insert (\ref{eq98}) into the system (\ref{eq97}) and cancel the common factors and get the following algebraic system
\begin{align}
\begin{pmatrix}
\omega & 1\\
-1 & \omega
\end{pmatrix}\boldsymbol{\alpha}=0.\label{eq99}
\end{align}
To have a nontrivial solution to (\ref{eq99}), the determinant of the matrix has to be zero. This condition leads to the equation
\begin{align}
1+\omega^2=0,\label{eq100}
\end{align}
which has two real solutions
\begin{align}
\omega_{1,2}=\pm i.\label{eq101}
\end{align}
Now we can find basis for the solutions space of (\ref{eq99}) corresponding to $\omega=\omega_{1,2}$ to be
\begin{align}
\boldsymbol{u}=
\begin{pmatrix}
i\\1
\end{pmatrix},\boldsymbol{u}=
\begin{pmatrix}
-i\\1
\end{pmatrix}=\boldsymbol{u}^*,\label{eq102}
\end{align}
respectively. The general solution to (\ref{eq97}) can be then written as
\begin{align}
\begin{pmatrix}
x_0\\y_0
\end{pmatrix}=A\boldsymbol{u}e^{it}+A^*\boldsymbol{u}^*e^{-it}=A\boldsymbol{u}e^{it}+(*),\label{eq103}
\end{align}
where the second free constant needs to be the complex conjugate of $A$ since we look for a real solution. Using this result in the right hand-side of the $\varepsilon^1$ order equation in (\ref{eq96}), we get
\begin{align}
\begin{pmatrix}
x_1'\\y_1'
\end{pmatrix}+\begin{pmatrix}
0 & 1\\
-1 & 0
\end{pmatrix}
\begin{pmatrix}
x_1\\y_1
\end{pmatrix}&=e^{2 i t}\begin{pmatrix}
-i A_0^2 \\
i A_0^2
\end{pmatrix}+(*).\label{eq104}
\end{align}
The particular solution is found in the form $\boldsymbol{\xi}e^{2it}$. Substituting this into (\ref{eq104}), collecting the unknowns and cancelling the common factor $e^{2it}$, we get a system for the vector $\boldsymbol{\xi}$ which we solve and find
\begin{align}
\begin{pmatrix}
x_1\\y_1
\end{pmatrix}&=e^{2 i t}\begin{pmatrix}
\left(-\frac{2}{3}+\frac{i}{3}\right) A_0^2\\
\left(\frac{2}{3}+\frac{i}{3}\right) A_0^2
\end{pmatrix}+\alpha_0\textbf{u}e^{it}+(*),\label{eq105}
\end{align}
where we also added the homogeneous solution which is the main feature of our modified method.

Let us insert (\ref{eq103}) and (\ref{eq105}) into the $\varepsilon^2$ order. Collecting the exponential terms, we get the following
\begin{align}
\begin{pmatrix}
x_2'\\y_2'
\end{pmatrix}+\begin{pmatrix}
0 & 1\\
-1 & 0
\end{pmatrix}
\begin{pmatrix}
x_2\\y_2
\end{pmatrix}&=e^{it}\begin{pmatrix}
\left(\frac{1}{3}+\frac{i}{3}\right) A_0^2 \left(A_0\right){}^* \\
\left(-\frac{1}{3}-\frac{i}{3}\right) A_0^2 \left(A_0\right){}^* 
\end{pmatrix}+e^{2it}\begin{pmatrix}
-2 i \alpha _0 A_0\\
2 i \alpha _0 A_0
\end{pmatrix}+e^{3it}\begin{pmatrix}
(1-i) A_0^3\\
(-1+i) A_0^3
\end{pmatrix}+(*).\label{eq106}
\end{align}
Each particular solution belonging to a different exponential is found in the form $\boldsymbol{\xi}e^{ijt}$ for $j=\pm2,\pm3$ except for $j=\pm 1$. In these cases we look for the solution in the form
\begin{align}
\begin{pmatrix}
u_1\\v_1
\end{pmatrix}=\left(
\begin{pmatrix}
a\\b
\end{pmatrix}
+ t
\begin{pmatrix}
c\\d
\end{pmatrix}\right)
e^{it},\label{eq107}
\end{align}
for some constants $a,b,c,d$. We  After substituting (\ref{eq107}) into the left-hand side of (\ref{eq106}) we collect the coefficients of the variable $t$, cancel the common exponential term and get
\begin{align}
\begin{pmatrix}
i a+b+c+t (d+i c)\\
-a+i b+d+t (-c+i d)
\end{pmatrix}&=\begin{pmatrix}
\left(\frac{1}{3}+\frac{i}{3}\right) A_0^2 \left(A_0\right){}^* \\
\left(-\frac{1}{3}-\frac{i}{3}\right) A_0^2 \left(A_0\right){}^* 
\end{pmatrix}.\label{eq108}
\end{align}
We equate the terms in the variable $t$ at both sides and end up with a system of algebraic equations for $a,b,c,d$. In the matrix form we have
\begin{align}
\left(
\begin{array}{cccc}
 i & 1 & 1 & 0 \\
 0 & 0 & i & 1 \\
 -1 & i & 0 & 1 \\
 0 & 0 & -1 & i \\
\end{array}
\right)
\begin{pmatrix}
a\\b\\c\\d
\end{pmatrix}=
\begin{pmatrix}
\left(\frac{1}{3}+\frac{i}{3}\right) A_0^2 \left(A_0\right){}^* \\
0\\
\left(-\frac{1}{3}-\frac{i}{3}\right) A_0^2 \left(A_0\right){}^* \\
0
\end{pmatrix}.\label{eq109}
\end{align}
By row-reducing this system, we get
\begin{align}
\left(
\begin{array}{cccc}
 1 & -i & 0 & 0 \\
 0 & 0 & 1 & 0 \\
 0 & 0 & 0 & 1 \\
 0 & 0 & 0 & 0 \\
\end{array}
\right)
\begin{pmatrix}
a\\b\\c\\d
\end{pmatrix}=
\begin{pmatrix}
\frac{A_0^2 A_0^*}{3}\\
\frac{A_0^2 A_0^*}{3}\\
-\frac{1}{3} i A_0^2A_0^*\\
0
\end{pmatrix}.\label{eq110}
\end{align}
It is straightforward now to compute $a,b,c,d$ and write down the solution. Notice, we get one free parameter which becomes our new homogeneous solution.
\begin{align}
\begin{pmatrix}
u_1\\v_1
\end{pmatrix}&=e^{it}\left(\begin{pmatrix}
\frac{A_0^2 A_0^*}{3}+i\beta_0\\
\beta_0
\end{pmatrix}+\begin{pmatrix}
\frac{A_0^2 A_0^*}{3}\\
-\frac{1}{3} i A_0^2A_0^*
\end{pmatrix}t\right),\label{eq111}
\end{align}
where $\beta_0$ is a free parameter. The rest of the particular solution is obtained in the same way as for the previous order. The full solution for the current order is
\begin{align}
\begin{pmatrix}
x_2\\y_2
\end{pmatrix}&=e^{it}\begin{pmatrix}
\frac{1}{3} \left(A_0\right){}^* A_0^2 t+\frac{1}{3} \left(A_0\right){}^* A_0^2+i \beta _0\\
\beta _0-\frac{1}{3} i A_0^2 \left(A_0\right){}^* t
\end{pmatrix}+e^{2it}\begin{pmatrix}
-\left(\frac{4}{3}-\frac{2 i}{3}\right) \alpha _0 A_0\\
\left(\frac{4}{3}+\frac{2 i}{3}\right) \alpha _0 A_0
\end{pmatrix}+e^{3it}\begin{pmatrix}
\left(-\frac{1}{2}-\frac{i}{4}\right) A_0^3\\
\left(\frac{1}{4}+\frac{i}{2}\right) A_0^3
\end{pmatrix}+(*).\label{eq112}
\end{align}

Moving on to the last order $\varepsilon^3$ in (\ref{eq96}) we insert the obtained functions into the right-hand side and get
\begin{align}
\begin{pmatrix}
x_3'\\y_3'
\end{pmatrix}+\begin{pmatrix}
0 & 1\\
-1 & 0
\end{pmatrix}
\begin{pmatrix}
x_3\\y_3
\end{pmatrix}&=e^{it}\begin{pmatrix}
\left(\frac{1}{3}+\frac{i}{3}\right) \left(\alpha _0\right){}^* A_0^2+\left(\frac{2}{3}+\frac{2 i}{3}\right) \alpha _0 \left(A_0\right){}^* A_0\\
\left(-\frac{2}{3}-\frac{2 i}{3}\right) \alpha _0 A_0 \left(A_0\right){}^*-\left(\frac{1}{3}+\frac{i}{3}\right) \left(\alpha _0\right){}^* A_0^2
\end{pmatrix}\nonumber\\
&+e^{2it}\begin{pmatrix}
-i \alpha _0^2-2 i A_0 \beta _0-\frac{2}{3} \left(A_0\right){}^* A_0^3 t-\left(\frac{1}{3}-\frac{i}{2}\right) \left(A_0\right){}^* A_0^3\\
i \alpha _0^2+2 i A_0 \beta _0+\frac{2}{3} \left(A_0\right){}^* A_0^3 t+\left(\frac{1}{3}-\frac{i}{2}\right) \left(A_0\right){}^* A_0^3
\end{pmatrix}\nonumber\\
&+e^{3it}\begin{pmatrix}
(3-3 i) \alpha _0 A_0^2\\
(-3+3 i) \alpha _0 A_0^2
\end{pmatrix}+e^{4it}\begin{pmatrix}
\frac{14}{9} A_0^4\\
\frac{1}{9} (-14) A_0^4
\end{pmatrix}+(*).\label{eq113}
\end{align}
The solution to (\ref{eq113}) is found in exactly the same way as in the previous order, except now the particular solution for the $e^{it}$ and $e^{2it}$ terms are sought in the same form as (\ref{eq107}). Since the $e^{it}$ term is a secular term, we also get a new homogeneous solution there with the free parameter $\gamma_0$. The solution for the $\varepsilon^3$ order is found to be
\begin{align}
\begin{pmatrix}
x_3\\y_3
\end{pmatrix}&=e^{it}\begin{pmatrix}
\frac{1}{3} \left(\alpha _0\right){}^* A_0^2+\frac{2}{3} \alpha _0 \left(A_0\right){}^* A_0+\frac{1}{3} \left(\alpha _0\right){}^* A_0^2 t+\frac{2}{3} \alpha _0 \left(A_0\right){}^* A_0 t+i \gamma _0\\
-\frac{1}{3} i \left(\alpha _0\right){}^* A_0^2 t-\frac{2}{3} i \alpha _0 \left(A_0\right){}^* A_0 t+\gamma _0
\end{pmatrix}\nonumber\\
&+e^{2it}\begin{pmatrix}
\left(-\frac{2}{3}+\frac{i}{3}\right) \alpha _0^2-\left(\frac{4}{3}-\frac{2 i}{3}\right) A_0 \beta _0+\left(\frac{2}{9}+\frac{4 i}{9}\right) \left(A_0\right){}^* A_0^3 t+\left(\frac{2}{27}+\frac{19 i}{54}\right) \left(A_0\right){}^* A_0^3\\
\left(\frac{2}{3}+\frac{i}{3}\right) \alpha _0^2+\left(\frac{4}{3}+\frac{2 i}{3}\right) A_0 \beta _0+\left(\frac{2}{9}-\frac{4 i}{9}\right) \left(A_0\right){}^* A_0^3 t+\left(\frac{4}{27}-\frac{5 i}{54}\right) \left(A_0\right){}^* A_0^3
\end{pmatrix}\nonumber\\
&+e^{3it}\begin{pmatrix}
-\left(\frac{3}{2}+\frac{3 i}{4}\right) \alpha _0 A_0^2\\
\left(\frac{3}{4}+\frac{3 i}{2}\right) \alpha _0 A_0^2
\end{pmatrix}+e^{4it}\begin{pmatrix}
\left(-\frac{14}{135}-\frac{56 i}{135}\right) A_0^4\\
\left(-\frac{14}{135}+\frac{56 i}{135}\right) A_0^4
\end{pmatrix}+(*).\label{eq114}
\end{align}

We are now ready to apply the renormalization group method on the naive expansion (\ref{eq95}) by introducing the splitting $t-t_0\to \tau+\xi$, where $\tau=t-\mu$ and $\xi=\mu-t_0$. Note that we have several amplitudes and free parameters as well. All of which need to be renormalized as
\begin{align}
A_0(t_0)&=\sum_{n=0}^\infty Z_n(\xi,\mu)\varepsilon^n A(\mu)\approx (Z_0+Z_1\varepsilon+Z_2\varepsilon^2+Z_3\varepsilon^3)A,\nonumber\\
\alpha_0(t_0)&=\sum_{n=0}^\infty U_n(\xi,\mu)\varepsilon^n \alpha(\mu)\approx (U_0+U_1\varepsilon+U_2\varepsilon^2)\alpha,\nonumber\\
\beta_0(t_0)&=\sum_{n=0}^\infty V_n(\xi,\mu)\varepsilon^n \beta(\mu)\approx (V_0+V_1\varepsilon)\beta,\nonumber\\
\gamma_0(t_0)&=\sum_{n=0}^\infty W_n(\xi,\mu)\varepsilon^n \gamma(\mu)\approx W_0\gamma.\label{eq115}
\end{align}
We have gone up to $\varepsilon^3$ order for the amplitude $A_0$ because it appears already at the 0-th order in the naive expansion. The free parameter $\alpha_0$ first appears in the $\varepsilon^1$ order term so we do not need to go higher than  2nd order in its renormalization constant. Recall that the constants $Z_0,Z_1,\ldots,W_0$ are chosen such that the naive expansion (\ref{eq95}) with $t-t_0\to\tau+\xi$, is $\xi$-free.

We perform the splitting and substituting the renormalizations from (\ref{eq115}). The 0-th order then becomes
\begin{align}
\begin{pmatrix}
i\\1
\end{pmatrix}Z_0Ae^{i(\mu+\xi)}+(*).\label{eq116}
\end{align}
To ensure the expression (\ref{eq116}) to be $\xi$-free, we need to choose
\begin{align}
Z_0&=e^{-i\xi},\label{eq117}
\end{align}
with its complex conjugate. Using this newly computed constant in the $\varepsilon^1$ order term from the naive expansion, the term becomes
\begin{align}
&x\text{-component}:\;\left(-\frac{2}{3}+\frac{i}{3}\right) A(\mu )^2 e^{2 i \tau}+i Z_1 A(\mu ) e^{i (\xi +\tau )}+i U_0 \alpha (\mu ) e^{i (\xi +\tau )}+(*),\nonumber\\
&y\text{-component}:\;\left(\frac{2}{3}+\frac{i}{3}\right) A(\mu )^2 e^{2 i \tau }+Z_1 A(\mu ) e^{i (\xi +\tau )}+U_0 \alpha (\mu ) e^{i (\xi +\tau )}+(*).\label{eq118}
\end{align}
The variable $\xi$ is not appearing other places than the exponents, so the natural choice for $Z_1$ and $U_0$ are
\begin{align}
U_0&=e^{-i\xi},Z_1=0.\label{eq119}
\end{align}
With these choices, the next order $\varepsilon^2$ looks as follows
\begin{align}
&x\text{-component}:\;e^{i \tau } \left(\frac{1}{3} A^* \xi  A(\mu )^2+\frac{1}{3} A^* \tau  A(\mu )^2+\frac{1}{3} A^* A(\mu )^2+i e^{i \xi } Z_2 A(\mu )+i \alpha  e^{i \xi } U_1+i \beta  e^{i \xi } V_0\right)\nonumber\\
&+\left(-\frac{4}{3}+\frac{2 i}{3}\right) \alpha  e^{2 i \tau } A(\mu )-\left(\left(\frac{1}{2}+\frac{i}{4}\right) e^{3 i \tau } A(\mu )^3\right)+(*),\nonumber\\
&y\text{-component}:\;e^{i \tau } \left(-\frac{1}{3} i A^* \xi  A(\mu )^2-\frac{1}{3} i A^* \tau  A(\mu )^2+e^{i \xi } Z_2 A(\mu )+\alpha  e^{i \xi } U_1+\beta  e^{i \xi } V_0\right)\nonumber\\
&+\left(\frac{4}{3}+\frac{2 i}{3}\right) \alpha  e^{2 i \tau } A(\mu )+\left(\frac{1}{4}+\frac{i}{2}\right) e^{3 i \tau } A(\mu )^3+(*).\label{eq120}
\end{align}
Terms containing $\xi$ are present in the factors of $e^{\pm i\tau}$. We use the free constants $Z_2,U_1$ and $V_0$ for the factor of $e^{i\tau}$ to remove the these terms. At the end, we get the following expressions for the renormalization constants
\begin{align}
Z_2&=\frac{1}{3} i e^{-i \xi } \xi  A A^*, U_1=0, V_0=e^{-i\xi}.\label{eq121}
\end{align}

The final order $\varepsilon^3$ now reads
\begin{align}
&x\text{-component}:\;\left(-\frac{14}{135}-\frac{56 i}{135}\right) A^4 e^{4 i \tau }-\left(\frac{3}{2}+\frac{3 i}{4}\right) \alpha  A^2 e^{3 i \tau }+e^{2 i \tau } \left(\left(-\frac{2}{3}+\frac{i}{3}\right) \alpha ^2\right.\nonumber\\
&\left.+\left(\frac{2}{9}+\frac{4 i}{9}\right) A^* A^3 \tau +\left(\frac{2}{27}+\frac{19 i}{54}\right) A^* A^3-\left(\frac{4}{3}-\frac{2 i}{3}\right) A \beta \right)+e^{i \tau } \left(\frac{1}{3} \alpha ^* A^2 \xi +\frac{1}{3} \alpha ^* A^2 \tau +\frac{\alpha ^* A^2}{3}\right.\nonumber\\
&\left.+\frac{2}{3} \alpha  A^* A \xi +\frac{2}{3} \alpha  A^* A \tau +\frac{2}{3} \alpha  A^* A+i A e^{i \xi } Z_3+i \alpha  e^{i \xi } U_2+i \beta  e^{i \xi } V_1+i e^{i \xi } W_0 \gamma (\mu )\right)+(*),\nonumber\\
&y\text{-component}:\;\left(-\frac{14}{135}+\frac{56 i}{135}\right) A^4 e^{4 i \tau }+\left(\frac{3}{4}+\frac{3 i}{2}\right) \alpha  A^2 e^{3 i \tau }+e^{2 i \tau } \left(\left(\frac{2}{3}+\frac{i}{3}\right) \alpha ^2\right.\nonumber\\
&\left.+\left(\frac{2}{9}-\frac{4 i}{9}\right) A^* A^3 \tau +\left(\frac{4}{27}-\frac{5 i}{54}\right) A^* A^3+\left(\frac{4}{3}+\frac{2 i}{3}\right) A \beta \right)+e^{i \tau } \left(-\frac{1}{3} i \alpha ^* A^2 \xi -\frac{1}{3} i \alpha ^* A^2 \tau\right.\nonumber\\
&\left. -\frac{2}{3} i \alpha  A^* A \xi -\frac{2}{3} i \alpha  A^* A \tau +A e^{i \xi } Z_3+\alpha  e^{i \xi } U_2+\beta  e^{i \xi } V_1+e^{i \xi } W_0 \gamma (\mu )\right)+(*).\label{eq122}
\end{align}
In order to remove the $\xi$ variable from this order, we can choose
\begin{align}
Z_3=\frac{2}{3} i \alpha  A^* e^{-i \xi } \xi +\frac{1}{3} i \alpha ^* A e^{-i \xi } \xi,U_2=0,V_1=0,W_0=e^{-i\xi}.\label{eq123}
\end{align}
Substituting all the renormalization constants into the whole naive expansion, we get a $\xi$-free, thus infinity-free solution. We must, however remember the renormalized free parameters arisen from the naive solution, namely $\alpha,\beta$ and $\gamma$. The right hand side is composed solely of the amplitude $A$. Therefore, it is reasonable to assume and write these free parameters as functions of the amplitude. Thus we write
\begin{align}
\alpha(\mu)&:=\alpha(A(\mu),A^*(\mu)),\nonumber\\
\beta(\mu)&:=\beta(A(\mu),A^*(\mu)),\nonumber\\
\gamma(\mu)&:=\gamma(A(\mu),A^*(\mu)).\label{eq124}
\end{align}
The derivatives of these functions are then computed accordingly. This is important when we derive the amplitude equation. The solutions we get are of the following form
\begin{align}
x(t;\mu)&=i A e^{i \tau }+\varepsilon  \left(i \alpha  e^{i \tau }+\left(-\frac{2}{3}+\frac{i}{3}\right) A^2 e^{2 i \tau }\right)+\varepsilon ^2 \left(\left(-\frac{1}{2}-\frac{i}{4}\right) A^3 e^{3 i \tau }+\frac{1}{3} A^* A^2 e^{i \tau }\right.\nonumber\\
&\left.+\frac{1}{3} A^* A^2 e^{i \tau } \tau -\left(\frac{4}{3}-\frac{2 i}{3}\right) \alpha  A e^{2 i \tau }+i \beta  e^{i \tau }\right)+\varepsilon ^3 \left(-\left(\frac{2}{3}-\frac{i}{3}\right) \alpha ^2 e^{2 i \tau }\right.\nonumber\\
&\left.+\left(-\frac{14}{135}-\frac{56 i}{135}\right) A^4 e^{4 i \tau }+\frac{1}{3} \alpha ^* A^2 e^{i \tau }+\frac{1}{3} \alpha ^* A^2 e^{i \tau } \tau -\left(\frac{3}{2}+\frac{3 i}{4}\right) \alpha  A^2 e^{3 i \tau }\right.\nonumber\\
&\left.+\frac{2}{3} \alpha  A^* A e^{i \tau }+\frac{2}{3} \alpha  A^* A e^{i \tau } \tau +\left(\frac{2}{27}+\frac{19 i}{54}\right) A^* A^3 e^{2 i \tau }+\left(\frac{2}{9}+\frac{4 i}{9}\right) A^* A^3 e^{2 i \tau } \tau\right.\nonumber\\
&\left. -\left(\frac{4}{3}-\frac{2 i}{3}\right) A \beta  e^{2 i \tau }+i \gamma  e^{i \tau }\right)+(*),\nonumber\\
y(t;\mu)&=A e^{i \tau }+\varepsilon  \left(\alpha  e^{i \tau }+\left(\frac{2}{3}+\frac{i}{3}\right) A^2 e^{2 i \tau }\right)+\varepsilon ^2 \left(\left(\frac{1}{4}+\frac{i}{2}\right) A^3 e^{3 i \tau }-\frac{1}{3} i A^* A^2 e^{i \tau } \tau\right.\nonumber\\
&\left. +\left(\frac{4}{3}+\frac{2 i}{3}\right) \alpha  A e^{2 i \tau }+\beta  e^{i \tau }\right)+\varepsilon ^3 \left(\left(\frac{2}{3}+\frac{i}{3}\right) \alpha ^2 e^{2 i \tau }+\left(-\frac{14}{135}+\frac{56 i}{135}\right) A^4 e^{4 i \tau }-\frac{1}{3} i \alpha ^* A^2 e^{i \tau } \tau\right.\nonumber\\
&\left. +\left(\frac{3}{4}+\frac{3 i}{2}\right) \alpha  A^2 e^{3 i \tau }-\frac{2}{3} i \alpha  A^* A e^{i \tau } \tau +\left(\frac{4}{27}-\frac{5 i}{54}\right) A^* A^3 e^{2 i \tau }+\left(\frac{2}{9}-\frac{4 i}{9}\right) A^* A^3 e^{2 i \tau } \tau\right.\nonumber\\
&\left. +\left(\frac{4}{3}+\frac{2 i}{3}\right) A \beta  e^{2 i \tau }+\gamma  e^{i \tau }\right)+(*).\label{eq125}
\end{align}
Since $\tau$ is defined as $\tau=t-\mu$ and the original problem does not include the variable $\mu$, we set $\left(\partial x/\partial \mu\right)_{\mu=t}=\left(\partial y/\partial \mu\right)_{\mu=t}=0$ which gives us the amplitude equation. This is the same procedure as for the scalar cases. Carrying out the calculation we obtain two equations in which the $A'$ and $(A^*)'$ are the unknowns. Thus we treat these equations as a $2\times 2$ linear system for $A',(A^*)'$. We used Kramer's rule to obtain the solution. The unknowns are thus of the form $\det M'/\det M$, where $M$ is the actual matrix and $M'$ is the modified matrix $M$, where the corresponding column is replaced by the right hand side. This makes the unknown a fraction where the denominator is consisting of an expression with different orders of $\varepsilon$. For this reason, we use Taylor expansion for the $1/\det M$ part in the small parameter $\varepsilon$. The final expression is then an algebraic expression with no fractions which we truncate and keep only the terms up to order $\varepsilon^3$. Doing so, we arrive at the equation
\begin{align}
A'&=i A+\varepsilon  \left(i \alpha +i A^* \frac{\partial \alpha }{\partial A^*}-i A \frac{\partial \alpha }{\partial A}\right)+\varepsilon ^2 \left(i A \frac{\partial \alpha }{\partial A^*} \frac{\partial \alpha ^*}{\partial A}-i A^* \frac{\partial \alpha }{\partial A^*} \frac{\partial \alpha ^*}{\partial A^*}+i \alpha ^* \frac{\partial \alpha }{\partial A^*}-i A^* \frac{\partial \alpha }{\partial A} \frac{\partial \alpha }{\partial A^*}\right.\nonumber\\
&\left.+i A^* \frac{\partial \beta }{\partial A^*}-\frac{1}{3} i A^* A^2+i A \left(\frac{\partial \alpha }{\partial A}\right)^2-i \alpha  \frac{\partial \alpha }{\partial A}-i A \frac{\partial \beta }{\partial A}+i \beta \right)+\varepsilon ^3 \left(-\frac{1}{3} i \alpha ^* A^2+i A \frac{\partial \alpha ^*}{\partial A} \frac{\partial \beta }{\partial A^*}\right.\nonumber\\
&\left.-i A^* \frac{\partial \alpha ^*}{\partial A^*} \frac{\partial \beta }{\partial A^*}+i \alpha ^* \frac{\partial \beta }{\partial A^*}-2 i A \frac{\partial \alpha }{\partial A^*} \frac{\partial \alpha ^*}{\partial A} \frac{\partial \alpha }{\partial A}+i A^* \frac{\partial \alpha }{\partial A^*} \frac{\partial \alpha ^*}{\partial A^*} \frac{\partial \alpha }{\partial A}-i \alpha ^* \frac{\partial \alpha }{\partial A^*} \frac{\partial \alpha }{\partial A}+i \alpha  \frac{\partial \alpha }{\partial A^*} \frac{\partial \alpha ^*}{\partial A}\right.\nonumber\\
&\left.-i A \frac{\partial \alpha }{\partial A^*} \frac{\partial \alpha ^*}{\partial A} \frac{\partial \alpha ^*}{\partial A^*}+i A^* \frac{\partial \alpha }{\partial A^*} \left(\frac{\partial \alpha ^*}{\partial A^*}\right)^2+i A^* \left(\frac{\partial \alpha }{\partial A^*}\right)^2 \frac{\partial \alpha ^*}{\partial A}-i \alpha ^* \frac{\partial \alpha }{\partial A^*} \frac{\partial \alpha ^*}{\partial A^*}+i A \frac{\partial \alpha }{\partial A^*} \frac{\partial \beta ^*}{\partial A}\right.\nonumber\\
&\left.-i A^* \frac{\partial \alpha }{\partial A^*} \frac{\partial \beta ^*}{\partial A^*}+i \beta ^* \frac{\partial \alpha }{\partial A^*}-i A^* \frac{\partial \alpha }{\partial A} \frac{\partial \beta }{\partial A^*}-i A^* \frac{\partial \alpha }{\partial A^*} \frac{\partial \beta }{\partial A}-\frac{2}{3} i \alpha  A A^*+i A^* \frac{\partial \alpha }{\partial A^*} \left(\frac{\partial \alpha }{\partial A}\right)^2\right.\nonumber\\
&\left.-\frac{1}{3} i A \left(A^*\right)^2 \frac{\partial \alpha }{\partial A^*}+i A^* \frac{\partial \gamma }{\partial A^*}+\frac{1}{3} i A^2 A^* \frac{\partial \alpha }{\partial A}-i \beta  \frac{\partial \alpha }{\partial A}+2 i A \frac{\partial \alpha }{\partial A} \frac{\partial \beta }{\partial A}-i \alpha  \frac{\partial \beta }{\partial A}-i A \left(\frac{\partial \alpha }{\partial A}\right)^3\right.\nonumber\\
&\left.+i \alpha  \left(\frac{\partial \alpha }{\partial A}\right)^2-i A \frac{\partial \gamma }{\partial A}+i \gamma \right).\label{eq126}
\end{align}
We will now see our method in action and use the homogeneous functions $\alpha,\beta,\gamma$ to remove terms in the amplitude equation (\ref{eq126}) leaving only a specific core.

Let us start with the $\varepsilon^1$ term. Here, there are no terms including only the amplitude $A$, so we are solving
\begin{align}
\alpha + A^* \frac{\partial \alpha }{\partial A^*}- A \frac{\partial \alpha }{\partial A}=0.\label{eq127}
\end{align}
This equation is exactly the same as in (\ref{eq25}), the homogeneous part, so that the general solution is
\begin{align}
\alpha(A,A^*)=c(AA^*)A=(c_0+c_1AA^*)A,\label{eq128}
\end{align}
for some arbitrary function $c$ which we assume to be a first order polynomial in its argument with free complex constants $c_0,c_1$. 

Using (\ref{eq128}) the $\varepsilon^2$ order becomes
\begin{align}
-\frac{1}{3} i A^* A^2+i A^* \frac{\partial \beta }{\partial A^*}-i A \frac{\partial \beta }{\partial A}+i \beta.\label{eq129}
\end{align}
We could solve the whole equation above and remove also the $-\frac{1}{3} i A^* A^2$ term, but this would mean getting rid of the whole $\varepsilon^2$ term and making the amplitude of the form $e^{it}$. The function $\beta$ would then include a $\Log(A)$ term which would be a growing term. We therefore declare $-\frac{1}{3} i A^* A^2$ to be the core of our amplitude equation and it is not going to be removed. Hence the next homogeneous function is of the same form as the previous one.
\begin{align}
\beta(A,A^*)=d(AA^*)A=(d_0+d_1AA^*)A,\label{eq130}
\end{align}
The $\varepsilon^3$ order term turns to
\begin{align}
i A^* \frac{\partial \gamma }{\partial A^*}-\frac{1}{3} 2 i \left(A^*\right)^2 A^3 \text{Re}[c_1]-\frac{2}{3} i A^* A^2 \text{Re}[c_0]-i A \frac{\partial \gamma }{\partial A}+i \gamma.\label{eq131}
\end{align}
It is straightforward to remove the non-$\gamma$ terms using the constants $c_0,c_1$. If we choose
\begin{align}
\text{Re}[c_0]=\text{Re}[c_1]&=0,\label{eq132}
\end{align}
we get rid of all the particular terms. We are then left with the classical homogeneous part $\gamma + A^* \partial_{A^*} \gamma- A\partial_A \gamma$. Since this was the last order, we let $\gamma=0$. The rest of the free complex constants $c_0,\ldots,d_1$ is set to 0. The amplitude equation (\ref{eq126}) is then
\begin{align}
A'=iA-\varepsilon^2\frac{1}{3} i A^* A^2.\label{eq133}
\end{align}
As for the scalar examples, we introduce a transformation that removes the fast oscillating term $iA$, namely $A=\tilde{A}e^{it}$. The amplitude equation (\ref{eq133}) becomes
\begin{align}
A'(t)=-\varepsilon^2\frac{1}{3} i A^* A^2,\label{eq134}
\end{align}
where we dropped the tilde signs. Finally, we get the overall solution   from (\ref{eq125}) in the limit $\mu\rightarrow t$ or $\tau\rightarrow 0$. Together with the amplitude transformation, the overall solution is
\begin{align}
x(t)&=i A e^{i t}-\left(\frac{2}{3}-\frac{i}{3}\right) A^2 e^{2 i t} \varepsilon+\varepsilon ^2 \left(\left(-\frac{1}{2}-\frac{i}{4}\right) A^3 e^{3 i t}+\frac{1}{3} A^* A^2 e^{i t}\right)\nonumber\\
& +\varepsilon ^3 \left(\left(-\frac{14}{135}-\frac{56 i}{135}\right) A^4 e^{4 i t}+\left(\frac{2}{27}+\frac{19 i}{54}\right) A^* A^3 e^{2 i t}\right)+(*),\nonumber\\
y(t)&=A e^{i t}+\left(\frac{2}{3}+\frac{i}{3}\right) A^2 e^{2 i t} \varepsilon+\left(\frac{1}{4}+\frac{i}{2}\right) A^3 e^{3 i t} \varepsilon ^2\nonumber\\
& +\varepsilon ^3 \left(\left(-\frac{14}{135}+\frac{56 i}{135}\right) A^4 e^{4 i t}+\left(\frac{4}{27}-\frac{5 i}{54}\right) A^* A^3 e^{2 i t}\right)+(*).\label{eq135}
\end{align}

\setcounter{equation}{0}
\numberwithin{equation}{section}
\section*{Appendix C, System of second order ODE's}
\renewcommand{\theequation}{C.\arabic{equation}}
The modified RG solution is obtained in the same manner as for the Lotka-Volterra system. We therefore skip presenting all the calculations involved in the process.  At the $\varepsilon^1$ order we find two homogeneous functions $\alpha,\beta$ and at the $\varepsilon^2$ order there are the functions $\gamma$ and $\delta$. The solution obtained solely from RG method with the homogeneous functions is
\begin{align}
x(t)&=A e^{i t}-B e^{2 i t}+\varepsilon  \left(\alpha -\frac{1}{2} A^2 e^{2 i t}+A^* A-\frac{1}{8} A B e^{3 i t}-\beta +\frac{2}{15} B^2 e^{4 i t}-2 B B^*\right)\nonumber\\
&+\varepsilon ^2 \left(\frac{1}{16} A^3 e^{3 i t}-A^2 B^*+\frac{1}{12} A^2 B e^{4 i t}-\frac{77}{960} A^* B^2 e^{3 i t}-\frac{37}{24} A^* A B e^{2 i t}-\frac{17 A B^2 e^{5 i t}}{2880}+2 \alpha ^* A e^{i t}\right.\nonumber\\
&\left.-\alpha  A e^{i t}-\frac{1}{8} A \beta  e^{i t}-\frac{2}{525} B^3 e^{6 i t}+\frac{29}{15} B^2 B^* e^{2 i t}-\frac{1}{8} \alpha  B e^{2 i t}-4 \beta ^* B e^{2 i t}+\frac{4}{15} \beta  B e^{2 i t}+\gamma -\delta \right)+(*),\nonumber\\
y(t)&=A e^{i t}+2 B e^{2 i t}+\varepsilon  \left(\alpha +A A^*-\frac{1}{8} A B e^{3 i t}+2 \beta +\frac{2}{15} B^2 e^{4 i t}-2 B^* B\right)\nonumber\\
&+\varepsilon ^2 \left(\frac{1}{16} A^3 e^{3 i t}-A^2 B^*+\frac{1}{12} A^2 B e^{4 i t}-\frac{77}{960} A^* B^2 e^{3 i t}-\frac{17 A B^2 e^{5 i t}}{2880}+2 \alpha ^* A e^{i t}-\frac{1}{8} A \beta  e^{i t}\right.\nonumber\\
&\left.-\frac{2}{525} B^3 e^{6 i t}-\frac{1}{8} \alpha  B e^{2 i t}-4 \beta ^* B e^{2 i t}+\frac{4}{15} \beta  B e^{2 i t}+\gamma +2 \delta \right)+(*),\label{eq138}
\end{align}
where all the homogeneous functions depend on $A,B,A^*,B^*$. This solution was already transformed with the amplitude transformations $\tilde{A}=Ae^{it},\tilde{B}=Be^{2it}$. The corresponding non-transformed amplitude equations are
\begin{align}
A'&=i A+\varepsilon  \left(i \alpha +i A^* \frac{\partial \alpha }{\partial A^*}-\frac{1}{2} i A^* B-i A \frac{\partial \alpha }{\partial A}+2 i B^* \frac{\partial \alpha }{\partial B^*}-2 i B \frac{\partial \alpha }{\partial B}\right)+\varepsilon ^2 \left(i A \frac{\partial \alpha }{\partial A^*} \frac{\partial \alpha ^*}{\partial A}\right.\nonumber\\
&\left.-i A^* \frac{\partial \alpha }{\partial A^*} \frac{\partial \alpha ^*}{\partial A^*}+i \alpha ^* \frac{\partial \alpha }{\partial A^*}-i A^* \frac{\partial \alpha }{\partial A} \frac{\partial \alpha }{\partial A^*}-\frac{1}{2} i A^* \beta -2 i B^* \frac{\partial \alpha }{\partial A^*} \frac{\partial \alpha ^*}{\partial B^*}-i A^* \frac{\partial \beta ^*}{\partial A^*} \frac{\partial \alpha }{\partial B^*}-\frac{1}{2} i A B^* \frac{\partial \alpha }{\partial A^*}\right.\nonumber\\
&\left.+2 i B \frac{\partial \alpha }{\partial A^*} \frac{\partial \alpha ^*}{\partial B}-i A^* \frac{\partial \beta }{\partial A^*} \frac{\partial \alpha }{\partial B}+\frac{1}{2} i A^* B \frac{\partial \alpha }{\partial A}+i A^* \frac{\partial \gamma }{\partial A^*}-\frac{1}{4} 7 i A^* A^2+i A \left(\frac{\partial \alpha }{\partial A}\right)^2-i \alpha  \frac{\partial \alpha }{\partial A}\right.\nonumber\\
&\left.+i A \frac{\partial \beta ^*}{\partial A} \frac{\partial \alpha }{\partial B^*}-2 i B^* \frac{\partial \alpha }{\partial A} \frac{\partial \alpha }{\partial B^*}+\frac{67}{16} i A B B^*+i A \frac{\partial \beta }{\partial A} \frac{\partial \alpha }{\partial B}+2 i B \frac{\partial \alpha }{\partial A} \frac{\partial \alpha }{\partial B}-i A \frac{\partial \gamma }{\partial A}+2 i B \frac{\partial \alpha }{\partial B^*} \frac{\partial \beta ^*}{\partial B}\right.\nonumber\\
&\left.-2 i B^* \frac{\partial \alpha }{\partial B^*} \frac{\partial \beta ^*}{\partial B^*}+2 i \beta ^* \frac{\partial \alpha }{\partial B^*}-2 i B^* \frac{\partial \alpha }{\partial B} \frac{\partial \beta }{\partial B^*}+2 i B^* \frac{\partial \gamma }{\partial B^*}-\frac{1}{2} i \alpha ^* B-2 i \beta  \frac{\partial \alpha }{\partial B}+2 i B \frac{\partial \alpha }{\partial B} \frac{\partial \beta }{\partial B}\right.\nonumber\\
&\left.-2 i B \frac{\partial \gamma }{\partial B}+i \gamma \right),\label{eq139}\\
B'&=2 i B+\varepsilon  \left(i A^* \frac{\partial \beta }{\partial A^*}-i A \frac{\partial \beta }{\partial A}+2 i \beta +2 i B^* \frac{\partial \beta }{\partial B^*}-2 i B \frac{\partial \beta }{\partial B}\right)+\varepsilon ^2 \left(i A \frac{\partial \alpha ^*}{\partial A} \frac{\partial \beta }{\partial A^*}-i A^* \frac{\partial \alpha ^*}{\partial A^*} \frac{\partial \beta }{\partial A^*}\right.\nonumber\\
&\left.+i \alpha ^* \frac{\partial \beta }{\partial A^*}-i A^* \frac{\partial \alpha }{\partial A^*} \frac{\partial \beta }{\partial A}-2 i B^* \frac{\partial \beta }{\partial A^*} \frac{\partial \alpha ^*}{\partial B^*}-i A^* \frac{\partial \beta ^*}{\partial A^*} \frac{\partial \beta }{\partial B^*}-\frac{1}{2} i A B^* \frac{\partial \beta }{\partial A^*}+2 i B \frac{\partial \beta }{\partial A^*} \frac{\partial \alpha ^*}{\partial B}\right.\nonumber\\
&\left.-i A^* \frac{\partial \beta }{\partial A^*} \frac{\partial \beta }{\partial B}+\frac{1}{2} i A^* B \frac{\partial \beta }{\partial A}+\frac{1}{6} i A A^* B+i A^* \frac{\partial \delta }{\partial A^*}-i \alpha  \frac{\partial \beta }{\partial A}+i A \frac{\partial \alpha }{\partial A} \frac{\partial \beta }{\partial A}-2 i B^* \frac{\partial \beta }{\partial A} \frac{\partial \alpha }{\partial B^*}\right.\nonumber\\
&\left.+i A \frac{\partial \beta ^*}{\partial A} \frac{\partial \beta }{\partial B^*}+2 i B \frac{\partial \beta }{\partial A} \frac{\partial \alpha }{\partial B}+i A \frac{\partial \beta }{\partial A} \frac{\partial \beta }{\partial B}-i A \frac{\partial \delta }{\partial A}+2 i B \frac{\partial \beta }{\partial B^*} \frac{\partial \beta ^*}{\partial B}\right.\nonumber\\
&\left.-2 i B^* \frac{\partial \beta }{\partial B^*} \frac{\partial \beta ^*}{\partial B^*}+2 i \beta ^* \frac{\partial \beta }{\partial B^*}-2 i B^* \frac{\partial \beta }{\partial B^*} \frac{\partial \beta }{\partial B}+2 i B^* \frac{\partial \delta }{\partial B^*}+2 i B \left(\frac{\partial \beta }{\partial B}\right)^2-2 i \beta  \frac{\partial \beta }{\partial B}-2 i B \frac{\partial \delta }{\partial B}\right.\nonumber\\
&\left.+2 i \delta \right).\label{eq140}
\end{align}
We are now going to apply the modified RG method and remove the terms in the orders of the two amplitude equations (\ref{eq139}), (\ref{eq140}) using the homogeneous functions. Let us start with the $\varepsilon^1$ order. The amplitude equation for $A$ includes the term $-\frac{1}{2} i A B^*$ while the equation for $B$ does not include any. We will set these orders to zero and solve for $\alpha$ and $\beta$. We begin with the amplitude $A$ equation.
\begin{align}
\alpha +A^* \frac{\partial \alpha }{\partial A^*}-A \frac{\partial \alpha }{\partial A}+2 B^* \frac{\partial \alpha }{\partial B^*}-2 B \frac{\partial \alpha }{\partial B}=\frac{1}{2} A^* B.\label{eq141}
\end{align}
This equation is different from the previous ones we encountered until now such as (\ref{eq128}) but it is still a quasi-linear first order PDE. Before we used the method of characteristics which we will deploy here also, although there are some major differences. First of all, the number of variables is 4 instead of 2. This complicates the calculations in such a way that it will be harder to invert from the characteristic variables $\tau,s$ into $A,A^*,B,B^*$. Let us start with setting up the equations for the characteristics.
\begin{align}
\frac{\mathrm{d}A}{\mathrm{d}s}&=-A,A(0,\tau)=a(\tau)\Rightarrow A(s,\tau)=a(\tau)e^{-s},\label{eq142}\\
\frac{\mathrm{d}A^*}{\mathrm{d}s}&=A^*,A^*(0,\tau)=a^*(\tau)\Rightarrow A^*(s,\tau)=a^*(\tau)e^{s},\label{eq143}\\
\frac{\mathrm{d}B}{\mathrm{d}s}&=-2B,B(0,\tau)=b(\tau)\Rightarrow B(s,\tau)=b(\tau)e^{-2s},\label{eq144}\\
\frac{\mathrm{d}B^*}{\mathrm{d}s}&=2B^*,B^*(0,\tau)=b^*(\tau)\Rightarrow B^*(s,\tau)=b^*(\tau)e^{2s},\label{eq145}\\
\frac{\mathrm{d}\alpha}{\mathrm{d}s}&=\frac{1}{2}A^*B-\alpha,\alpha(0,\tau)=\varphi(\tau),\label{eq146}
\end{align}
where $a,a^*,b,b^*,\varphi$ are some free functions to be determined by the initial conditions. The equations corresponding to the variables were readily solved. We stress again that the variable pairs $A,A^*$ and $B,B^*$ need not to be strictly complex conjugates of each other. Substituting them into (\ref{eq146}) we get
\begin{align}
\frac{\mathrm{d}\alpha}{\mathrm{d}s}=-\alpha+\frac{1}{2}a^*be^{-s}.\label{eq147}
\end{align}
The solution $\alpha$ is split into homogeneous $\alpha_h$ and a particular part $\alpha_p$. The homogeneous part is
\begin{align}
\alpha_h(s,\tau)=\varphi(\tau)e^{-s},\label{eq148}
\end{align}
for some arbitrary function $\varphi$. The right hand side of (\ref{eq147}) contains the homogeneous solution thus we seek for the particular solution in the form $\alpha_p=Kse^{-s}$ for some constant $K$. We compute this constant by substituting $\alpha_p$ into (\ref{eq147}) and solve for $K$. We find $\alpha_p=\frac{1}{2}a^*bse^{-s}$. The full solution then becomes
\begin{align}
\alpha(s,\tau)=\varphi(\tau)e^{-s}+\frac{1}{2}a^*(\tau)b(\tau)se^{-s}.\label{eq149}
\end{align}
Normally, at this point, the variables $s,\tau$ would be expressed in terms of the original ones $A,A^*,\ldots$. We observe that multiplying (\ref{eq142}) and (\ref{eq143}) we get $AA^*=a(\tau)a^*(\tau)$ and the variable $\tau$ can be expressed by taking the inverse of the right hand side. But there is more than one way to express $\tau$, for example $A^2B^*=a^2(\tau)b^*(\tau)$. How can we get all the ways to express $\tau$? Luckily, there is an easy process how to generate all the combinations that give us $\tau$. Take the first equation for $A$ (\ref{eq142}) and divide it by the other 3 equations for $A^*,B$ and $B^*$. Let us take the first division.
\begin{align}
\frac{\frac{\mathrm{d}A}{\mathrm{d}s}}{\frac{\mathrm{d}A^*}{\mathrm{d}s}}=\frac{\mathrm{d}A}{\mathrm{d}A^*}&=-\frac{A}{A^*}\nonumber\\
\int \frac{\mathrm{d}A}{A}&=-\int\frac{\mathrm{d}A^*}{A^*}\nonumber\\
\ln A&=\ln\frac{1}{A^*}+c_1\nonumber\\
A&=\frac{1}{A^*}e^c_1\nonumber\\
AA^*&=c_2,\label{eq150}
\end{align}
where $e^c_1=c_2$ for some constants $c_1,c_2$. Note that $AA^*$ gives us one way to get $\tau$ as we have seen. So does taking any function of the equation (\ref{eq150}) since the constant can be redefined, for example $1/(AA^*)=1/c_2=c_3$ for some new constant $c_3$. Let us take the next division which is (\ref{eq142}) by (\ref{eq144}).
\begin{align}
\frac{\frac{\mathrm{d}A}{\mathrm{d}s}}{\frac{\mathrm{d}B}{\mathrm{d}s}}=\frac{\mathrm{d}A}{\mathrm{d}B}&=\frac{A}{2B}\nonumber\\
\int \frac{\mathrm{d}A}{A}&=-\int\frac{\mathrm{d}B}{2B}\nonumber\\
\ln A^2&=\ln B+d_1\nonumber\\
\frac{A^2}{B}&=d_2.\label{eq151}
\end{align}
Observe, that also combining (\ref{eq150}) with (\ref{eq151}) gives us a way to express $\tau$. Similarly, dividing (\ref{eq142}) by (\ref{eq145}) we get $A^2B^*=e_1$. To summarize, the equation for $A$ gives us the following constants which we can also call \textit{monomers}:
\begin{align}
\mathrm{d}A\longrightarrow AA^*,\frac{A^2}{B},A^2B^*.\label{eq152}
\end{align}
Similarly, we find from the other 3 equations the following monomers
\begin{align}
\mathrm{d}A^*&\longrightarrow AA^*,\frac{(A^*)^2}{B^*},(A^*)^2B^,\label{eq153}\\
\mathrm{d}B&\longrightarrow \frac{B}{A^2},B(A^*)^2,BB^*,\label{eq154}\\
\mathrm{d}B^*&\longrightarrow \frac{B^*}{(A^*)^2},B^*A^2,BB^*.\label{eq155}
\end{align}
Recall that all of these monomers are equal to a constant and a way to express $\tau$. If we remove all the repeats, we get the following list of distinct monomers,
\begin{align}
AA^*,BB^*,\frac{A^2}{B},A^2B^*,\frac{(A^*)^2}{B^*},(A^*)^2B,\frac{B}{A^2},\frac{B^*}{(A^*)^2}.\label{eq156}
\end{align}
Since we can manipulate each of them by applying any function to them and still get a valid monomer, some of these can be expressed by combining other ones, for example $\frac{B}{A^2}\cdot A^2B^*=BB^*$. In this case, we would disregard $BB^*$ or any of the 3 monomers from that equation. Hence, this list is reducible. The goal is then to reduce this list (\ref{eq156}) such that we get a unique list of monomers that is irreducible. In theory, we could be satisfied with the current list already, but we wish to make our lives easier and get the simplest possible solution. The most obvious way to reduce this list is to go through the monomers one by one and see whether it can be expressed as a product with different exponents of the other ones. For example, let's take $BB^*$.
\begin{align}
(BB^*)^b&=(AA^*)^{a_1}\left(\frac{A^2}{B}\right)^{a_2}\left( A^2B^*\right)^{a_3}\left(\frac{(A^*)^2}{B^*}\right)^{a_4}\left((A^*)^2B\right)^{a_5}\left(\frac{B}{A^2}\right)^{a_6}\left(\frac{B^*}{(A^*)^2}\right)^{a_7}.\label{eq157}
\end{align}
We collect the exponent on both sides, compare and get the system for the exponents
\begin{align}
\left(
\begin{array}{cccccccc}
 1 & 2 & 2 & 0 & 0 & -2 & 0 & 0 \\
 1 & 0 & 0 & 2 & 2 & 0 & -2 & 0 \\
 0 & -1 & 0 & 0 & 1 & 1 & 0 & -1 \\
 0 & 0 & 1 & -1 & 0 & 0 & 1 & -1 \\
\end{array}
\right)\begin{pmatrix}
a_1\\a_2\\a_3\\a_4\\a_5\\a_6\\a_7\\b
\end{pmatrix}=\begin{pmatrix}
0\\0\\0\\0
\end{pmatrix}.\label{eq158}
\end{align}
The nullspace of this matrix is
\begin{align}
\text{Null}=\left\{\left(
\begin{array}{c}
 0 \\
 -1\\
 1 \\
 0\\
 0\\
 0 \\
 0\\
 1 \\
\end{array}
\right),\left(\begin{array}{c}
 2  \\
 0  \\
 -1  \\
 0  \\
 0  \\
 0  \\
 1  \\
 0  \\
\end{array}
\right),\left(\begin{array}{c}
 0  \\
 1  \\
 0 \\
 0 \\
 0\\
 1  \\
 0  \\
 0  \\
\end{array}
\right),\left(\begin{array}{c}
 -2  \\
 1 \\
 0 \\
 0  \\
 1  \\
 0 \\
 0 \\
 0  \\
\end{array}
\right),\left(\begin{array}{c}
 -2 \\
  0 \\
1 \\
 1 \\
 0 \\
 0 \\
0 \\
0 \\
\end{array}
\right)\right\}.\label{eq159}
\end{align}
The first vector is the only one containing $b=1$, so in the process of constructing a combination of these vectors in the nullspace, the first vector must be included, otherwise, $b=0$ and the left-hand side in (\ref{eq157}) vanishes.  However, we are not interested in constructing any combinations to get a solution for the problem (\ref{eq157}). We are only interested in whether or not it is possible to express $BB^*$ as a combination of the rest. In other words, whether $\text{dim}(\text{Null})\neq 0$ together with the existence of such a vector where $b\neq 0$. If the nullspace dimension is bigger than zero and there is a vector with $b\neq 0$, the list is still reducible by removing the monomer in question. In this case we have $\text{dim}(\text{Null})=5$ and the first vector has $b=1$, thus we can remove $BB^*$ from the list (\ref{eq156}). And the process goes on. We remove $BB^*$ and repeat the process for the next monomer, but now with a shorter list. At the end, we find that the list $AA^*,\frac{B}{A^2},A^2B^*$ is irreducible. Of course, one could arrive at a different list based on the choices which monomer to remove. For example from $\frac{B}{A^2}\cdot A^2B^*=BB^*$ which corresponds only to the first vector from the nullspace without the rest, we could remove $\frac{B}{A^2}$ and not $BB^*$, which would still constitute a valid choice. However, in the end, we would still have an irreducible list of monomers of length 3, although a different one. We argue that it doesn't matter which one we end up with. Let us take our choice $AA^*,\frac{B}{A^2},A^2B^*$. Each one of them is producing $\tau$. From the first monomer, we get $AA^*=c_1(\tau)\Rightarrow \tau=c_1^{-1}(AA^*)$ for some function $c_1$. From the second and the third we get a similar result $\tau=c_2^{-1}\left(\frac{B}{A^2}\right)$ and $\tau=c_3^{-1}(A^2B^*)$, for some functions $c_2,c_3$. The variable $\tau$ thus depends on all of these 3 monomers at the same time. We can then write
\begin{align}
\tau=f\left(AA^*,\frac{B}{A^2},A^2B^*\right),\label{eq160}
\end{align}
for some arbitrary function $f$ that depends on 3 variables. If we chose another irreducible list of monomers, they would find their way as the arguments of the function $f$. This being an arbitrary function, it could scramble up the monomers and produce any of the irreducible lists, or even the longer reducible lists. That is why it doesn't matter which list we choose.

The variable $\tau$ is taken care of. The variable $s$ is also necessary to express in terms of the original variables $A,A^*,\ldots$. For this, we can use any of the 4 equations (\ref{eq142})-(\ref{eq146}). The question is, which one to choose. As it turns out, the way we obtained $\tau$, we also made this choice irrelevant. Let's express $s$ both from (\ref{eq142}) and (\ref{eq144}). We get $e^{-s}=A/a=\sqrt{B/b}\Rightarrow \frac{A^2}{B}=d(\tau)$ for some function $d$ which is exactly one of the monomers being a function of $\tau$. In fact, if we express $s$ from any of the 4 equations, they are all equivalent. To illustrate this point even further, we compute the solution using 2 ways to express $s$ and compare. Let us first use equation (\ref{eq142}).
\begin{align}
A&=a(\tau)e^{-s},\nonumber\\
s&=\Log(a)-\Log(A).\label{eq161}
\end{align}
Then using it in the solution (\ref{eq149}), we get
\begin{align}
\alpha_1&=\varphi(\tau)\frac{A}{a(\tau)}+\frac{1}{2}A^*e^{-s}Be^{2s}\left(\Log(a(\tau))-\Log(A)\right)e^{-s}\nonumber\\
&=f_1(\tau)A+\frac{1}{2}A^*B\Log(a(\tau))-\frac{1}{2}A^*B\Log(A)=f_1(\tau)A+f_2(\tau)A^*B-\frac{1}{2}A^*B\Log(A)\nonumber\\
&=f_3\left(AA^*,\frac{B}{A^2},A^2B^*\right)A+f_4\left(AA^*,\frac{B}{A^2},A^2B^*\right)A\left(\frac{B}{A^2}AA^*\right)-\frac{1}{2}A^*B\Log(A)\nonumber\\
&=f_3\left(AA^*,\frac{B}{A^2},A^2B^*\right)A+f_5\left(AA^*,\frac{B}{A^2},A^2B^*\right)A-\frac{1}{2}A^*B\Log(A)\nonumber\\
&=g_1\left(AA^*,\frac{B}{A^2},A^2B^*\right)A-\frac{1}{2}A^*B\Log(A),\label{eq162}
\end{align}
for some functions $f_1,\ldots,g_1$. Next, we obtain the solution in a similar manner but now using equation (\ref{eq144}) to express $s$. The expression for $s$ now becomes $s=\frac{1}{2}\left(\Log(b)-\Log(B)\right)$ and the solution reads
\begin{align}
\alpha_2&=\varphi(\tau)\sqrt{\frac{B}{b(\tau)}}+\frac{1}{4}A^*e^{-s}Be^{2s}\left(\Log(b(\tau))-\Log(B)\right)e^{-s}\nonumber\\
&=f_1(\tau)\sqrt{B}+\frac{1}{4}A^*B\Log(b(\tau))-\frac{1}{4}A^*B\Log(B)=f_1(\tau)\sqrt{B}+f_2(\tau)A^*B-\frac{1}{4}A^*B\Log(B)\nonumber\\
&=f_1(\tau)A\sqrt{\frac{B}{A^2}}+f_2(\tau)A\left(\frac{B}{A^2}AA^*\right)-\frac{1}{4}A^*B\Log(B)=f_3(\tau)A+f_4(\tau)A-\frac{1}{4}A^*B\Log(B)\nonumber\\
&=g_2\left(AA^*,\frac{B}{A^2},A^2B^*\right)A-\frac{1}{4}A^*B\Log(B),\label{eq163}
\end{align}
for some functions $f_1,\ldots,g_2$. The first terms are essentially the same in both solutions (\ref{eq162}) and (\ref{eq163}), but the second terms are not. However, using the functions $g_1,g_2$, these solutions can be made equal. Comparing them both we get
\begin{align}
g_1\left(AA^*,\frac{B}{A^2},A^2B^*\right)A-\frac{1}{2}A^*B\Log(A)&=g_2\left(AA^*,\frac{B}{A^2},A^2B^*\right)A-\frac{1}{4}A^*B\Log(B),\nonumber\\
g_3\left(AA^*,\frac{B}{A^2},A^2B^*\right)A&=\frac{1}{2}A^*B\left(\Log(A)-\frac{1}{2}\Log(B)\right),\nonumber\\
g_3\left(AA^*,\frac{B}{A^2},A^2B^*\right)&=\frac{1}{4}\frac{A^*B}{A}\Log\left(\frac{A^2}{B}\right)=\frac{1}{4}AA^*\frac{B}{A^2}\Log\left(\frac{1}{\frac{B}{A^2}}\right),\label{eq164}
\end{align}
which shows that the function $g_3$ can be used to transform $\alpha_1$ into $\alpha_2$ writing $g_1\longrightarrow g_4+\frac{1}{4}\frac{A^*B}{A}\Log\left(\frac{A^2}{B}\right)$ for some function $g_4$.
\begin{align}
\alpha_1&=g_1\left(AA^*,\frac{B}{A^2},A^2B^*\right)A-\frac{1}{2}A^*B\Log(A)=g_4A+\frac{1}{4}A\frac{A^*B}{A}\Log\left(\frac{A^2}{B}\right)-\frac{1}{2}A^*B\Log(A)\nonumber\\
&=g_4A+\frac{1}{2}A^*B\Log(A)-\frac{1}{4}A^*B\Log(B)-\frac{1}{2}A^*B\Log(A)\nonumber\\
&=g_4\left(AA^*,\frac{B}{A^2},A^2B^*\right)A-\frac{1}{4}A^*B\Log(B)=\alpha_2,\label{eq165}
\end{align}
which concludes that both $s$ and $\tau$ are consistently derived even when starting with more than 2 equations for them. The final solution to (\ref{eq141}) is thus
\begin{align}
\alpha(A,A^*,B,B^*)=f\left(AA^*,\frac{B}{A^2},A^2B^*\right)A-\frac{1}{2}A^*B\Log(A),\label{eq166}
\end{align}
for any function $f$.

The obtained solution (\ref{eq166}) contains a logarithm of the amplitude $A$ and we would like to avoid this. Since it solves the $\varepsilon^1$ term in (\ref{eq139}) for the $A$ amplitude equation, we declare the term $-\frac{1}{2}iA^*B$ to be a core term. Similarly, we set the $\varepsilon^1$ order for the $B$ amplitude equation (\ref{eq140}) to be zero and solve for $\beta$.
\begin{align}
\beta(A,A^*,B,B^*)=g\left(AA^*,\frac{B}{A^2},A^2B^*\right)A^2,\label{eq167}
\end{align}
for some function $g$. This removes the whole $\varepsilon^1$ order. 

The next step is to use these free functions $f$ and $g$ to remove the $\varepsilon^2$ order terms up to the core terms in both amplitude equations. Notice, that the $B$ amplitude equation doesn't have any core terms yet. Since all the terms not containing the homogeneous functions $\alpha,\ldots,\delta$ are polynomial expression of the amplitudes, we search assume the following form of the free functions $f,g$.
\begin{align}
f\left(AA^*,\frac{B}{A^2},A^2B^*\right)&=\sum_{i,j,k=0}^1 a_{i,j,k}\left(AA^*\right)^i\left(\frac{B}{A^2}\right)^j\left(A^2B^*\right)^k,\label{eq168}\\
g\left(AA^*,\frac{B}{A^2},A^2B^*\right)&=\sum_{i,j,k=0}^1 b_{i,j,k}\left(AA^*\right)^i\left(\frac{B}{A^2}\right)^j\left(A^2B^*\right)^k,\label{eq169}
\end{align}
where the constants $a_{i,j,k},b_{i,j,k}$ are complex. The bounds for the summation indices were chosen such that it is sufficient to remove the non-desired terms in the current order of $\varepsilon^2$. Using these functions in the second orders, they turn to
\begin{align}
A:&\quad-\frac{1}{2} i \left(A^*\right)^4 A B^2 \left(a_{1,0,1}\right){}^*-\frac{1}{2} i \left(A^*\right)^3 B^2 \left(a_{0,0,1}\right){}^*-\frac{i B B^* \left(a_{0,1,0}\right){}^*}{2 A^*}-\frac{1}{2} i \left(A^*\right)^2 A B^2 B^* \left(a_{1,1,1}\right){}^*\nonumber\\
&+\frac{1}{2} i A^* B^2 B^* a_{0,1,1}-\frac{1}{2} i A^* B^2 B^* \left(a_{0,1,1}\right){}^*-\frac{1}{2} i \left(A^*\right)^2 A B \left(a_{1,0,0}\right){}^*+\frac{1}{2} i A^* B a_{0,0,0}-\frac{1}{2} i A^* B \left(a_{0,0,0}\right){}^*\nonumber\\
&+a_{1,0,0} \left(i A \left(A^*\right)^2 B-\frac{1}{2} i A^3 B^*\right)+a_{1,1,1} \left(i A \left(A^*\right)^2 B^2 B^*-\frac{1}{2} i A^3 B \left(B^*\right)^2\right)-\frac{i A^* B^2 a_{0,1,0}}{2 A^2}\nonumber\\
&+\frac{3}{2} i A^* A^2 B B^* a_{0,0,1}+a_{1,0,1} \left(2 i A^3 \left(A^*\right)^2 B B^*-\frac{1}{2} i A^5 \left(B^*\right)^2\right)-\frac{1}{2} i A B B^* a_{1,1,0}-\frac{1}{2} i A B B^* \left(a_{1,1,0}\right){}^*\nonumber\\
&-\frac{1}{2} i \left(A^*\right)^2 A B b_{1,1,0}-\frac{1}{2} i A^* B b_{0,1,0}-\frac{1}{2} i \left(A^*\right)^2 A^5 B^* b_{1,0,1}-\frac{1}{2} i A^* A^4 B^* b_{0,0,1}-\frac{1}{2} i \left(A^*\right)^2 A^3 B B^* b_{1,1,1}\nonumber\\
&-\frac{1}{2} i \left(A^*\right)^2 A^3 b_{1,0,0}-\frac{1}{2} i A^* A^2 B B^* b_{0,1,1}-\frac{1}{2} i A^* A^2 b_{0,0,0}+i A^* \frac{\partial \gamma }{\partial A^*}-\frac{7}{4} i A^* A^2+\frac{67}{16} i A B B^*-i A \frac{\partial \gamma }{\partial A}\nonumber\\
&+2 i B^* \frac{\partial \gamma }{\partial B^*}-2 i B \frac{\partial \gamma }{\partial B}+i \gamma,\label{eq170}\\
B:&\quad i A^* A B^2 B^* b_{0,1,1}+i A^* A B b_{0,0,0}+2 i A^* A^3 B B^* b_{0,0,1}+b_{1,1,0} \left(\frac{1}{2} i \left(A^*\right)^2 B^2-\frac{1}{2} i A^2 B B^*\right)\nonumber\\
&+b_{1,0,1} \left(\frac{5}{2} i A^4 \left(A^*\right)^2 B B^*-\frac{1}{2} i A^6 \left(B^*\right)^2\right)+b_{1,0,0} \left(\frac{3}{2} i A^2 \left(A^*\right)^2 B-\frac{1}{2} i A^4 B^*\right)\nonumber\\
&+b_{1,1,1} \left(\frac{3}{2} i A^2 \left(A^*\right)^2 B^2 B^*-\frac{1}{2} i A^4 B \left(B^*\right)^2\right)+\frac{1}{6} i A^* A B+i A^* \frac{\partial \delta }{\partial A^*}-i A \frac{\partial \delta }{\partial A}+2 i B^* \frac{\partial \delta }{\partial B^*}\nonumber\\
&-2 i B \frac{\partial \delta }{\partial B}+2 i \delta.\label{eq171}
\end{align}
In both expressions, the terms we are interested to remove are those that stand without the constants $a,b$ since the other ones can be removed just by setting $a$ and $b$ to zero. From the $A$-equation, the unwanted terms are $A^*A^2$ and $ABB^*$. The $B$-equation includes the term $A^*AB$. Before we do anything, it should be checked if some of these terms cannot be removed using the function $\gamma$ itself. Remember, the homogeneous functions play an active role both in removing terms in the current order, not only the next one. If we set the $\gamma$ terms together with the unwanted terms to zero and solve it, as we did with (\ref{eq141}), we get
\begin{align}
\gamma=h\left(A A^*,\frac{B}{A^2},A^2 B^*\right)A-\frac{7}{4} A^* A^2 \Log (A)+\frac{67}{16} A B B^* \Log (A),\label{eq172}
\end{align}
for some function $h$. Observe that both terms are unwanted and cannot be removed using $\gamma$. The constants $a,b$ need to be used in order to remove them. We therefore collect all the factors of the terms $A^*A^2,ABB^*$.
\begin{align}
A B B^* \left(-\frac{1}{2} i a_{1,1,0}-\frac{1}{2} i a^*_{1,1,0}+\frac{67 i}{16}\right),\label{eq173}\\
A^2 A^* \left(-\frac{1}{2} i b_{0,0,0}-\frac{7i}{4}\right).\label{eq174}
\end{align}
It is easy to see, that these terms vanish if we choose
\begin{align}
a_{1,1,0}&=\frac{67}{16},b_{0,0,0}=-\frac{7}{2}.\label{eq175}
\end{align}
This choice, in turn, affects the $\varepsilon^2$ order in the $B$-equation (\ref{eq171}). Substituting (\ref{eq175}) into (\ref{eq171}) we find that the factor of the unwanted term $A^*AB$ doesn't have any constants $a,b$, so it cannot be removed and neither using the function $\delta$. However, the $B$-equation doesn't have any core terms yet. Therefore we declare the term $A^*AB$ to be the core of the $B$-equation. All the rest of the order can be removed by setting the rest of the constants $a,b$ to be zero. We also need to solve for $\delta$, but we just set it to zero, as well as the function $h$ in (\ref{eq172}) since we don't have any higher orders. At the end, the amplitude equations (\ref{eq139}), (\ref{eq140}) become
\begin{align}
A'&=i A-\frac{1}{2} i \varepsilon  A^* B,\label{eq176}\\
B'&=2 i B-\frac{10}{3} i \varepsilon ^2 AA^*B,\label{eq177}
\end{align}
together with the homogeneous functions
\begin{align}
\alpha&=\frac{67}{16}A^*B,\label{eq178}\\
\beta&=-\frac{7}{2}A^2,\label{eq179}\\
\gamma&=0,\label{eq180}\\
\delta&=0.\label{eq181}
\end{align}
Remembering the amplitude transformations $\tilde{A}=Ae^{it},\tilde{B}=Be^{2it}$, the final form of the amplitude equations are
\begin{align}
A'&=-\frac{1}{2} i \varepsilon  A^* B,\label{eq182}\\
B'&=-\frac{10}{3} i \varepsilon ^2 AA^*B,\label{eq183}
\end{align}
where we dropped the tilde signs. Using (\ref{eq178})-(\ref{eq181}),  the solution (\ref{eq138}) turns to
\begin{align}
x(t)&=A e^{i t}-B e^{2 i t}+\varepsilon  \left(3 A^2 e^{2 i t}+\frac{67}{16} A^* B e^{i t}+A^* A-\frac{1}{8} A B e^{3 i t}+\frac{2}{15} B^2 e^{4 i t}-2 B B^*\right)\nonumber\\
&+\varepsilon ^2 \left(\frac{1}{2} A^3 e^{3 i t}+\frac{171 A^2 B^*}{16}-\frac{17}{20} A^2 B e^{4 i t}-\frac{1159 A^* B^2 e^{3 i t}}{1920}-\frac{275}{48} A^* A B e^{2 i t}+\frac{171}{16} \left(A^*\right)^2 B\right.\nonumber\\
&\left.-\frac{17 A B^2 e^{5 i t}}{2880}-\frac{2}{525} B^3 e^{6 i t}+\frac{29}{15} B^2 B^* e^{2 i t}\right)+(*),\nonumber\\
y(t)&=A e^{i t}+2 B e^{2 i t}+\varepsilon  \left(-7 A^2 e^{2 i t}+\frac{67}{16} A^* B e^{i t}+A^* A-\frac{1}{8} A B e^{3 i t}+\frac{2}{15} B^2 e^{4 i t}-2 B B^*\right)\nonumber\\
&+\varepsilon ^2 \left(\frac{1}{2} A^3 e^{3 i t}+\frac{171 A^2 B^*}{16}-\frac{17}{20} A^2 B e^{4 i t}-\frac{1159 A^* B^2 e^{3 i t}}{1920}+\frac{171}{16} \left(A^*\right)^2 B-\frac{17 A B^2 e^{5 i t}}{2880}\right.\nonumber\\
&\left.-\frac{2}{525} B^3 e^{6 i t}\right)+(*).\label{eq184}
\end{align}

\setcounter{equation}{0}
\numberwithin{equation}{section}
\section*{Appendix D, Selkov model (Hopf bifurcation)}
\renewcommand{\theequation}{D.\arabic{equation}}
The RG solution to the system (\ref{eq193}) with homogeneous solutions introduced at each order is
\begin{align}
u&=\omega  Ae^{i t \omega }+\varepsilon\left[\frac{1}{3} \sqrt{2} \left(2 i \omega ^2+2 \omega -i\right) \sqrt{\omega ^2+1} A^2 e^{2 i t \omega }+\frac{\left(2 \omega ^2-i \omega -1\right) \sqrt{\omega ^2+1} A e^{i t \omega }}{\sqrt{2} \omega  (\omega -i)}+\omega  \alpha\right]\nonumber\\
&+\varepsilon^2\left[\frac{A e^{i t \omega }}{12 \omega ^3 (\omega -i) \sqrt{\omega ^2+1}}\left(8 \sqrt{2} \alpha  (\omega -i)^2 \left(2 i \omega ^3+i \omega +1\right) \omega ^3\right.\right.\nonumber\\
&\left.\left.+2 A A^* \left(8 i \omega ^6-6 \omega ^5+2 i \omega ^4+6 \omega ^3-7 i \omega ^2+3 \omega +2 i\right) \sqrt{\omega ^2+1} \omega ^2\right.\right.\nonumber\\
&\left.\left.+3 \left(-4 i \omega ^6-4 \omega ^5+3 i \omega ^4-2 \omega ^3-2 i \omega ^2+4 \omega +i\right) \sqrt{\omega ^2+1}\right)\right.\nonumber\\
&\left.+\frac{A^2 \left(4 \omega ^6+2 i \omega ^5+7 i \omega ^3+15 \omega ^2-i \omega -5\right) e^{2 i t \omega }}{9 \omega ^2}\right.\nonumber\\
&\left.+\frac{A^3 \left(-8 \omega ^6+14 i \omega ^5+6 \omega ^4+10 i \omega ^3+15 \omega ^2-7 i \omega -2\right) e^{3 i t \omega }}{8 \omega }\right]+\varepsilon^3(\ldots)+\varepsilon^4(\ldots)+(*).\nonumber\\
v&=A (-\omega +i) e^{i t \omega }+\varepsilon\left[\frac{A^2 \sqrt{\omega ^2+1} \left(-4 i \omega ^3-6 \omega ^2+4 i \omega +1\right) e^{2 i t \omega }}{3 \sqrt{2} \omega }-\alpha  (\omega -i)\right]\nonumber\\
&+\varepsilon^2\left[\frac{\sqrt{2} A \sqrt{\omega ^2+1} \left(\alpha ^* \left(6 \omega ^2-3\right)+\alpha  \left(-4 i \omega ^3-6 \omega ^2+4 i \omega +1\right)\right) e^{i t \omega }}{3 \omega }\right.\nonumber\\
&\left.+\frac{A^2 \left(-4 \omega ^7+6 i \omega ^6+8 \omega ^5-10 i \omega ^4-14 \omega ^3+i \omega ^2+\omega -i\right) e^{2 i t \omega }}{9 \omega ^3}\right.\nonumber\\
&\left.+\frac{A^3 \left(24 \omega ^7-50 i \omega ^6-32 \omega ^5-24 i \omega ^4-55 \omega ^3+36 i \omega ^2+13 \omega -2 i\right) e^{3 i t \omega }}{24 \omega ^2}\right]+\varepsilon^3(\ldots)+\varepsilon^4(\ldots)+(*),\label{eq195}
\end{align}
where the $\varepsilon^3$ and the $\varepsilon^4$ order terms were omitted. Each order of $\varepsilon$ contains one homogeneous function. In total, there are 4 of them: $\alpha,\beta,\gamma$ and $\delta$. The 0-th order part of the system (\ref{eq193}) has eigenvalues $\pm i\omega$ and the corresponding eigenvectors $(\omega,\pm i-\omega)$. Therefore, solution was transformed using $\tilde{A}=Ae^{i\omega t}$.

The amplitude equation we get, reads
\begin{align}
A'(t)&=i A \omega+\varepsilon\left[i \alpha  \omega +i A^* \omega  \frac{\partial \alpha }{\partial A^*}-i A \omega  \frac{\partial \alpha }{\partial A}+\frac{A \left(-2 \omega ^2+i \omega +1\right) \sqrt{\omega ^2+1}}{\sqrt{2} \omega }\right]+\varepsilon^2\frac{1}{12 \omega ^3 \sqrt{\omega ^2+1}}\nonumber\\
&\left[6 \omega ^2 \frac{\partial \alpha }{\partial A^*} \left(2 i \alpha ^* \sqrt{\omega ^2+1} \omega ^2+A^* \left(-2 i \sqrt{\omega ^2+1} \omega ^2 \frac{\partial \alpha ^*}{\partial A^*}-2 i \sqrt{\omega ^2+1} \omega ^2 \frac{\partial \alpha }{\partial A}\right.\right.\right.\nonumber\\
&\left.\left.\left.+\sqrt{2} \left(2 \omega ^4+i \omega ^3+\omega ^2+i \omega -1\right)\right)+2 i A \sqrt{\omega ^2+1} \omega ^2 \frac{\partial \alpha ^*}{\partial A}\right)\right.\nonumber\\
&\left.+A \left(-i \sqrt{\omega ^2+1} \left(16 A A^* \omega ^8+12 i A A^* \omega ^7+4 \left(A A^*+3\right) \omega ^6-12 i \left(A A^*-1\right) \omega ^5\right.\right.\right.\nonumber\\
&\left.\left.\left.+\left(3-14 A A^*\right) \omega ^4-6 i \left(A A^*+2\right) \omega ^3+4 \left(A A^*-3\right) \omega ^2-6 i \omega +3\right)+12 i \sqrt{\omega ^2+1} \omega ^4 \left(\frac{\partial \alpha }{\partial A}\right)^2\right.\right.\nonumber\\
&\left.\left.+6 \sqrt{2} \left(2 \omega ^4-i \omega ^3+\omega ^2-i \omega -1\right) \omega ^2 \frac{\partial \alpha }{\partial A}\right)\right.\nonumber\\
&\left.-6 \alpha  \left(2 i \sqrt{\omega ^2+1} \omega ^4 \frac{\partial \alpha }{\partial A}+\sqrt{2} \left(2 \omega ^4-i \omega ^3+\omega ^2-i \omega -1\right) \omega ^2\right)+i \beta  \omega+i A^* \omega  \frac{\partial \beta }{\partial A^*}-i A \omega  \frac{\partial \beta }{\partial A}\right]\nonumber\\
&+\varepsilon^3\left[\ldots+i \gamma  \omega+i A^* \omega  \frac{\partial \gamma }{\partial A^*}-i A \omega  \frac{\partial \gamma }{\partial A}\right]+\varepsilon^4\left[\ldots+i \delta  \omega+i A^* \omega  \frac{\partial \delta }{\partial A^*}-i A \omega  \frac{\partial \delta }{\partial A}\right],\label{eq196}
\end{align}
where the $\varepsilon^3$ and the $\varepsilon^4$ order terms were omitted.

The modified RG method is now ready to be applied. Let us investigate the $\varepsilon$ order term of the amplitude equation (\ref{eq196}). It is nice and linear and doesn't contain any term that could be assigned as the core. Thus the function $\alpha(A,A^*)$ can be still held as free and used later if necessary, but we determine its form first. Let us set the $\varepsilon$ order to zero.
\begin{align}
i \alpha  \omega +i A^* \omega  \frac{\partial \alpha }{\partial A^*}-i A \omega  \frac{\partial \alpha }{\partial A}+\frac{A \left(-2 \omega ^2+i \omega +1\right) \sqrt{\omega ^2+1}}{\sqrt{2} \omega }=0.\label{eq197}
\end{align}
This equation is of the same form as (\ref{eq25}) or (\ref{eq127}) and the solution can be obtained in the same way as before. It is easy to see that the last term which is proportional to $A$ is a solution to the homogeneous part of the equation and, hence, is a source of logarithm of $A$. Therefore we exclude it form the equation and solve the rest. Skipping the details, the solution is
\begin{align}
\alpha(A,A^*)=c\left(AA^*\right)A=A\left( c_0+c_1AA^*\right),\label{eq198}
\end{align}
where $c$ is an arbitrary function and we assumed a polynomial up to the first order. $c_0$ and $c_1$ are complex constants. The $\varepsilon^2$ order term then becomes
\begin{align}
&i \beta  \omega +i A^* \omega  \frac{\partial \beta }{\partial A^*}-i A \omega  \frac{\partial \beta }{\partial A}+A^2A^*\left(-\frac{8 i \omega ^6-6 \omega ^5+2 i \omega ^4+6 \omega ^3-7 i \omega ^2+3 \omega +2 i}{6 \omega }\right.\nonumber\\
&\left.+\frac{\sqrt{2} \left(2 \omega ^4+\omega ^2-1\right)}{\omega  \sqrt{\omega ^2+1}}c_1\right)-A\frac{i \left(4 \omega ^6+4 i \omega ^5+\omega ^4-4 i \omega ^3-4 \omega ^2-2 i \omega +1\right)}{4 \omega ^3}=0.\label{eq199}
\end{align}
Here again, both the inhomogeneous terms are solutions to the homogeneous part and so they produce logarithms. But only the term proportional to $A^2A^*$ is nonlinear and is suitable to be called the core of the amplitude equation. Therefore it will be not removed and we solve only the homogeneous part once more and get a similar expression as before.
\begin{align}
\beta(A,A^*)=d\left(AA^*\right)A=A\left( d_0+d_1AA^*\right),\label{eq200}
\end{align}
where $d$ is an arbitrary function and we assumed a polynomial up to the first order. $d_0$ and $d_1$ are complex constants. We proceed to the next order term $\varepsilon^3$ in (\ref{eq196}) and get
\begin{align}
&i \gamma  \omega +i A^* \omega  \frac{\partial \gamma }{\partial A^*}-i A \omega  \frac{\partial \gamma }{\partial A}+A^3\left(A^*\right)^2\left(c_1^2 \left(-\frac{2 \sqrt{2} \omega }{\sqrt{\omega ^2+1}}+\frac{2 \sqrt{2}}{\omega  \sqrt{\omega ^2+1}}-\frac{4 \sqrt{2} \omega ^3}{\sqrt{\omega ^2+1}}\right)\right.\nonumber\\
&\left.+c_1 \left(\left(c_1\right){}^* \left(-\frac{\sqrt{2} \omega }{\sqrt{\omega ^2+1}}+\frac{\sqrt{2}}{\omega  \sqrt{\omega ^2+1}}-\frac{2 \sqrt{2} \omega ^3}{\sqrt{\omega ^2+1}}\right)-\frac{1}{3} 4 i \omega ^5-\omega ^4-\frac{i \omega ^3}{3}+\omega ^2+\frac{7 i \omega }{6}-\frac{i}{3 \omega }+\frac{1}{2}\right)\right.\nonumber\\
&\left.+\left(c_1\right){}^* \left(-\frac{1}{3} 4 i \omega ^5+\omega ^4-\frac{i \omega ^3}{3}-\omega ^2+\frac{7 i \omega }{6}-\frac{i}{3 \omega }-\frac{1}{2}\right)\right)+\frac{A^2A^*}{18 \sqrt{2} \omega ^3 \sqrt{\omega ^2+1}}\left(-208 \omega ^{10}+336 i \omega ^9\right.\nonumber\\
&\left.-264 \omega ^8-18 i \omega ^7+332 \omega ^6-279 i \omega ^5+167 \omega ^4-15 i \omega ^3-183 \omega ^2+54 i \omega +38+c_1 \left(-2 \omega ^2+\frac{1}{\omega ^2}+2\right)\right.\nonumber\\
&\left.+c_0 \left(-c_1\frac{\sqrt{2} \left(2 \omega ^4+\omega ^2-1\right)}{\omega  \sqrt{\omega ^2+1}}-\frac{8 i \omega ^6-6 \omega ^5+2 i \omega ^4+6 \omega ^3-7 i \omega ^2+3 \omega +2 i}{6 \omega }\right)\right.\nonumber\\
&\left.+d_1\frac{\sqrt{2} \left(2 \omega ^4+\omega ^2-1\right)}{\omega  \sqrt{\omega ^2+1}}-c_0^*\frac{8 i \omega ^6-6 \omega ^5+2 i \omega ^4+6 \omega ^3-7 i \omega ^2+3 \omega +2 i}{6 \omega }\right)\nonumber\\
&+A\frac{i \left(12 \omega ^8+\omega ^6-15 \omega ^4-\omega ^2+3\right)}{4 \sqrt{2} \omega ^4 \sqrt{\omega ^2+1}}=0.\label{eq201}
\end{align}
At this current order, we need to remove some terms, namely $A^3\left(A^*\right)^2$ using $c_1=0$. The term $A^2A^*$ can also be removed, although it is not necessary since it is the core term. However, in many cases, it may be wiser to refrain from using the free constants to remove terms that don't need to be removed, as later on, these constants can be useful. In any case, we will solve (\ref{eq201}) taking only the homogeneous part, since the inhomogeneous parts are either linear, core terms, or nonlinear terms that we have removed. The solution is again similar to the previous ones.
\begin{align}
\gamma(A,A^*)=e\left(AA^*\right)A=A\left( e_0+e_1AA^*+e_2\left(AA^*\right)^2\right),\label{eq202}
\end{align}
where $e$ is an arbitrary function and we assumed a polynomial up to the second order and $e_0,e_1$ and $e_2$ are complex constants. We decided to go up one more order in this polynomial because it is needed based on the number of terms in the next order in $\varepsilon$. Let us proceed with the last order $\varepsilon^4$ with $c_1=0$.
\begin{align}
&i \delta  \omega +i A^* \omega  \frac{\partial \delta }{\partial A^*}-i A \omega  \frac{\partial \delta }{\partial A}+A^3\left(A^*\right)^2\left(\frac{1}{432 \omega ^3 \left(\omega ^2+1\right)}\left(-320 i \omega ^{14}+1920 \omega ^{13}-2964 i \omega ^{12}+480 \omega ^{11}\right.\right.\nonumber\\
&\left.\left.+848 i \omega ^{10}-4272 \omega ^9+2560 i \omega ^8-1344 \omega ^7+1147 i \omega ^6+1992 \omega ^5-394 i \omega ^4+264 \omega ^3-669 i \omega ^2\right.\right.\nonumber\\
&\left.\left.-240 \omega +76 i\right)-\left( d_1+d_1^*\right)\frac{i \left(8 \omega ^6-6 i \omega ^5+2 \omega ^4+6 i \omega ^3-7 \omega ^2+3 i \omega +2\right)}{6 \omega }\right.\nonumber\\
&\left.+e_2\frac{2 \sqrt{2} \sqrt{\omega ^2+1} \left(2 \omega ^2-1\right)}{\omega }\right)+A^2A^*\left(-\frac{i}{216 \omega ^5 \left(\omega ^2+1\right)}\left(8672 \omega ^{14}+9024 i \omega ^{13}+5872 \omega ^{12}-2664 i \omega ^{11}\right.\right.\nonumber\\
&\left.\left.-6110 \omega ^{10}-16788 i \omega ^9-9003 \omega ^8-1044 i \omega ^7+171 \omega ^6+7830 i \omega ^5+1567 \omega ^4+258 i \omega ^3+901 \omega ^2-924 i \omega\right.\right.\nonumber\\
&\left.\left. +14\right)-\left( d_0+d_0^*\right)\frac{8 i \omega ^6-6 \omega ^5+2 i \omega ^4+6 \omega ^3-7 i \omega ^2+3 \omega +2 i}{6 \omega }+d_1\left(-2 \omega ^2+\frac{1}{\omega ^2}+2\right)\right.\nonumber\\
&\left.+e_1\frac{\sqrt{2} \sqrt{\omega ^2+1} \left(2 \omega ^2-1\right)}{\omega }-\frac{c_0^*}{18 \sqrt{2} \omega ^3 \sqrt{\omega ^2+1}}\left(208 \omega ^{10}-336 i \omega ^9+264 \omega ^8+18 i \omega ^7-332 \omega ^6\right.\right.\nonumber\\
&\left.\left.+279 i \omega ^5-167 \omega ^4+15 i \omega ^3+183 \omega ^2-54 i \omega -38\right)+c_0\left[\frac{1}{18 \sqrt{2} \omega ^3 \sqrt{\omega ^2+1}}\left(-208 \omega ^{10}+336 i \omega ^9\right.\right.\right.\nonumber\\
&\left.\left.\left.-264 \omega ^8-18 i \omega ^7+332 \omega ^6-279 i \omega ^5+167 \omega ^4-15 i \omega ^3-183 \omega ^2+54 i \omega +38\right)\right.\right.\nonumber\\
&\left.\left.-c_0^*\frac{8 i \omega ^6-6 \omega ^5+2 i \omega ^4+6 \omega ^3-7 i \omega ^2+3 \omega +2 i}{6 \omega }+d_1\frac{\sqrt{2} \left(1-2 \omega ^2\right) \sqrt{\omega ^2+1}}{\omega }\right]\right)\nonumber\\
&-A\frac{i \left(16 \omega ^{12}+72 \omega ^{10}-59 \omega ^8-60 \omega ^6+30 \omega ^4+8 \omega ^2+1\right)}{32 \omega ^7}=0,\label{eq203}
\end{align}
If we analyse the above equation, we can see that the term $A^3\left(A^*\right)^2$ can be easily removed using $e_2$. Again, we could also remove the term $A^2A^*$ both from $\varepsilon^3$ and $\varepsilon^4$ order, but we choose not to do it. The rest of the free constants are $c_0,d_0,d_1,e_1$. Choosing
\begin{align}
e_2&=\frac{1}{288 \sqrt{2} \omega ^2 \left(1-2 \omega ^2\right)^2 \sqrt{\omega ^2+1}}\left(2432 i \omega ^{14}-1280 \omega ^{13}+120 i \omega ^{12}+1600 \omega ^{11}-3620 i \omega ^{10}+1408 \omega ^9\right.\nonumber\\
&\left.+1052 i \omega ^8-1936 \omega ^7+2162 i \omega ^6+160 \omega ^5-2429 i \omega ^4+328 \omega ^3+861 i \omega ^2-80 \omega -76 i\right),\label{eq204}
\end{align}
we have removed the $A^3\left(A^*\right)^2$ term. The $\varepsilon^4$ order equation then becomes
\begin{align}
&i \delta  \omega +i A^* \omega  \frac{\partial \delta }{\partial A^*}-i A \omega  \frac{\partial \delta }{\partial A}-\frac{i A \left(16 \omega ^{12}+72 \omega ^{10}-59 \omega ^8-60 \omega ^6+30 \omega ^4+8 \omega ^2+1\right)}{32 \omega ^7}\nonumber\\
&+A^2A^*f\left(c_0,d_0,d_1,e_1;\omega\right)=0,\label{eq205}
\end{align}
where the function $f$ represents the factor at the $A^2A^*$ term which we can see in (\ref{eq203}). Since we don't have any next order, we can safely set the last homogeneous function to zero $\delta=0$.

To summarize, with all the choices for the free constants, the homogeneous functions turned to
\begin{align}
\alpha(A,A^*)&=Ac_0,\nonumber\\
\beta(A,A^*)&=A\left(d_0+AA^*d_1\right),\nonumber\\
\gamma(A,A^*)&=A\left(e_0+AA^*e_1\right)+\frac{A^3\left(A^*\right)^2}{864 \sqrt{2} \omega ^2 \left(\omega ^2+1\right)^{3/2} \left(2 \omega ^2-1\right)}\left[d_1\left(576 \omega ^{10}-432 i \omega ^9+720 \omega ^8-360 \omega ^6\right.\right.\nonumber\\
&\left.\left.+648 i \omega ^5-360 \omega ^4+216 i \omega ^3+144 \omega ^2\right)+d_1^*\left(576 \omega ^{10}+432 i \omega ^9+720 \omega ^8-360 \omega ^6-648 i \omega ^5\right.\right.\nonumber\\
&\left.\left.-360 \omega ^4-216 i \omega ^3+144 \omega ^2\right)+320 \omega ^{14}+1920 i \omega ^{13}+2964 \omega ^{12}+480 i \omega ^{11}-848 \omega ^{10}-4272 i \omega ^9\right.\nonumber\\
&\left.-2560 \omega ^8-1344 i \omega ^7-1147 \omega ^6+1992 i \omega ^5+394 \omega ^4+264 i \omega ^3+669 \omega ^2-240 i \omega -76\right],\nonumber\\
\delta(A,A^*)&=0.\label{eq206}
\end{align}
The amplitude equation, which is the our main focus, simplifies to
\begin{align}
A'(t)&=A\left[-\varepsilon\frac{\sqrt{\omega ^2+1} \left(2 \omega ^2-i \omega -1\right)}{\sqrt{2} \omega }-\varepsilon^2\frac{i \left(4 \omega ^6+4 i \omega ^5+\omega ^4-4 i \omega ^3-4 \omega ^2-2 i \omega +1\right)}{4 \omega ^3}\right.\nonumber\\
&\left.+\varepsilon^3\frac{i\left(12 \omega ^8+\omega ^6-15 \omega ^4-\omega ^2+3\right)}{4 \sqrt{2} \omega ^4 \sqrt{\omega ^2+1}}-\varepsilon^4\frac{i \left(16 \omega ^{12}+72 \omega ^{10}-59 \omega ^8-60 \omega ^6+30 \omega ^4+8 \omega ^2+1\right)}{32 \omega ^7}\right]\nonumber\\
&+A^2 A^*\left[\varepsilon^2\frac{ \left(-8 i \omega ^6+6 \omega ^5-2 i \omega ^4-6 \omega ^3+7 i \omega ^2-3 \omega -2 i\right)}{6 \omega }+\varepsilon^3\left(-\frac{1}{18 \sqrt{2} \omega ^3 \sqrt{\omega ^2+1}}\left(208 \omega ^{10}\right.\right.\right.\nonumber\\
&\left.\left.\left.-336 i \omega ^9+264 \omega ^8+18 i \omega ^7-332 \omega ^6+279 i \omega ^5-167 \omega ^4+15 i \omega ^3+183 \omega ^2-54 i \omega -38\right)\right.\right.\nonumber\\
&\left.\left.+g(c_0,d_1)\right)+\varepsilon^4\left(-\frac{i}{216 \omega ^5 \left(\omega ^2+1\right)}\left(8672 \omega ^{14}+9024 i \omega ^{13}+5872 \omega ^{12}-2664 i \omega ^{11}-6110 \omega ^{10}\right.\right.\right.\nonumber\\
&\left.\left.\left.-16788 i \omega ^9-9003 \omega ^8-1044 i \omega ^7+171 \omega ^6+7830 i \omega ^5+1567 \omega ^4+258 i \omega ^3+901 \omega ^2-924 i \omega\right.\right.\right.\nonumber\\
&\left.\left.\left. +14\right)+h(c_0,d_0,d_1,e_1)\right)\right],\label{eq207}
\end{align}
where the amplitude was transformed using $\tilde{A}=Ae^{i\omega t}$ and the tilde sign was dropped. The functions $g$ and $h$ include more terms, but such that $g(0,0)=h(0,0,0,0)=0$. The amplitude equation has now a less complicated form without the $A^3\left(A^*\right)^2$ term.

\end{appendices}
\newpage
\bibliographystyle{unsrt}
\bibliography{paper}

\end{document}